\documentclass[twocolumn,english,prb]{revtex4-2}
\usepackage{color}
\usepackage[utf8]{inputenc}
\usepackage{graphicx}
\usepackage[T1]{fontenc}
\usepackage{float}
\usepackage{textcomp}
\usepackage{amsmath}
\usepackage{amssymb}
\usepackage{graphicx}
\usepackage{esint}
\usepackage{xcolor}
\usepackage{multirow}
\usepackage{dsfont}
\usepackage{physics}
\usepackage{bbold}

\usepackage[colorlinks=true,hyperfootnotes=true,breaklinks=true]{hyperref}
\usepackage[capitalise]{cleveref}

\begin{document}

\title{Topological superconductivity of a two-dimensional electron gas at the (001) LaAlO\textsubscript{3}/SrTiO\textsubscript{3} interface} 

\author{Piotr Żeberek}
\affiliation{AGH University of Krakow, Faculty of Physics and Applied Computer Science, al. A. Mickiewicza 30, 30-059 Krakow, Poland}

\author{Paweł W\'ojcik}
\email{pawel.wojcik@fis.agh.edu.pl}
\affiliation{AGH University of Krakow, Faculty of Physics and Applied Computer Science, al. A. Mickiewicza 30, 30-059 Krakow, Poland}

\begin{abstract}
We investigate the emergence of topological superconductivity and Majorana zero modes in the two-dimensional electron gas formed at the LaAlO$_3$/SrTiO$_3$ (001) interface. Using a realistic multiband tight-binding model that incorporates the $t_{2g}$ orbital structure together with atomic and Rashba spin–orbit couplings, we determine the topological phase diagrams for both fully two-dimensional and quasi-one-dimensional geometries. In the two-dimensional limit, we show that a finite out-of-plane magnetic-field component is required to drive a topological phase transition. In this case, the critical field is strongly band dependent, and for higher-lying bands, it is controlled by the interplay of spin and orbital Zeeman effects, as well as atomic spin–orbit coupling. Although a purely in-plane field is insufficient to induce the topological transition in a full 2D system, we demonstrate that a lateral confinement relaxes this constraint. In this case, the character of the edge modes depends sensitively on the field orientation, with out-of-plane fields producing conventional counterpropagating chiral modes and transverse in-plane fields giving rise to co-propagating antichiral modes. Finally, Majorana zero modes in LAO/STO nanowires with varying widths are analyzed. We demonstrate that subbands predominantly composed of $d_{yz/xz}$ orbitals exhibit exceptionally long localization lengths, which may preclude the observation of Majorana bound states in nanowires of typical experimental dimensions.
\end{abstract}

\date{\today}
\maketitle

\section{Introduction}
In recent years, topological superconductors~\cite{Sato_2017, BernevigHughes2013, FuKane2008,Pientka2013}, which can host Majorana zero modes (MZMs) at their surfaces or edges, have attracted increasing interest owing to the non-Abelian braiding statistics of Majorana states~\cite{Alicea2010}, which are predicted to play a central role in the realization of fault-tolerant topological quantum computers~\cite{Alicea2010,Sau2010}. 

Although in the Kitaev model~\cite{Kitaev2003} the emergence of Majorana states strictly requires superconducting pairing with p‑wave symmetry, which typically emerges only in exotic materials, current research is largely directed toward alternative schemes that exploit conventional s-wave superconductivity in combination with strong spin–orbit (SO) coupling and an external magnetic field~\cite{Lutchyn2010}. Among such approaches, hybrid superconductor–semiconductor heterostructures are widely regarded as one of the most promising and technologically advanced platforms for MZMs engineering ~\cite{Oreg2010,Mourik2012,Lutchyn2010}. 
Note, however, that direct braiding of MZMs in the commonly employed one-dimensional hybrid nanowires is effectively unfeasible, since Majorana quasiparticles, forming a pair of particle and antiparticle, have to remain spatially separated throughout the entire braiding protocol to prevent their anihilation. This fundamental constraint motivates the exploration and design of novel, scalable material platforms capable of hosting extended networks of MZMs~\cite{FuKane2008,FuKane2009,Hell2017, Pientka2017}. In this context, two-dimensional systems seem to be particularly important, as they naturally facilitate the implementation of elementary topological quantum operations (braiding) in a controllable fashion while avoiding undesired collisions between Majoranas~\cite{Hell2017,Pientka2017}.

One particularly promising yet still insufficiently explored class of systems is based on interfaces between transition-metal oxides. The two-dimensional electron gas (2DEG) formed at oxide interfaces, such as LaAlO$_3$/SrTiO$_3$ (LAO/STO) and LaAlO$_3$/KTaO$_3$ (LAO/KTO), is characterized by a unique combination of high carrier mobility \cite{Ohtomo_Hwang_2004}, strong SO coupling \cite{Diez2015,Rout2017,Caviglia2010,Shalom2010,Yin2020,Singh2017,Hurand2015}, superconductivity \cite{Reyren2007, joshua2012universal,maniv2015strong,
Biscaras, Monteiro2019, Monteiro2017, Wojcik_dome_schrodinger_poisson_sc_2024}, magnetic ordering \cite{Karen2012,Li2011,Brinkman2007,Bert2011}, and ferroelectricity \cite{Gastiasoro2020,Kanasugi2018,Kanasugi2019,Honig2013}. The coexistence of these phenomena within a single material platform, as realized in transition metal oxide heterostructures, enables access to a variety of unconventional superconducting phases, including topological superconductivity.

To date, oxide-based 2DEGs have not been employed as platforms for the realization of topological superconductivity, primarily due to the unavoidable presence of interfacial disorder. Note, however, that recent studies demonstrate that appropriate tuning of the stoichiometry of La/Al in the LAO film can drive these systems into a clean limit \cite{Singh2024}. This observation, together with recent reports of superconductivity with a critical temperature $T_c$ of up to 2~K in LAO/KTO heterostructures \cite{Ren2022, Liu2021, Chen2021, Liu2023, Zhang2025,Liu2023, Chen2021_110, Hua2022}, raises the question of the potential use of oxide-based 2DEGs as platforms for the realization of topological superconductivity and MZMs. Notably, recent advances in the fabrication of quantum dots based on the LAO/STO 2DEG \cite{Jespersen2020,Jouan2020} have additionally opened a promising avenue for the implementation of Kitaev chains in these materials. Despite their promising properties, research on oxide interfaces as topological superconductors is still at a very early stage \cite{Fukaya2018,Maiellaro2023,Stornaiuolo2017,GUARCELLO2024115596,Perroni2019}, and a more detailed analysis of how their  multiband character, inherently integrated into this platform, affects topological properties is still needed.

In the present study, we examine the topological characteristics of the 2DEG formed at the (001) LAO/STO interface in the presence of an external magnetic field. Our analysis is based on a realistic electronic structure and incorporates a superconducting pairing amplitude on the order of the experimentally observed value, approximately 20 \(\mu\)eV~\cite{Stornaiuolo2017, Zegrodnik_2020}. Given that the electronic structure of LAO/STO 2DEG is governed by \(d\) orbitals, the system is intrinsically multiband in nature, and its topological characteristics exhibit a pronounced sensitivity to the detailed orbital composition of the individual bands. In this study, we first present a comprehensive characterization of the topological phases of fully 2D system  (LAO/STO 2DEG). We then impose appropriate boundary conditions to confine the system to a quasi-one-dimensional geometry and perform a systematic analysis of edge states and MZMs arising in one-dimensional LAO/STO nanowires.

The manuscript is structured as follows. In Sec. \ref{sec:Theory}, we introduce the theoretical model employed to describe the superconducting 2DEG formed at the (001) LAO/STO interface. In Sec. \ref{sec:Results}, we present the analysis of the topological phase for a fully two-dimensional system and analyze the crossover to one-dimensional nanowires. A summary is provided in Sec. \ref{sec:Summary}.

\section{Theoretical model}
\label{sec:Theory}

The electronic structure of the 2DEG at the (001) LAO/STO interface originates from the Ti $t_{2g}$ orbitals ($d_{xy}$, $d_{yz}$, and $d_{xz}$), hybridized via the oxygen $2p$ states~\cite{Popovic2008,Pavlenko2012,Pentcheva2006,Diez2015}. Within the tight-binding approximation, the system is modeled on a square lattice with three orbitals per site, where confinement at the interface lowers the $d_{xy}$ band by approximately $47$~meV relative to higher bands composed mainly of the $d_{yz}$ and $d_{xz}$ orbitals~\cite{maniv2015strong}.

In the basis of $d$-orbitals $(d_{xy}^{\uparrow}, d_{xy}^{\downarrow}, d_{xz}^{\uparrow}, d_{xz}^{\downarrow}, d_{yz}^{\uparrow}, d_{yz}^{\downarrow})^T$ , the Hamiltonian of the (001) LAO/STO 2DEG is given by
\begin{equation}
    \hat{H}=\hat{H}_{0}+\hat{H}_{RSO}+\hat{H}_{SO}+\hat{H}_{B}.
    \label{eq:Hamiltonian_general}
\end{equation}
where the subsequent parts correspond to the kinetic energy, Rashba SO coupling, atomic $L \cdot S$ spin-orbit interaction, and the coupling of the orbital and spin magnetic moments to the external magnetic field.

The kinetic Hamiltonian $\hat{H}_0$ accounts for the dispersions of the three distinct orbitals, as well as the hybridization between them, and can be expressed as
\begin{equation}
\hat{H}_{0}=
\left(
\begin{array}{ccc}
 \epsilon^{xy}_{\mathbf{k}} & 0 & 0\\
 0 & \epsilon^{xz}_{\mathbf{k}} &  \epsilon^h_{\mathbf{k}} \\
 0 & \epsilon^h_{\mathbf{k}}  &  \epsilon^{yz}_{\mathbf{k}}
\end{array} \right) \otimes \hat {\sigma} _0\;,
\label{eq:H00}
\end{equation}
where
\begin{equation}
\begin{split}
    \epsilon^{xy}_{\mathbf{k}}&=4t_l-2t_l\cos{k_xa}-2t_l\cos{k_ya}-\Delta_E,\\
    \epsilon^{xz}_{\mathbf{k}}&=2t_l+2t_h-2t_l\cos{k_xa}-2t_h\cos{k_ya},\\
    \epsilon^{yz}_{\mathbf{k}}&=2t_l+2t_h-2t_h\cos{k_xa}-2t_l\cos{k_ya},\\
    \epsilon^h_{\mathbf{k}}&=2t_d\sin{k_xa}\sin{k_ya}.
\end{split}
\label{eq:H0}
\end{equation}
In the above equations, the hybridization between the $d_{xz}$ and $d_{yz}$ orbitals is determined by the parameter $t_d$. The remaining parameters, $t_l$ and $t_h$, are the hopping energies for the light and heavy bands, respectively, and $a$ is a lattice constant $a=0.39$~nm. In the calculations, we adopted the TBA parameters from  Ref. \onlinecite{maniv2015strong,wojcik2021impact}: $t_l=875\;$meV, $t_h=40\;$meV, $t_d=40\;$meV and $\Delta_E=47\;$meV.

The LAO/STO-based 2DEG is characterized by strong spin--orbit coupling, arising primarily from two contributions: the  atomic $\mathbf{L} \cdot \mathbf{S}$ interaction, which plays a pivotal role for the $d$ orbitals~\cite{Khalsa2013}
\begin{equation}
\hat{H}_{SO}= \frac{\Delta_{SO}}{3}
\left(
\begin{array}{ccc}
0 & i \sigma _x & -i \sigma _y\\
-i \sigma _x & 0 & i \sigma _z \\
i \sigma _y & -i \sigma _z & 0
\end{array} \right) \;,
\label{eq:hso}
\end{equation}
and the Rashba spin-orbit coupling associated with the breaking of mirror symmetry at the interface, where a strong electric field is present
\begin{equation}
\hat{H}_{RSO}= \Delta_{RSO}
\left(
\begin{array}{ccc}
0 & i \sin{k_ya} & i \sin{k_xa}\\
-i \sin{k_ya} & 0 & 0 \\
-i \sin{k_xa} & 0 & 0
\end{array} \right) \otimes \hat {\sigma} _0\;,
\label{eq:rso}
\end{equation}
where $\Delta_{\mathrm{SO}}$ and $\Delta_{\mathrm{RSO}}$ determine the strengths of the atomic and Rashba SO couplings, respectively, and $\sigma_{x}, \sigma_{y}, \sigma_{z}$ are the Pauli matrices.
In the calculations, the SO coupling parameters are taken as $\Delta_{\mathrm{SO}} = 10$~meV and $\Delta_{\mathrm{RSO}} = 20$~meV, corresponding to the experimentally measured values~\cite{Caviglia2010,Yin2020}.

For $d$ electrons, the coupling to an external magnetic field is described by both the spin and orbital Zeeman effects
\begin{equation}
\hat{H}_B=\mu_B(\mathbf{L}\otimes \sigma_0+g\mathds{1}_{3\times 3} \otimes \mathbf{S})\cdot \mathbf{B}/\hbar,
\label{eq:Hb}
\end{equation}  
where $\mu_B$ is the Bohr magneton, $g$ is the Land\'e factor (we assume $g=3$~\cite{Jespersen2020}), $\mathbf{S}=\hbar \pmb{\sigma}/2$ with $\pmb{\sigma}=(\sigma_x,\sigma_y,\sigma_z)$ and $\mathbf{L}=(L_x,L_y,L_z)$ with
\begin{equation}
\begin{split}
 L_x&= \left ( 
 \begin{array}{ccc}
  0 & i & 0 \\
  -i & 0 & 0 \\
  0 & 0 & 0 
 \end{array}
 \right ), 
 L_y= \left ( 
 \begin{array}{ccc}
  0 & 0 & -i \\
  0 & 0 & 0 \\
  i & 0 & 0 
 \end{array}
 \right ), 
 L_z= \left ( 
 \begin{array}{ccc}
  0 & 0 & 0 \\
  0 & 0 & i \\
  0 & -i & 0 
 \end{array}
 \right ).
 \end{split}
\end{equation}

It is well known that under appropriate electrostatic gating, the LAO/STO 2DEG becomes superconducting with a critical temperature of approximately $300$~mK~\cite{Reyren2007, joshua2012universal,maniv2015strong, Biscaras, Monteiro2019, Monteiro2017}. Although the microscopic origin of pairing in this compound remains under debate~\cite{Trevisan2019,Trevisan2020}, experiments indicate a relatively small superconducting gap on the order of 20~$\mu$eV~\cite{Stornaiuolo2017}, exhibiting multiband characteristics~\cite{Diez2015,Trevisan2018} and a nodeless symmetry \cite{singh2022,joshua2012universal,maniv2015strong}.

Within the standard mean-field approach, the Hamiltonian of the superconducting LAO/STO 2DEG can be expressed in Nambu space as
\begin{equation}
\begin{split}
 \mathbf{\hat{H}}_{\mathbf{k}}=\sigma _z \otimes H_0'+ \sigma_z \otimes& H_{RSO}+ \sigma_z \otimes H_{SO}\\
 +& \sigma _z \otimes H_B - i \sigma _y \otimes 
\Delta_{6\times 6},
\end{split}
\label{eq:matrix_H}
\end{equation}
where the prime of $H_0$ in Eq.~(\ref{eq:matrix_H}) indicates that in the diagonal elements $\epsilon^l_{\mathbf{k}}$, Eqs. (\ref{eq:H0}), we include the chemical potential term, i.e the diagonal elements of Eq.~(\ref{eq:matrix_H}) are replaced by $\xi^ l_{\mathbf{k}}=\epsilon^l_{\mathbf{k}}-\mu$, where $l=xy,xz,yz$ and $\mu$ is the chemical potential.

The last term of $\mathbf{\hat{H}}_{\mathbf{k}}$ associated with superconductivity is given by
\begin{equation}
\Delta_{6\times6}=\left(\begin{array}{cccccc}
 0 & \Delta & 0 & 0 & 0 & 0\\
-\Delta & 0 & 0 & 0 & 0 & 0\\
0 & 0 & 0 & \Delta & 0 & 0 \\
0 & 0 & -\Delta & 0 & 0 & 0 \\
0 & 0 & 0 & 0 & 0 & \Delta \\
0 & 0 & 0 & 0 & -\Delta & 0 \\
\end{array} \right)\;,
\label{eq:matrix}
\end{equation}
where we assume an $s$-wave spin-singlet pairing symmetry with an amplitude $\Delta = 20~\mu\mathrm{eV}$, consistent with experimental measurements~\cite{Stornaiuolo2017}, and equal for all bands.\\

\begin{widetext}
To analyze the topological properties of nanowires embedded in the LAO/STO and to study the emergence of Majorana states, it is necessary to express the Hamiltonian (\ref{eq:matrix_H}) in real space. In the corresponding tight-binding model, the real-space Hamiltonian of the LAO/STO interface is defined on a square lattice with lattice constant $a$. Within this framework, the Hamiltonian (\ref{eq:matrix_H}) takes the form
\begin{equation}
\begin{split}
    \hat{H}=&\sum _{\mu,\nu} \hat{C}^\dagger_{\mu, \nu} ( \sigma _z \otimes\hat{H}^{0}+ \sigma _z \otimes \hat{H}_{SO} + \sigma _z \otimes \hat{H}_B - i \sigma _y \otimes 
    \Delta_{6\times 6} ) \hat{C}_{\mu, \nu} + 
    \sum _{\mu,\nu} \hat{C}^\dagger_{\mu+1, \nu} \hat{H}^{x} \hat{C}_{\mu, \nu}  + \sum _{\mu,\nu} \hat{C}^\dagger_{\mu, \nu+1} \hat{H}^{y} \hat{C}_{\mu, \nu} +  \\ 
    & \sum _{\mu,\nu} \hat{C}^\dagger_{\mu+1, \nu-1} \hat{H}_{mix} \hat{C}_{\mu, \nu}-
    \sum _{\mu,\nu} \hat{C}^\dagger_{\mu+1, \nu+1} \hat{H}_{mix} \hat{C}_{\mu, \nu}  
    + h.c.,
\end{split}
\label{eq:realspace}
\end{equation}
where $\hat{C}_{\mu, \nu}$=($\hat{c}_{\mu, \nu,xy}^{\uparrow}$, $\hat{c}_{\mu, \nu, xy}^{\downarrow}$, $\hat{c}_{\mu, \nu, xz}^{\uparrow}$, $\hat{c}_{\mu, \nu, xz}^{\downarrow}$, $\hat{c}_{\mu, \nu, yz}^{\uparrow}$, $\hat{c}_{\mu, \nu, yz}^{\downarrow}$, $\hat{c}_{\mu, \nu,xy}^{ \uparrow \dagger}$, $\hat{c}_{\mu, \nu, xy}^{\downarrow \dagger}$, $\hat{c}_{\mu, \nu, xz}^{\uparrow \dagger}$, $\hat{c}_{\mu, \nu, xz}^{\downarrow \dagger}$, $\hat{c}_{\mu, \nu, yz}^{\uparrow \dagger }, \hat{c}_{\mu, \nu, yz}^{\downarrow \dagger}$)$^{T}$ corresponds to the vector in the Nambu space where $\hat{c}_{\mu, \nu, l}^{\sigma}$ is the annihilation operator of an electron with spin $\sigma=\uparrow,\downarrow$ on the orbital $l$ and the position on the real-space lattice determined by the indexes $( \mu,\nu )$.

The on-site energies and the hopping amplitudes are defined by the operators 
\begin{equation}
\hat{H}^{0}=
\left(
\begin{array}{ccc}
 4t_l-\Delta _E & 0 & 0\\
 0 & 2t_l+2t_h & 0 \\
 0 & 0  &  2t_l+2t_h
\end{array} \right) \otimes \hat {\sigma} _0 
+
\left(
\begin{array}{ccc}
 V_{\mu,\nu} & 0 & 0\\
 0 & V_{\mu,\nu} & 0 \\
 0 & 0  &  V_{\mu,\nu}
\end{array} \right) \otimes \hat {\sigma} _0 ,
\label{eq:Hamiltonian_real_H0}
\end{equation}
\begin{equation}
\hat{H}^{x}= \sigma_z \otimes
\left(
\begin{array}{ccc}
 -t_l & 0 & 0\\
 0 & -t_l & 0 \\
 0 & 0  &  -t_h
\end{array} \right) \otimes \hat {\sigma} _0
+
\sigma_0 \otimes
\frac{\Delta_{RSO}}{2}
\left(
\begin{array}{ccc}
 0 & 0 & -1 \\
 0 & 0 & 0 \\
 1 & 0  &  0
\end{array} \right) \otimes \hat {\sigma} _0\;,
\label{eq:real_space_Hx}
\end{equation}
\begin{equation}
\hat{H}^{y}= \sigma_z \otimes
\left(
\begin{array}{ccc}
 -t_l & 0 & 0\\
 0 & -t_h & 0 \\
 0 & 0  &  -t_l
\end{array} \right) \otimes \hat {\sigma} _0
+
\sigma_0 \otimes
\frac{\Delta_{RSO}}{2}
\left(
\begin{array}{ccc}
 0 & -1 & 0 \\
 1 & 0 & 0 \\
 0 & 0  &  0
\end{array} \right) \otimes \hat {\sigma} _0,
\label{eq:real_space_Hy}
\end{equation}
and 
\begin{equation}
\hat{H}_{mix}= \sigma_z \otimes \frac{t_d}{2}
\left(
\begin{array}{ccc}
 0 & 0 & 0 \\
 0 & 0 & 1 \\
 0 & 1  &  0
\end{array} \right) \otimes \hat {\sigma} _0.
\label{eq:real_space_hybrid}
\end{equation}
In Eq.~(\ref{eq:realspace}), $\hat{H}_{SO}$ and $\hat{H}_B$ have the same form as in the wave vector space formulation, given by Eqs.~(\ref{eq:hso}) and (\ref{eq:Hb}).
\end{widetext}

\section{Results}
\label{sec:Results}

In this section, we first analyze the topological phases of the LAO/STO 2DEG in the presence of an external magnetic field, and then, by imposing suitable boundary conditions, we effectively reduce the system to a quasi-one-dimensional geometry and carry out a systematic study of MZMs in one-dimensional LAO/STO nanowires.

\subsection{LAO/STO 2DEG} 

Lets consider the superconducting 2DEG at the LAO/STO interface in the presence of an external magnetic field applied in different directions. For simplicity, we restrict our theoretical analysis to account solely for the Zeeman splitting. The breaking of time-reversal symmetry related to the magnetic field, in conjunction with the inherent particle-hole symmetry of the superconducting state, assigns the system to symmetry class D within the Altland-Zirnbauer classification~\cite{Altland}. For a 2D system in this class, the topological phase is characterized by an integer-valued Chern number, $C$. 
\begin{figure}[!t]
\includegraphics[]{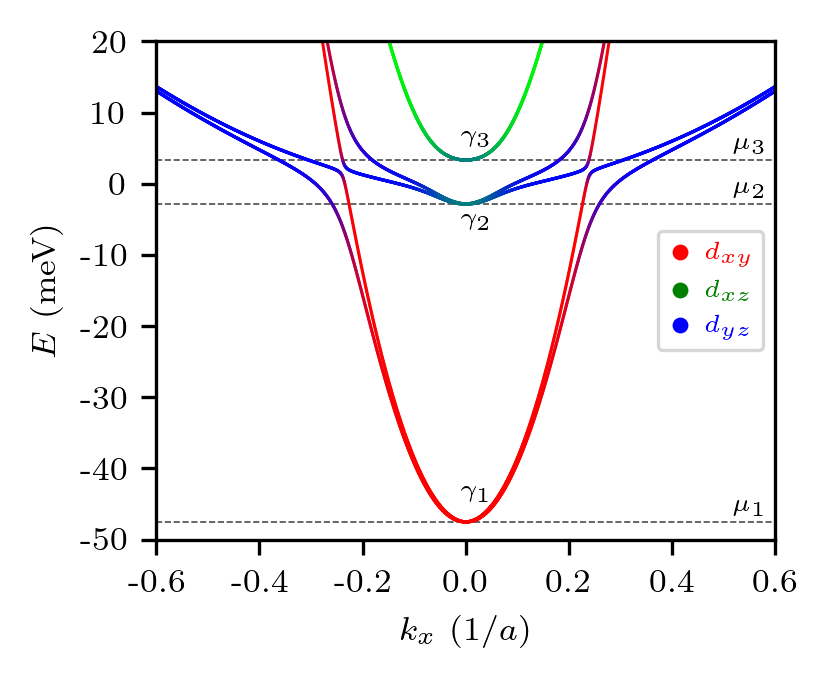}
\put(-156,142){(a)} \\
\includegraphics[]{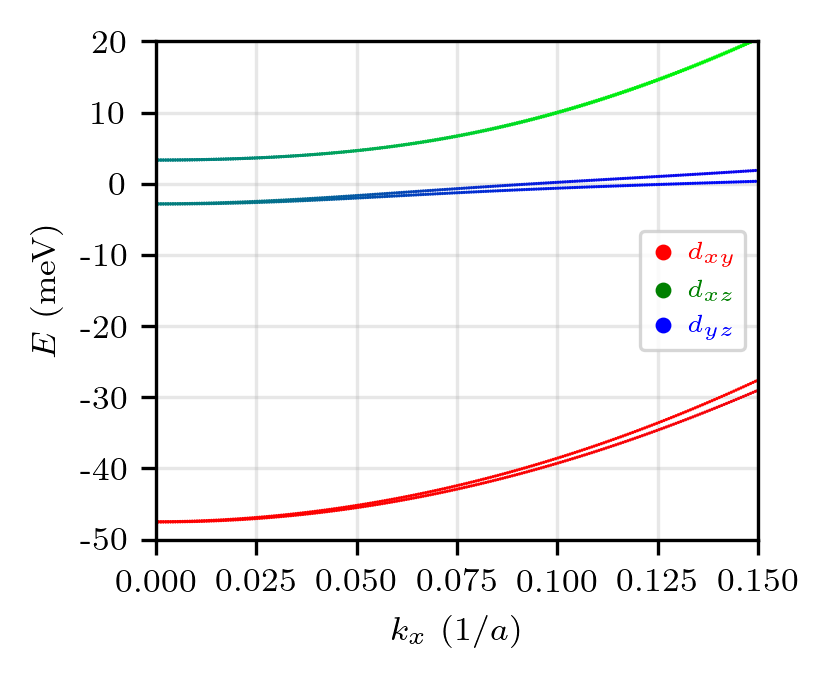}
\put(-156,142){(b)} \\
\caption{(a) Dispersion relation $E(k_x, k_y=0)$ for the LAO/STO 2DEG. The horizontal dashed lines indicate the helical band minima, denoted by $\mu_l$ with $l = 1,2,3$, which are used as reference chemical potentials for determining the corresponding topological phase diagram. (b) Enlarged view of panel (a), highlighting the spin splitting of the band induced by SO coupling. In both panels, the contributions of the $d_{xy}$, $d_{xz}$, and $d_{yz}$ orbitals are represented using the RGB color scheme, such that the resulting color reflects their relative mixture.}
\label{fig:disp_NM}
\end{figure}
\begin{figure}[!t]
\includegraphics[]{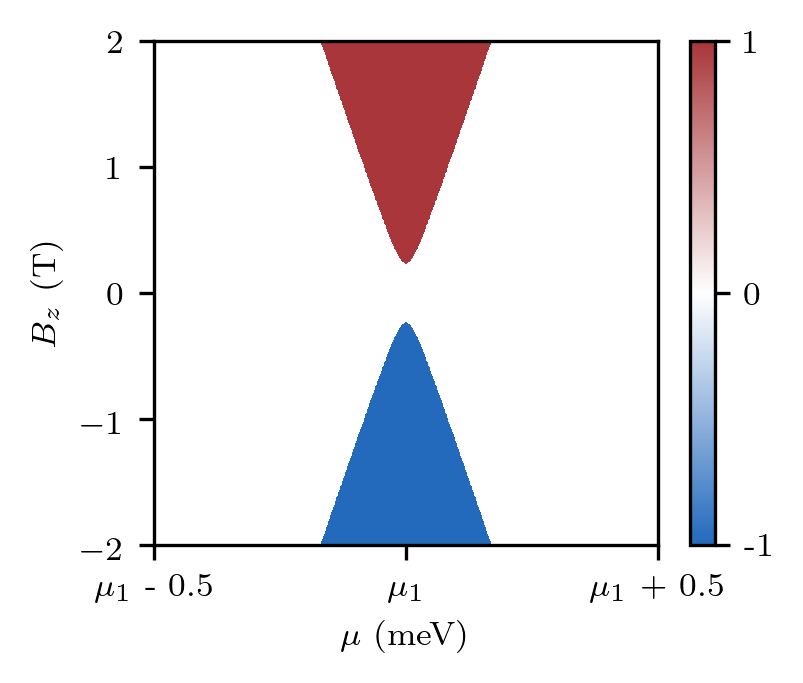}
\put(-150,140){(a)} \\
\includegraphics[]{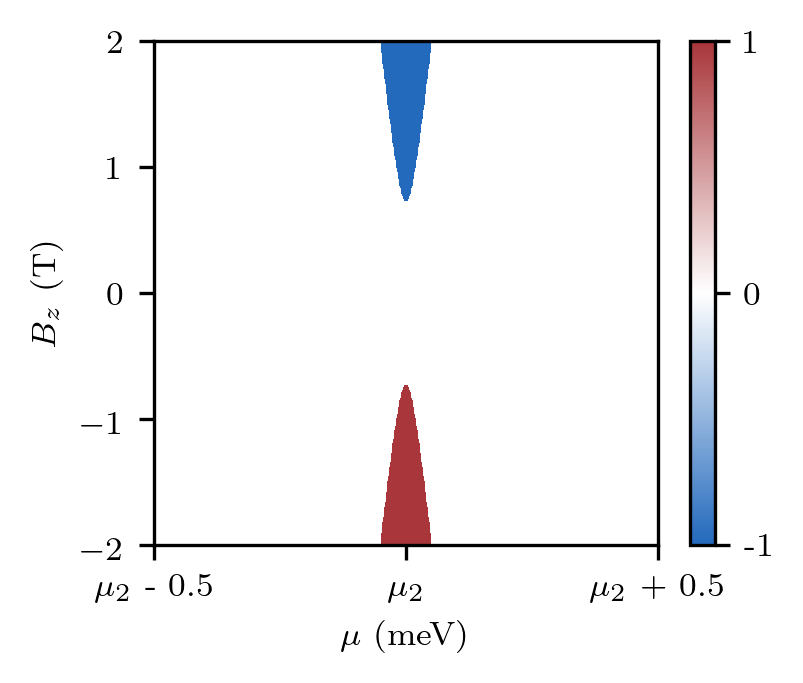}
\put(-150,140){(b)} \\
\includegraphics[]{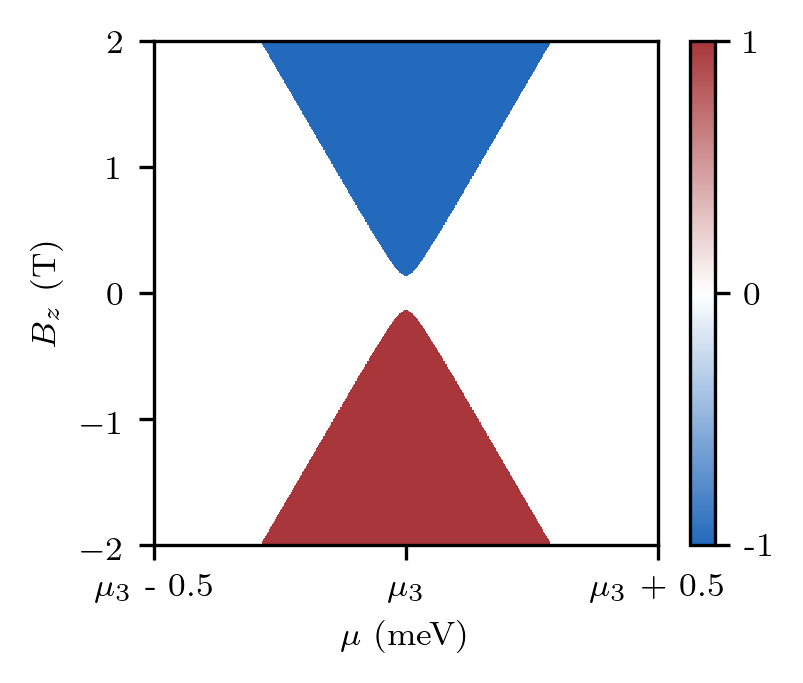}
\put(-150,140){(c)} \\
\caption{Chern number as a function of the chemical potential $\mu$ and the out-of-plane magnetic field $B_z$, evaluated in the vicinity of the bottom of the helical bands denoted as $\mu_l$ with $l = 1,2,3$ in Fig.~\ref{fig:disp_NM}(a).}
\label{fig:chern}
\end{figure}

Here, the Chern number associated with the occupied bands is determined by employing the Wilson loop formalism. This approach connects the Berry phase accumulated along a closed contour \(\Gamma\) in the Brillouin zone to the overlaps between neighboring Bloch states \( | u_{n,\vb{k}} \rangle \). For a discretized loop \(\{\vb{k}_0, \vb{k}_1, \dots, \vb{k}_N = \vb{k}_0\}\), the Wilson loop matrix is defined as~\cite{Fukui2005}
\begin{equation}
    W(\Gamma) = \prod_{i=0}^{N-1} M^{\vb{k}_i \to \vb{k}_{i+1}}, 
\end{equation}
where
\begin{equation}
M^{\vb{k}_i \to \vb{k}_{i+1}}_{mn} = \langle u_{m,\vb{k}_i} | u_{n,\vb{k}_{i+1}} \rangle.
\end{equation}
Due to the discretization of the Brillouin zone, the matrix \(M\) is not exactly unitary. To enforce unitarity, we perform a singular value decomposition \(M = U \Sigma V^\dagger\) and define the unitarized matrix \(F = U V^\dagger\). The Wilson loop is then given by
\begin{equation}
    W(\Gamma) = \prod_{i=0}^{N-1} F^{\vb{k}_i \to \vb{k}_{i+1}}.
\end{equation}
and the Berry phase accumulated by all occupied bands along the closed contour \(\Gamma\) is expressed as
\begin{equation}
    \gamma^\Gamma = -\Im{\ln(\det W(\Gamma))}.
\end{equation}
By dividing the Brillouin zone into small plaquettes labeled by $j$, evaluating \(\gamma^{\Gamma_j}\) around their perimeters, and summing the results, one effectively integrates Berry curvature over the Brillouin zone and obtains the associated Chern number, which is given by 
\begin{equation}
    C = \frac{1}{2\pi}\sum_j \gamma^{\Gamma_j}
\end{equation}

\begin{figure}[!t]
\includegraphics[]{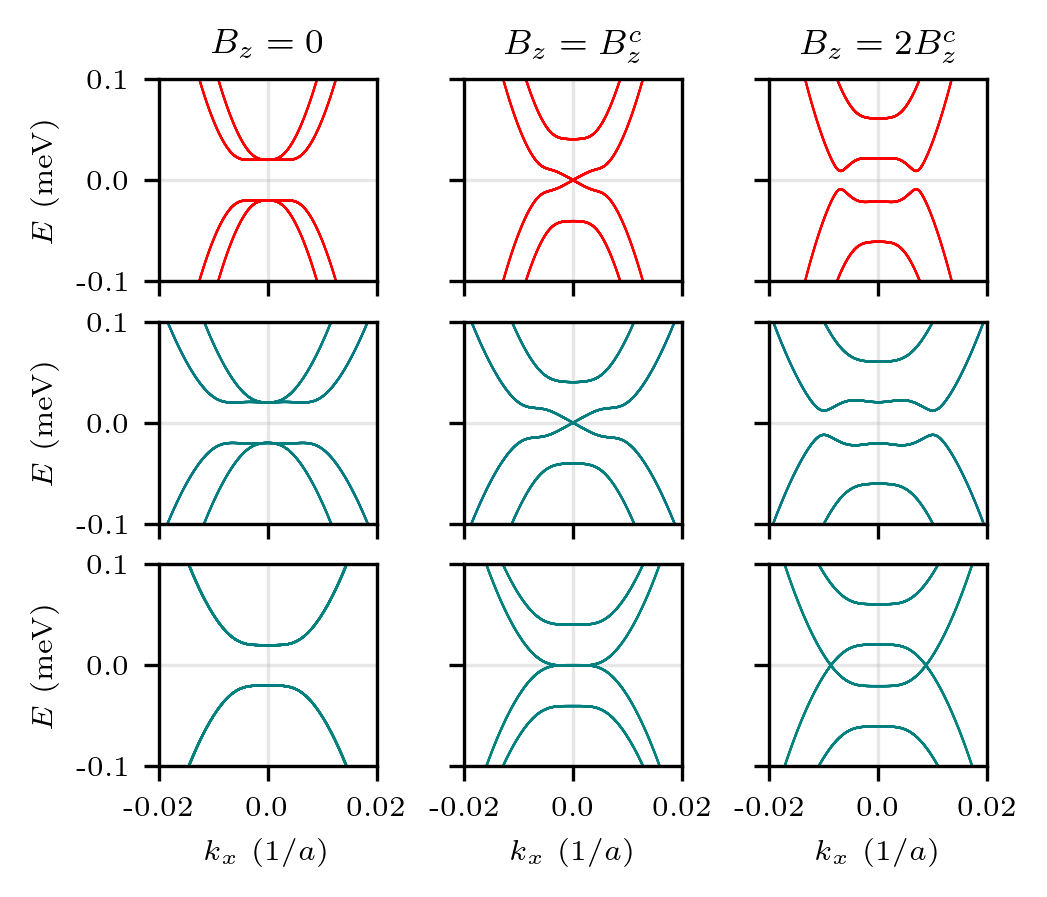}
\put(-213,170){(a)} 
\put(-139,170){(b)} 
\put(-65 ,170){(c)} 
\put(-213,112){(d)} 
\put(-139,112){(e)} 
\put(-65 ,112){(f)} 
\put(-213,54 ){(g)} 
\put(-139,54 ){(h)} 
\put(-65 ,54 ){(i)} 
\caption{Dispersion relations of the superconducting LAO/STO 2DEG determined along the $k_x$ direction for (a-c) $\mu_{1}$, (d-f) $\mu_{2}$, and (g-i) $\mu_{3}$. In each row, the subsequent panels present the results for three distinct values of the magnetic field $B_z$: $B_z = 0$, $B^c_z$ corresponding to the critical field at $\mu=\mu_l$, and $B_z>B^c_z$. Critical field values are presented in Tab.~\ref{tab1}. As in Fig.~\ref{fig:disp_NM}(a), the contributions of the $d_{xy}$ , $d_{xz}$ , and $d_{yz}$ orbitals are represented using the RGB
color scheme.} 
\label{fig:disp_sc}
\end{figure}
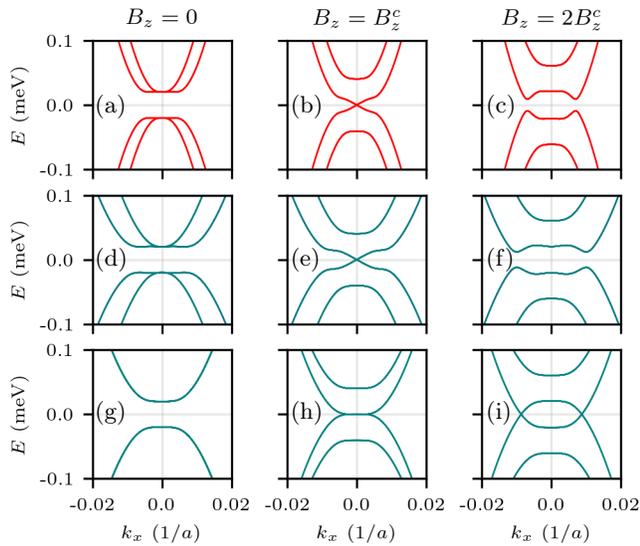

In this section, when analyzing the LAO/STO 2DEG, we restrict ourselves to the range of chemical potentials for which the system can be driven into an spinless regime. This corresponds to values close to the bottom of the electronic bands, where a helical gap can open under an applied magnetic field. Fig.~\ref{fig:disp_NM}(a) presents the dispersion relations of the LAO/STO 2DEG determined along the $k_x$ direction. The chemical potential values corresponding to the helical band minima are marked by dashed horizontal lines and labeled as $\mu_l$, with $l = 1,2,3$. Note that, at $k = 0$, only the lowest energy band $\gamma_1$ is purely derived from the $d_{xy}$ orbital, whereas the two higher-lying bands $\gamma_2$ and $\gamma_3$ are composed of an almost equal mixture of the $d_{yz}$ and $d_{xz}$ states, mainly induced by the $d_{xz}/d_{yz}$ hybridization (determined by the parameter $t_d$), and the atomic SO coupling [Eq.~(\ref{eq:hso})]. Fig.~\ref{fig:disp_NM}(b) shows an enlarged view of panel (a), highlighting the SO band splitting. It clearly demonstrates that the resulting SO coupling is substantially stronger in the bands $\gamma_1$ and $\gamma_2$ than in the higher-lying band $\gamma_3$, for which it remains finite but extremely weak and not discernible in the panel.\\

Since the SO coupling in the LAO/STO interface locks the electron spins within the plane of the interface, the formation of a helical gap induced by a solely in-plane magnetic field is prohibited, maintaining the system in a topologically trivial phase. Indeed, the computed Chern number is found to be $C = 0$ when the external magnetic field is oriented in-plane. In contrast, a perpendicular orientation of the magnetic field opens a helical gap at $k=0$, enabling the stabilization of a topological phase when the chemical potential is tuned in the vicinity of $\mu_l$ ($l = 1,2,3$). The corresponding phase diagrams, determined near the three chemical potentials $\mu_l$, shown in Fig.~\ref{fig:chern}, demonstrates that the critical field \(B^c_{z}\) is band dependent; its value is different for different $\gamma_l$. For individual bands, the corresponding values of the critical field, determined for $\mu=\mu_l$, are summarized in Table~\ref{tab1}. 
        
\begin{figure}[!t]
\includegraphics[]{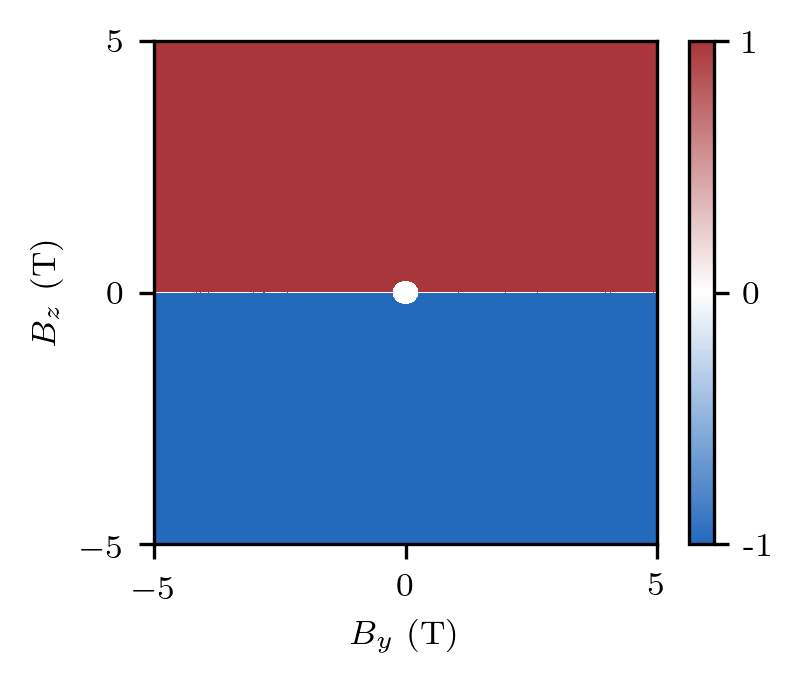}
\put(-150,140){(a)} \\
\includegraphics[]{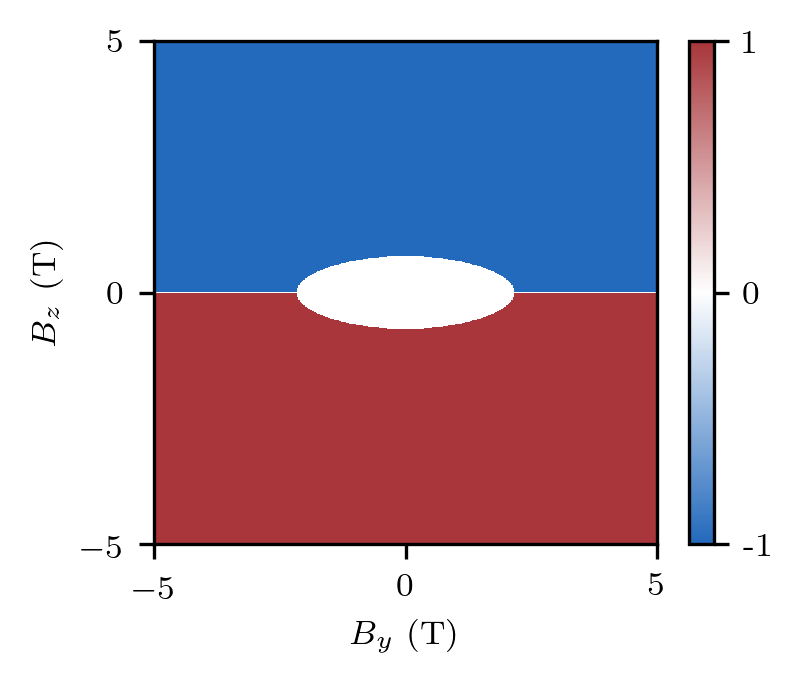}
\put(-150,140){(b)} \\
\includegraphics[]{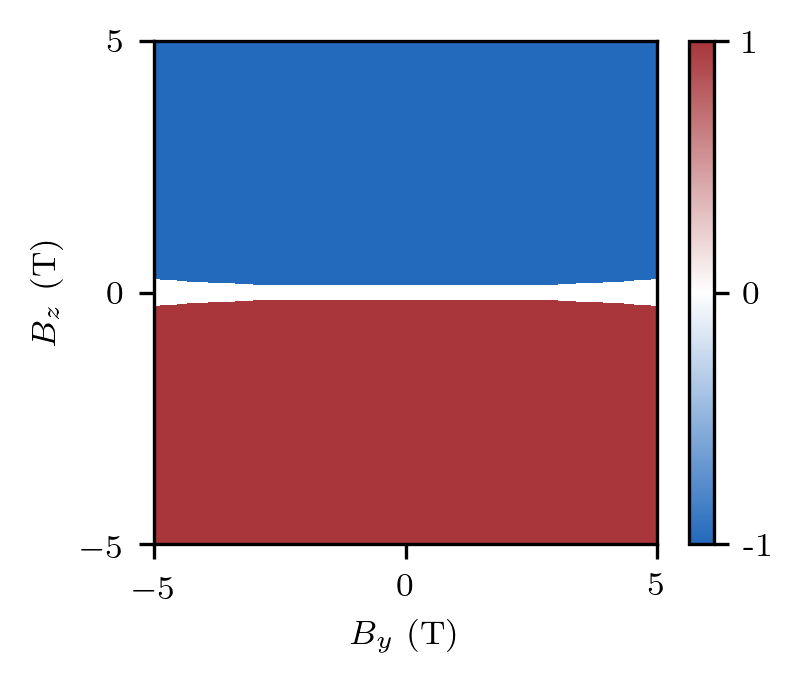}
\put(-150,140){(c)} \\
\caption{Chern number as a function of the in-plane and out-of-plane magnetic field components $(B_y,B_z)$, evaluated for $\mu_l$ with $l = 1, 2, 3$ marked in Fig.~\ref{fig:disp_NM}(a).} 
\label{fig:CByBz}
\end{figure}

To elucidate the topological phase transition, in Fig.~\ref{fig:disp_sc} we present the energy dispersions calculated for the three selected values of \(\mu_l\) and representative values of the magnetic field: zero field, the critical field \(B^c_z\) at which the transition occurs, and a field larger than this critical value. For all bands, the superconducting gap closes at \(k = 0\) when the magnetic field reaches \(B^c_z\), and then reopens for \(B_z > B^c_z\), which is clearly visible in the helical bands $\gamma_1$ and $\gamma_2$. For the band $\gamma_3$, the corresponding topological gap cannot be clearly resolved from the dispersion alone, due to its very small amplitude and the finite numerical resolution. Nevertheless, because in the considered parameter range \(C \neq 0\) [Fig.~\ref{fig:chern}(c)], a finite topological gap must be present, even though its magnitude is too small to be unambiguously distinguished in the plotted spectra. Based on the more accurate numerical calculations, we estimate this gap to be on the order of \(10^{-4}\,\text{meV}\). Such an exceedingly small value arises from the weak effective SO coupling in this helical band, as demonstrated in Fig.~\ref{fig:disp_NM}(b).
Interestingly, Fig.~\ref{fig:chern} shows  that the Chern number corresponding to the helical band $\gamma_1$ is opposite to that related to the bands $\gamma_2$ and $\gamma_3$, and this sign reverses when the direction of the magnetic field is inverted. This behavior originates from the opposite helicities of the bands, which are related to the opposite spin textures on the Fermi surfaces, determined by the SO coupling.\\
\begin{table}[H]
            \centering
            
            \begin{tabular}{c|c|c}
                \hline
                $\mu$      & $E$ (meV) & $B^c_z$ (T) \\
                \hline
                $\mu_{1}$ & -47.5     & 0.24             \\
                $\mu_{2}$ & -2.83     & 0.73             \\
                $\mu_{3}$ & 3.33      & 0.14             \\
                \hline
            \end{tabular}
            \caption{Table of the helical band bottom energy $\mu_l$ where $l=1,2,3$ and the corresponding critical field $B^c_z$ calculated at $\mu=\mu_l$.}
            \label{tab1}
\end{table}

Due to the multiband character of the LAO/STO 2DEG, an analytical determination of the critical field for individual bands is nontrivial which arises from the fact that the gap closing at \(k = 0\) is governed not only by the spin Zeeman splitting, but also by the orbital Zeeman effect and the atomic SO interaction. To address this issue, we employ a simplified model in which the parts of the Hamiltonian for the \(d_{xy}\) and \(d_{xz}/d_{yz}\) orbitals can be considered separately. This approximation is justified as the \(d_{xy}\) orbital is shifted by the energy offset \(\Delta E\) relative to the hybridized \(d_{xz}/d_{yz}\) states. Within this model, the lowest helical band $\gamma_1$, in the vicinity of \(k = 0\), can be reduced to a purely parabolic band with SO coupling in the Rashba form \(\alpha_R [\sigma_y \sin(k_xa) - \sigma_x\sin (k_ya)]\), where the Rashba energy is given by \(\alpha_R = \Delta_{\mathrm{SO}}\Delta_{\mathrm{RSO}}/3\Delta E \) - for the full derivation, see Appendix~\ref{sec:A1}. In this case, the critical magnetic field is described by the well-known expression~\cite{Lutchyn2010}
\[
B_z^c = \frac{1}{g\mu_b} \sqrt{(\mu-\mu_{1})^2 + \Delta^2}.
\]
For a chosen pairing potential $\Delta = 0.02$ meV and the chemical potential $\mu=\mu_{1}$, the critical field $B^c_z = 0.24$~T, in agreement with the value obtained from numerical simulations and reported in Table~\ref{tab1}.

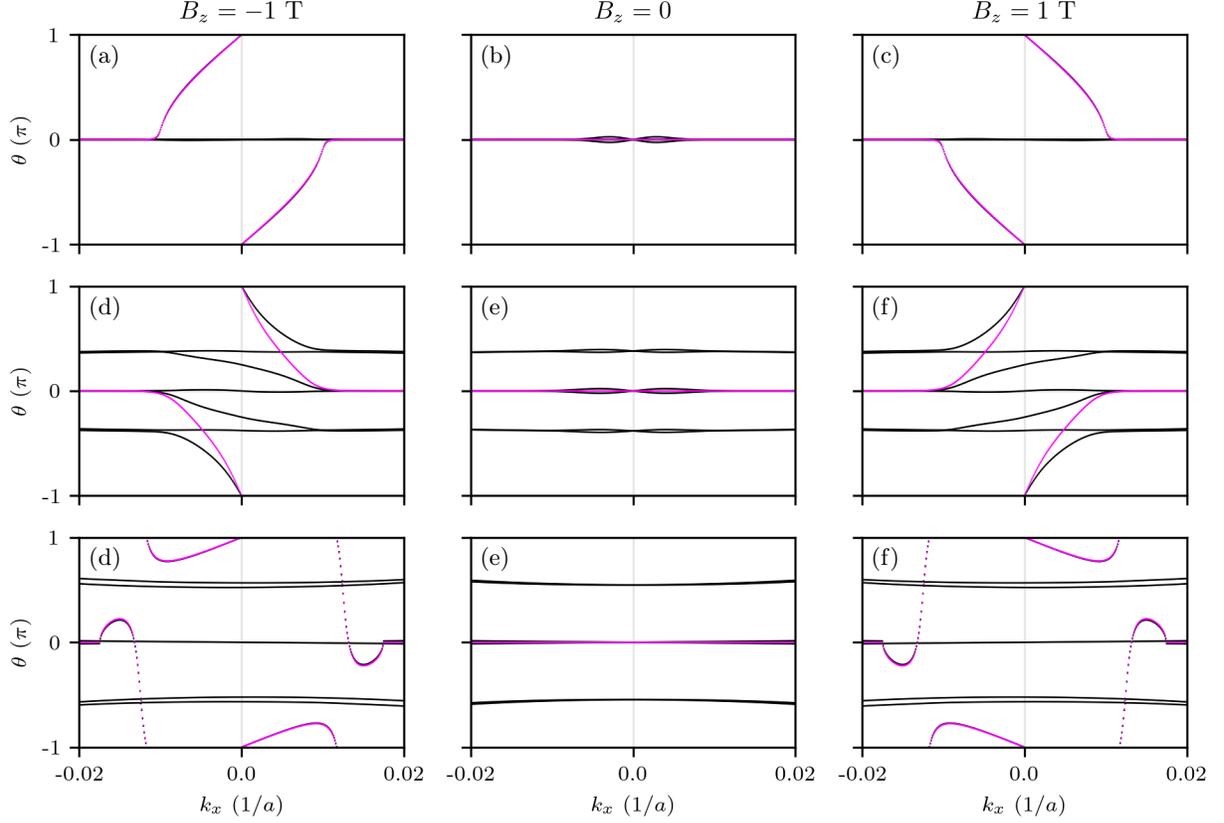
\begin{figure*}[!t]
\includegraphics[]{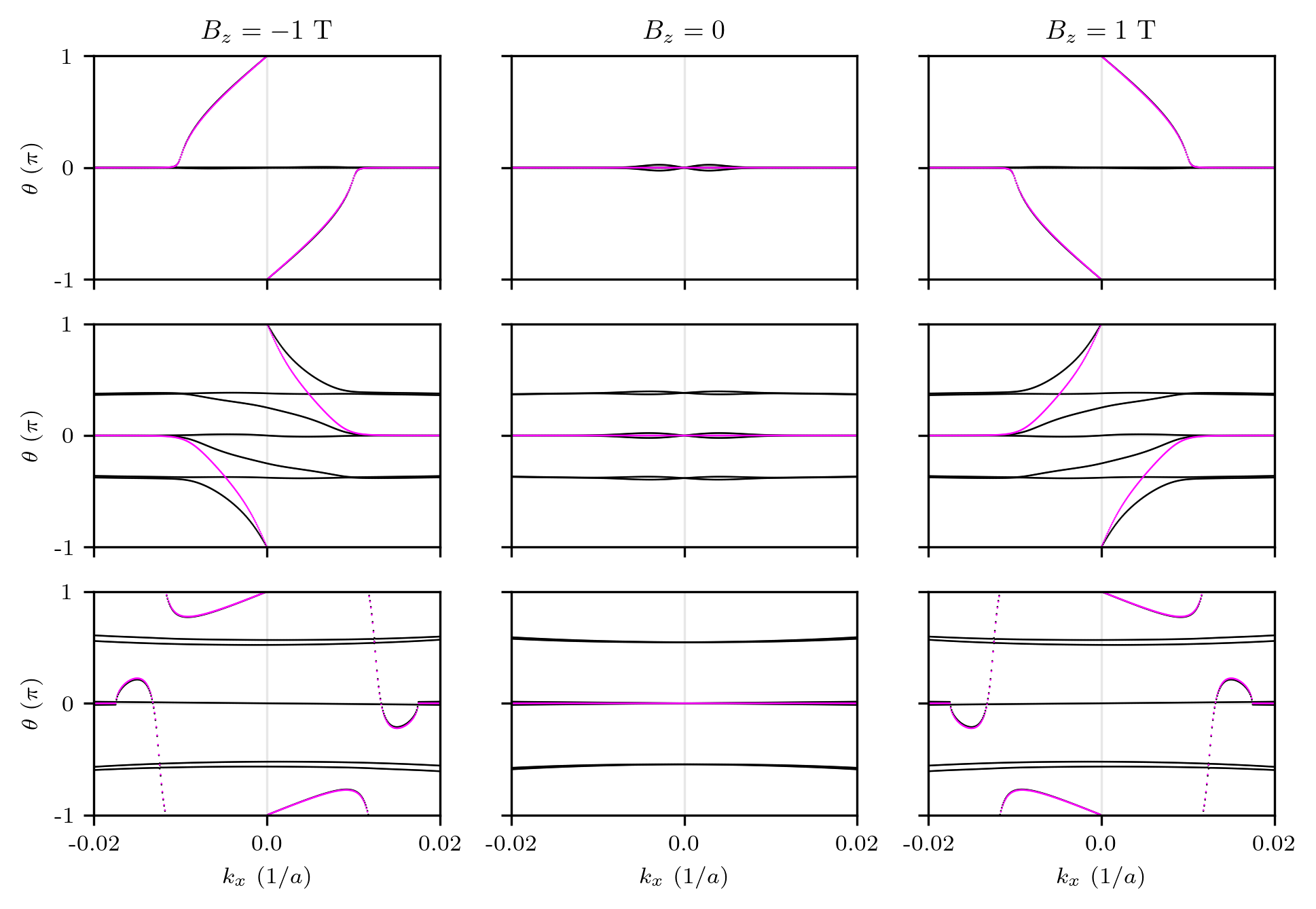}
\put(-430, 292){(a)}
\put(-282, 292){(b)}
\put(-134, 292){(c)}
\put(-430, 197){(d)}
\put(-282, 197){(e)}
\put(-134, 197){(f)}
\put(-430, 102){(d)}
\put(-282, 102){(e)}
\put(-134, 102){(f)}

\caption{The Wannier center flow as $\theta$ along $k_x$ for individual bands (black) and the sum over them (magenta) proportional to the electronic polarization. Results for the chosen chemical potentials (a-c) $\mu_{1}$, (d-f) $\mu_{2}$ and (g-i) $\mu_{3}$ and the magnetic field (a,d,g) $B_z = -1\ \mathrm{T}$, (b,e,h) $B_z=0$, and (c,f,i) $B_z=1\ \mathrm{T}$.
} 
\label{fig:WCC2D}
\end{figure*}

The analytical determination of the critical field for the hybridized helical bands $\gamma_2$ and $\gamma_3$ can be carried out using a Hamiltonian reduced exclusively to the \(d_{yz}/d_{xz}\) orbitals. Within this approximation, the interband hybridization between the \(d_{xz/yz}\)  and the lower-lying \(d_{xy}\) orbitals, induced by the atomic SO coupling, is neglected.   In this case, \(B^{c}\) can be determined by identifying the condition under which the eigenvalue of the Hamiltonian 
\begin{equation}
\hat{H}_{xz/yz}^{SC} =
\begin{pmatrix}
H_{xz/yz}(\mathbf{k}=0) & \Delta_{4\times4} \\
-\Delta_{4\times4} & - H_{xz/yz}^*(\mathbf{k}=0)
\end{pmatrix} \;,
\label{eq:HBdGk0}
\end{equation}
is equal to zero, where 
\begin{widetext}
\begin{equation}
H_{xz/yz}(\mathbf{k}=0) =
\begin{pmatrix}
\xi^{xz}_{\mathbf{k}=0} + \frac{g}{2} \mu_B B_z & 0 & i \left(\mu_B B_z + \frac{\Delta_{\mathrm{SO}}}{3} \right) & 0 \\
0 & \xi^{xz}_{\mathbf{k}=0} - \frac{g}{2} \mu_B B_z & 0 & i \left(\mu_B B_z -\frac{\Delta_{\mathrm{SO}}}{3} \right)  \\
-i \left( \mu_B B_z + \frac{\Delta_{\mathrm{SO}}}{3} \right) & 0 & \xi^{yz}_{\mathbf{k}=0} + \frac{g}{2} \mu_B B_z & 0 \\
0 & -i \left(\mu_B B_z -\frac{\Delta_{\mathrm{SO}}}{3}  \right) & 0 & \xi^{yz}_{\mathbf{k}=0} - \frac{g}{2} \mu_B B_z
\end{pmatrix}
\label{eq:Hnormal}
\end{equation}
\end{widetext} 
and $\Delta_{4\times 4}$ exhibits the same structure as Eq.~(\ref{eq:matrix}), but  reduced to a $4 \times 4$ dimensional representation. \\
The critical magnetic field for the helical bands $\gamma_{2(3)}$ is given by
\begin{equation}
    B_z^c=\frac{\sqrt{\left ( \mu \mp \frac{\Delta_{\mathrm{SO}}}{3} \right )^2  +\Delta^2}}{\mu_B \left (  \frac{g}{2}\pm 1\right )},
\end{equation}
which, for $\mu=\mu_{3}$, yields $B^c_z = 0.14~\mathrm{T}$, in agreement with Table~\ref{tab1}. For $\mu = -3.33~\mathrm{meV}$, corresponding to minima of the helical band $\gamma_2$ in the model reduced to the $d_{xz/yz}$ orbitals [Eq.~(\ref{eq:HBdGk0})], one obtains $B^c_z = 0.7~\mathrm{T}$, identical to the value found for $\mu_{2}$ (see Table~\ref{tab1}). 
Note that the minimum of the band $\gamma_2$ is predicted at a lower energy when calculated within the reduced model than when obtained from the full Hamiltonian. This deviation originates from the atomic SO coupling present in the full model, which hybridizes the $d_{xz/yz}$ orbitals with the lower-lying $d_{xy}$ states, thereby shifting their energies upward.

Although we have demonstrated above that only the out-of-plane component \(B_z\) is able to induce the topological transition, the in-plane magnetic field can also contribute to the emergence of a topological phase. In the latter case, however, a finite out-of-plane component is still required to drive the topological transition. In Fig.~\ref{fig:CByBz}, we present the Chern number, calculated for the chemical potential \(\mu_l\), as a function of the in-plane \(B_y\) and the out-of-plane \(B_z\) magnetic field components. In panel (a), we can see that the topological phase transition associated with the band $\gamma_1$ is isotropic with respect to the magnetic field direction, which is attributed to the isotropic dispersion of the orbital $d_{xy}$ arising from the equality of the effective mass in both the \(k_x\) and \(k_y\) directions [Eq.(\ref{eq:H0})]. In contrast, the topological transition exhibits pronounced anisotropy for the remaining bands $\gamma_{2}$ and $\gamma_{3}$. Specifically, for the chemical potential \(\mu_{2}\), the critical  field along the \(y\) direction (in the presence of a small \(B_z\)), is almost three times larger than the corresponding critical field along the \(z\) axis [Fig.~\ref{fig:CByBz}(b)]. Above a critical value \(B^c_y=2.17\)~T, even a negligible small \(B_z\) component is sufficient to drive the system into the topological regime. Conversely, for the  band \(\gamma_3\), the topological transition is nearly independent of \(B_y\), and within the range \(B_y \in [-3,3]\)~T, the emergence of the topological phase requires an out-of-plane field component exceeding the critical value \(B^c_z = 0.7\) T. An analogous behavior is observed when applying \(B_x\), and it is related to the fact that the highest band $\gamma_3$ is, to a large extent, insensitive to the in-plane magnetic field. This insensitivity originates from a nontrivial interplay between the spin and orbital Zeeman effects, which substantially cancel each other; see the Appendix \ref{sec:A2} for additional results.

Finally, to resolve the topological nature of the system, we analyze the evolution of the hybrid Wannier charge centers (HWCCs) as a function of momentum, which provides a direct measure of the band topology.
For 2D systems, the HWCCs can be evaluated by treating one momentum component, e.g., $k_x$, as a parameter and performing a Wilson loop along $k_y$. In this framework 
\begin{equation}
\bar{x}_n(k_x) = \frac{1}{2\pi} \int_{-\pi}^{\pi} \langle u_{n,k_x,k_y} | i \partial_{k_x} | u_{n,k_x,k_y} \rangle \, dk_y.
\end{equation}
The Wilson loop matrix is a unitary operator whose eigenvalue phases are directly linked to the positions of the Wannier centers within the unit cell. Specifically, the eigenvalues of this matrix are expressed as
\begin{equation}
\lambda_n(k_x) = e^{-i \theta(k_x)}, \:\:\: \text{where} \:\:\: \theta=2\pi \bar{x}_n(k_x).
\end{equation}
Note that the values of individual HWCCs typically exhibit a dependence on the specific choice of gauge. Nevertheless, the aggregate of these centers is gauge invariant and is directly correlated with the macroscopic electric polarization.
\begin{equation}
      P^h_e(k_x) = e\sum_{n}\bar{x}_n(k_x) = -\frac{e}{2\pi}\Im{\ln(\det(W( k_x)))}.
      \label{eq:polarization_hybrid_wannier_centers}
\end{equation}

\begin{figure*}[!t]
\includegraphics[]{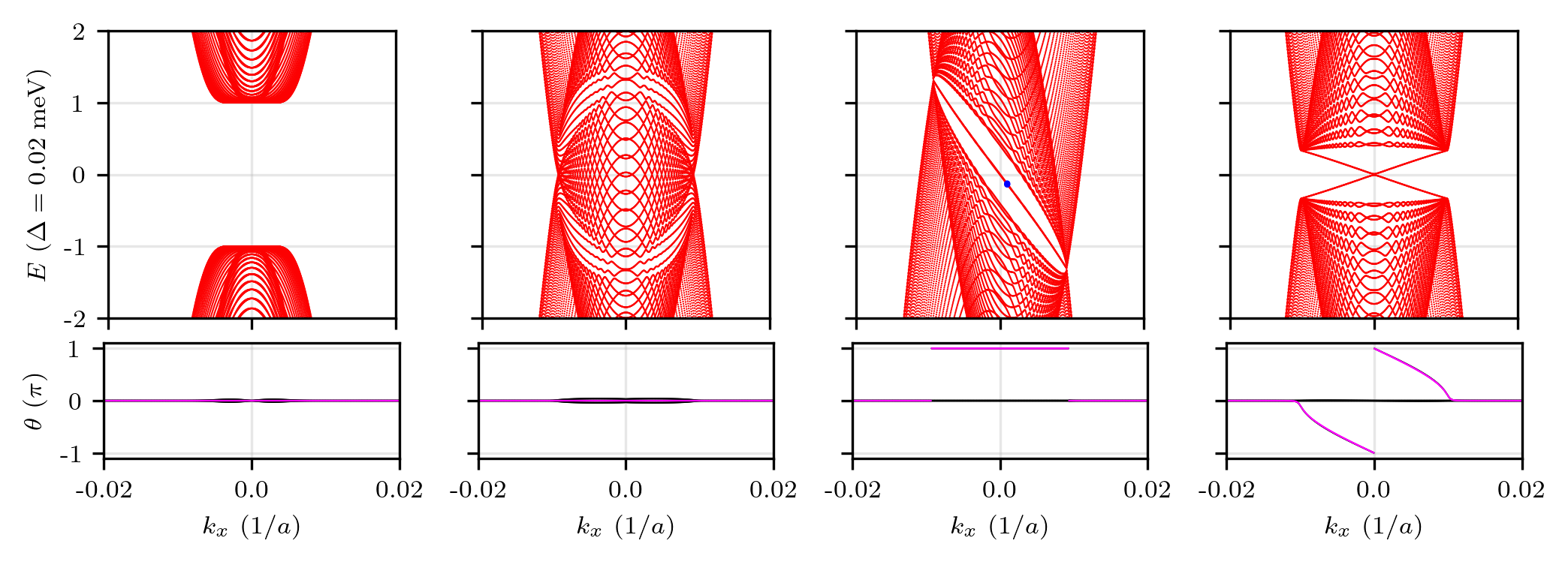}
\put(-470,175){(a)}
\put(-350,175){(b)}
\put(-230,175){(c)}
\put(-110,175){(d)}\\
\includegraphics[]{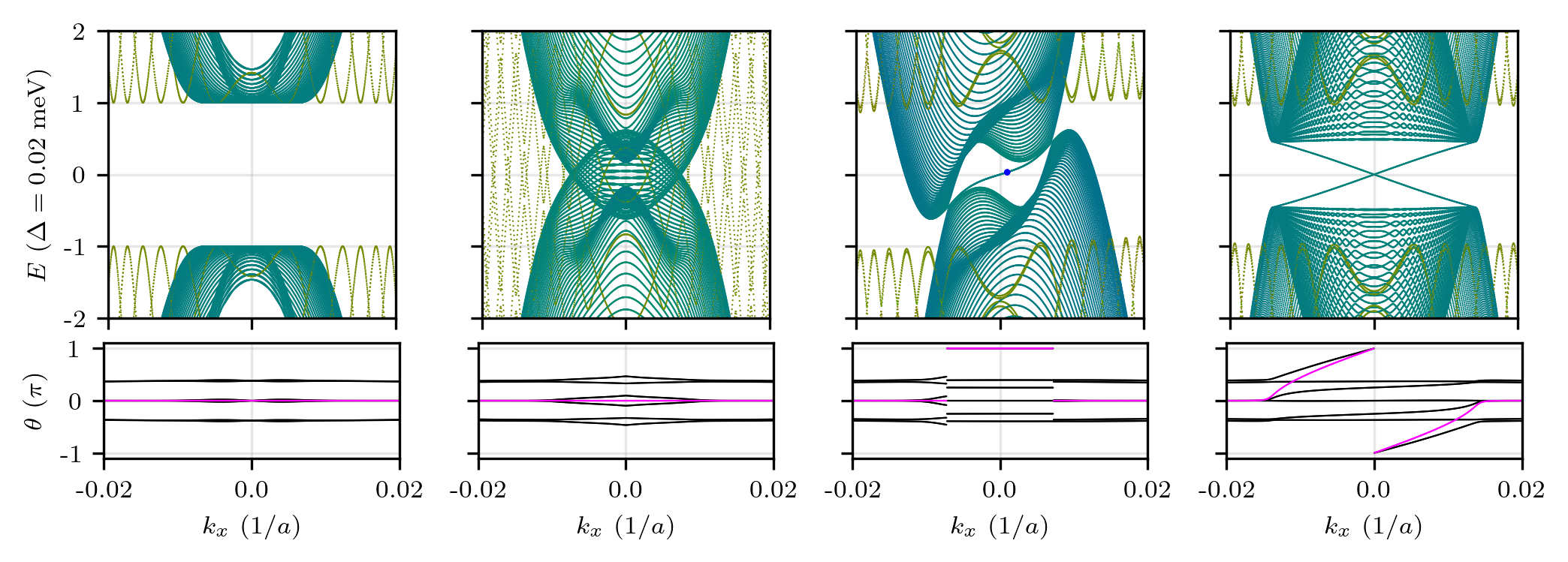}
\put(-470,175){(e)}
\put(-350,175){(f)}
\put(-230,175){(g)}
\put(-110,175){(h)}\\
\includegraphics[]{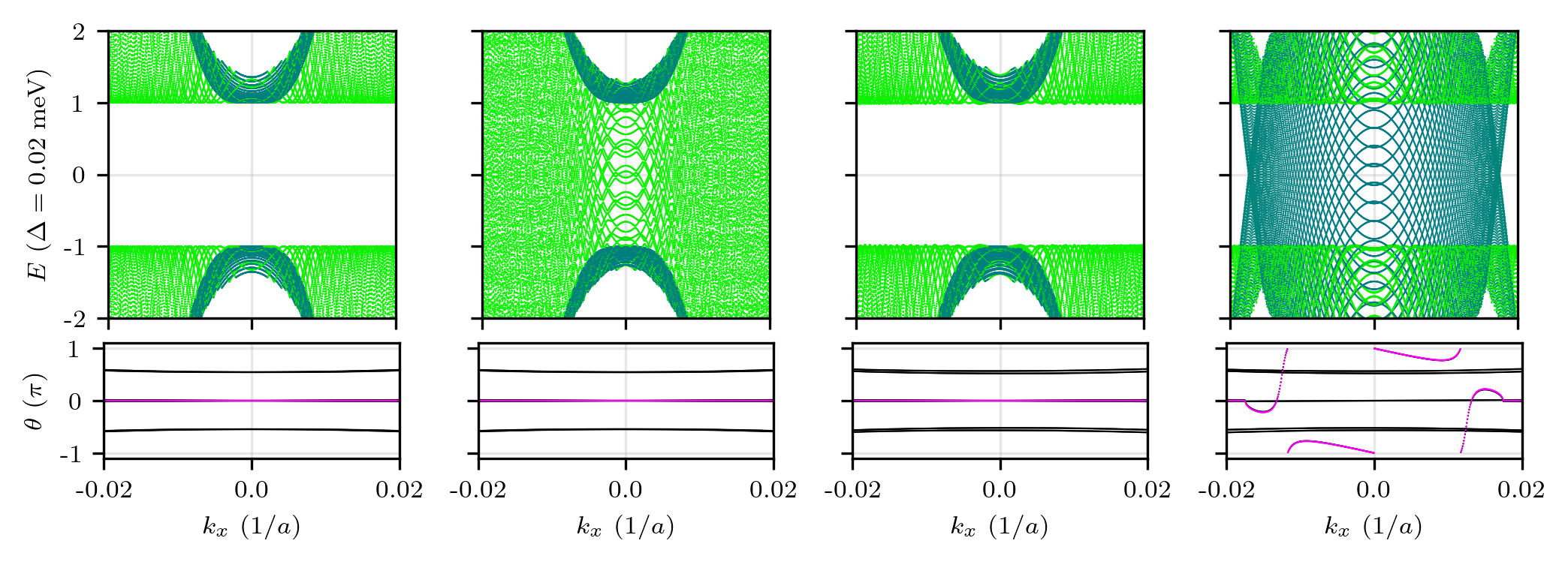}
\put(-470,175){(i)}
\put(-350,175){(j)}
\put(-230,175){(k)}
\put(-110,175){(l)}\\

\caption{Dispersion relations \(E(k_x)\) of a quasi-two dimensional superconducting LAO/STO 2DEG ($n_y=10000$) for three values of the chemical potential: (a–d) \(\mu_1\), (e–h) \(\mu_2\), and (i–l) \(\mu_3\). The panels correspond to different orientations of the applied magnetic field: (b, f, j) along the \(x\)-axis, (c, g, k) along the \(y\)-axis, and (d, h, l) along the \(z\)-axis. In each case, the magnitude of the magnetic field \(B\) is chosen to exceed the corresponding critical field, namely (b–d) \(B = 1\) T, (f–h) \(B = 3\) T, and (j–l) \(B = 1\) T. The lower panels display the evolution of the Wannier charge centers as \(\theta\) along \(k_x\), resolved for the individual bands (black curves) and for their total (magenta curves). The panels on the left (a,e,i) present spectrum for $B$=0. As in Fig.~\ref{fig:disp_NM}(a), the contributions of the $d_{xy}$ , $d_{xz}$ , and $d_{yz}$ orbitals are represented using the RGB color scheme.} 
\label{fig:Ekx1D}
\end{figure*}
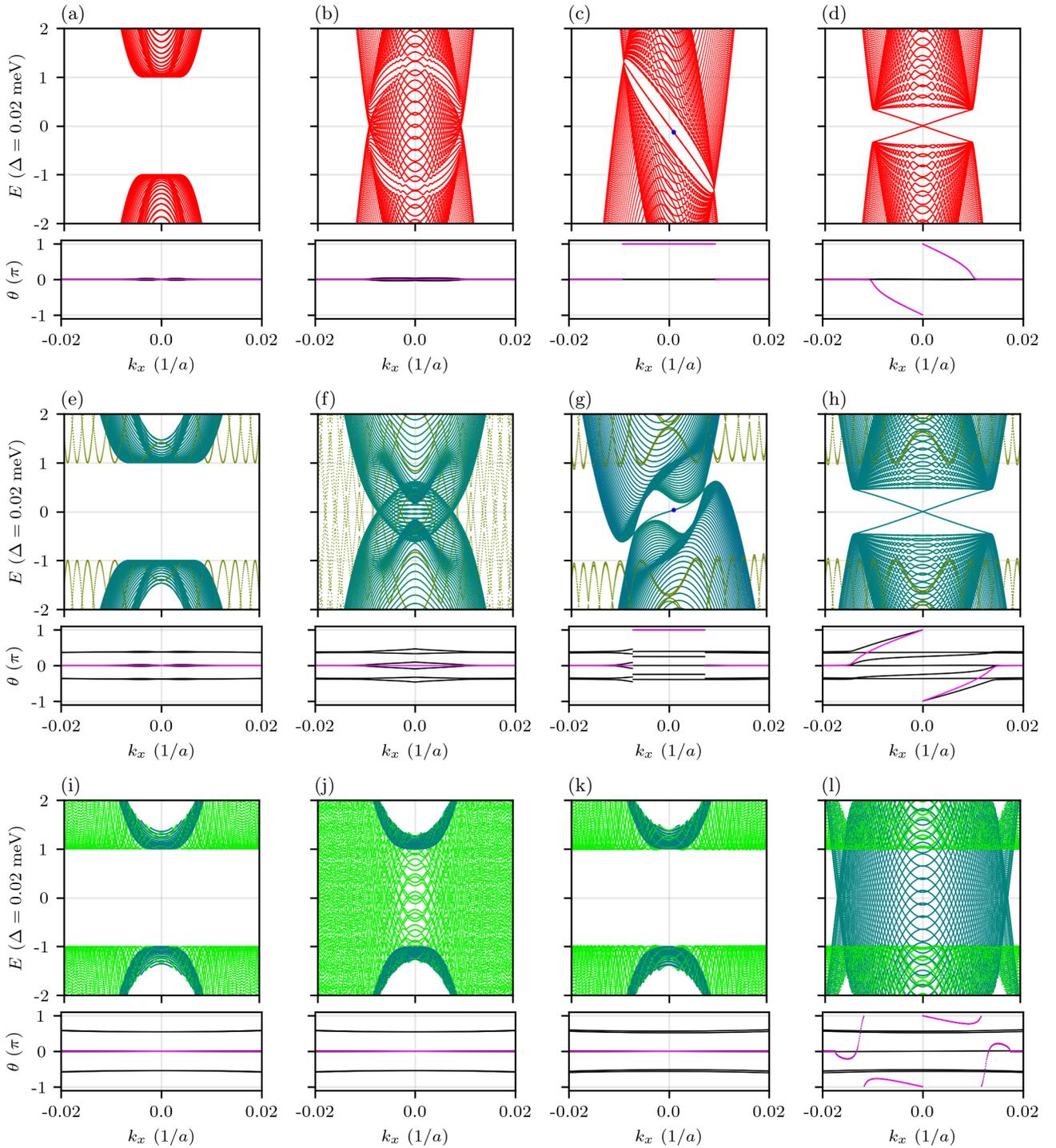

In Fig.~\ref{fig:WCC2D} we present the evolution of the Wannier center, $\theta$, along $k_x$ for a chosen $\mu_l$ and the magnetic fields $B_z = -1\ \mathrm{T}, 0,$ and $1\ \mathrm{T}$.  For $B_z = \pm 1\ \mathrm{T}$, the system is expected to be in a topological state with Chern number $|C| = 1$. The  phases summed over the bands proportional to the polarization are indicated in magenta.
A common feature of all three datasets is the flow of one Wannier center into the neighboring unit cell and the single winding of the polarization for $B_z = \pm 1\ \mathrm{T}$, indicating a Chern number of magnitude $|C| = 1$. This change occurs within a very narrow range $|k_x| < 0.02\ 1/a$. The values of $k_x$ at which the polarization begins to change correspond to the $k_x$-values where, in Figs.~\ref{fig:disp_sc}, the topological gap is open (for $\mu = \mu_{1}$ and $\mu_{2}$). For $B_z = 0$, no winding of the polarization occurs, yielding $C = 0$.
Interestingly, for $\mu = \mu_{3}$ [Fig.~\ref{fig:WCC2D}(d-f)], the Wannier center flow is the most complex (compared to the other two bands) due to the abrupt change in $\theta$  and the partially return of the Wannier center to the original unit cell after crossing at $k_x = 0$. Note that in this band, as reported above, the topological phase is highly fragile, as it is characterized by a weak topological gap on the order of $10^{-4}$~meV.

\subsection{Towards 1D system}
\begin{figure*}[!t]
\includegraphics[]{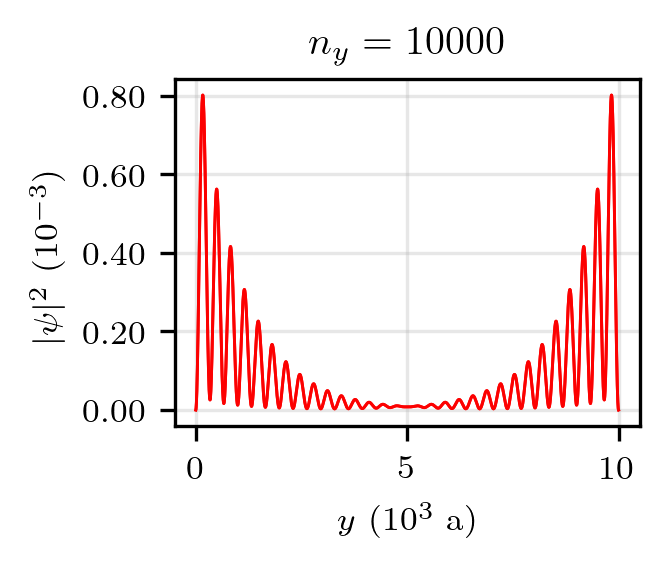}
\put(-120, 124){(a)}
\includegraphics[]{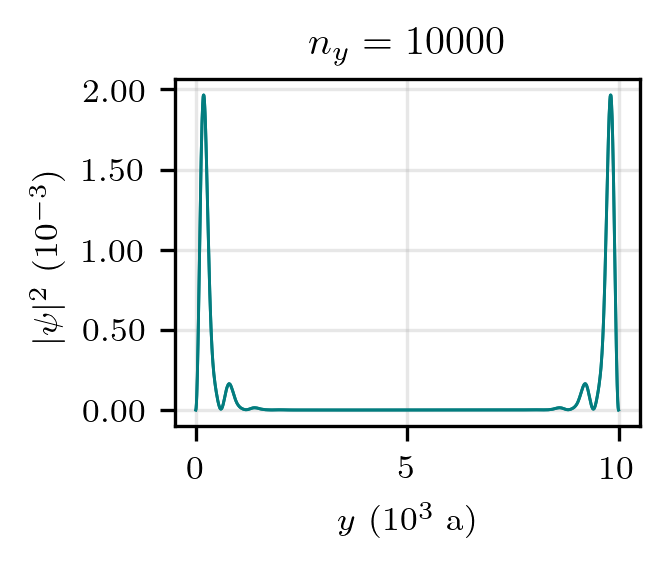}
\put(-120, 124){(b)}
\includegraphics[]{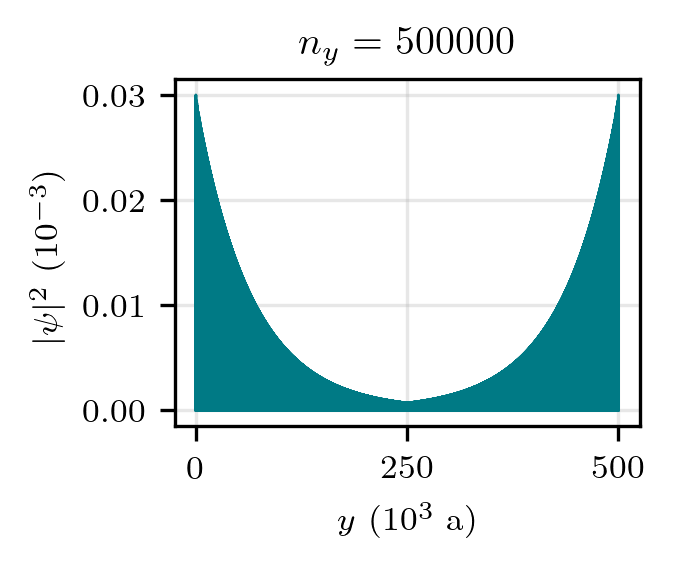}
\put(-120, 124){(c)}
\caption{Squared modulus of the wave function of the edge states determined for the parameters (a) $\mu=\mu_1$ and $B_y=1$~T, (b) $\mu=\mu_2$ and $B_y=3$~T, (c) $\mu=\mu_3$ and $B_z=1$~T. For (a) and (b), the corresponding electron state for which the wave function is represented is indicated by dots in panels~\ref{fig:Ekx1D} (c,g), respectively. As in Fig.~\ref{fig:disp_NM}(a), the contributions of the $d_{xy}$ , $d_{xz}$ , and $d_{yz}$ orbitals are represented using the RGB color scheme.} 
\label{fig:waves}
\end{figure*}
Let us now study the conditions for edge-state formation as one spatial dimension is reduced and the system is gradually transformed into a quasi-one-dimensional nanowire. The analysis in the previous subsection demonstrates that, in the fully 2D regime, the topological transition driven by an in-plane magnetic field is strongly band dependent. Importantly, in this case, to induce a topological phase, a weak but non-zero perpendicular magnetic field component is required. 
The reduction of the system dimension fundamentally relaxes this directional constraint since the wave vector along the confinement direction is no longer a good quantum number with $\langle k_\perp \rangle = 0$. The lateral confinement modifies the SO field such that, in addition to out-of-plane magnetic fields ($B_z$), purely in-plane fields are also sufficient to drive the system into a topological superconducting phase.

To investigate the crossover from a fully 2D system to an effectively 1D geometry, we impose open boundary conditions (OBC) along the $y$-direction while maintaining periodic boundary conditions (PBC) along the $x$-axis. In Fig.~\ref{fig:Ekx1D}, we present the dispersion relations in the superconducting state for the three selected values of $\mu_l$ [see Fig.~\ref{fig:disp_NM}(a)], with the magnetic field applied along different axes. In each panel, the field magnitude is chosen to exceed the critical value above which the system enters the topological phase. For the data shown in Fig.~\ref{fig:Ekx1D}, we assume $n_y = 10000$ lattice sites along the transverse direction, which is sufficiently large to regard the system as quasi-two-dimensional, in the sense that the energy level spacing between transverse subbands is so small that the chemical potential associated with the band minima remains effectively unchanged, with values close to the $\mu_l$ reported in Table~\ref{tab1}. To support the interpretation  we additionally show, beneath each panel in Fig.~\ref{fig:Ekx1D}, the evolution of the corresponding Wannier centers along $k_x$.

In Fig.~\ref{fig:Ekx1D}, we can see that as the transverse dimension is reduced, the originally continuous bands are quantized into a discrete set of subbands, which manifest as a dense manifold of states forming the spectrum. A global superconducting gap is clearly resolved in the left panels presenting results for \( B = 0 \). We can observe that, for the energy bands $\gamma_1$ and $\gamma_2$, the dispersion relations display two intersecting edge modes for magnetic fields applied along the $y$ and $z$ axes [Fig.~\ref{fig:Ekx1D}(c,d,g,h)]. Interestingly, the nature of the edge modes for these two magnetic field directions is qualitatively different.

When a magnetic field is applied perpendicular to the plane, the two intersecting edge modes possess opposite group velocities, propagating clockwise and anticlockwise along the edges [Fig.~\ref{fig:Ekx1D} (d,h)]. This behavior is characteristic of fully gapped two-dimensional topological superconductors with broken time-reversal symmetry, as evidenced by the continuous evolution of a Wannier center into the neighboring unit cell and the associated single winding of the polarization (see panels below).
In contrast, when the magnetic field is oriented along the confinement direction [Fig.~\ref{fig:Ekx1D} (c,g)], the spectrum is not fully gapped and exhibits two asymmetric Dirac-like cones, in the sense that one crosses the zero-energy axis from below, while the other crosses it from above. These cones are connected by antichiral edge modes that propagate in the same direction, with identical group velocity, along the edges perpendicular to the magnetic field. Within the range of crystal momenta $k_x$ for which the antichiral edge modes are present, the Wannier charge center exhibits an abrupt (step-like) shift of its localization toward the boundary of the Brillouin zone. Similar co-propagating antichiral states have been recently predicted in a Rashba 2DEG~\cite{Ruiz2025} and, earlier, in a modified Haldane model~\cite{Colomes2018}.
\begin{figure*}[!t]
\includegraphics[]{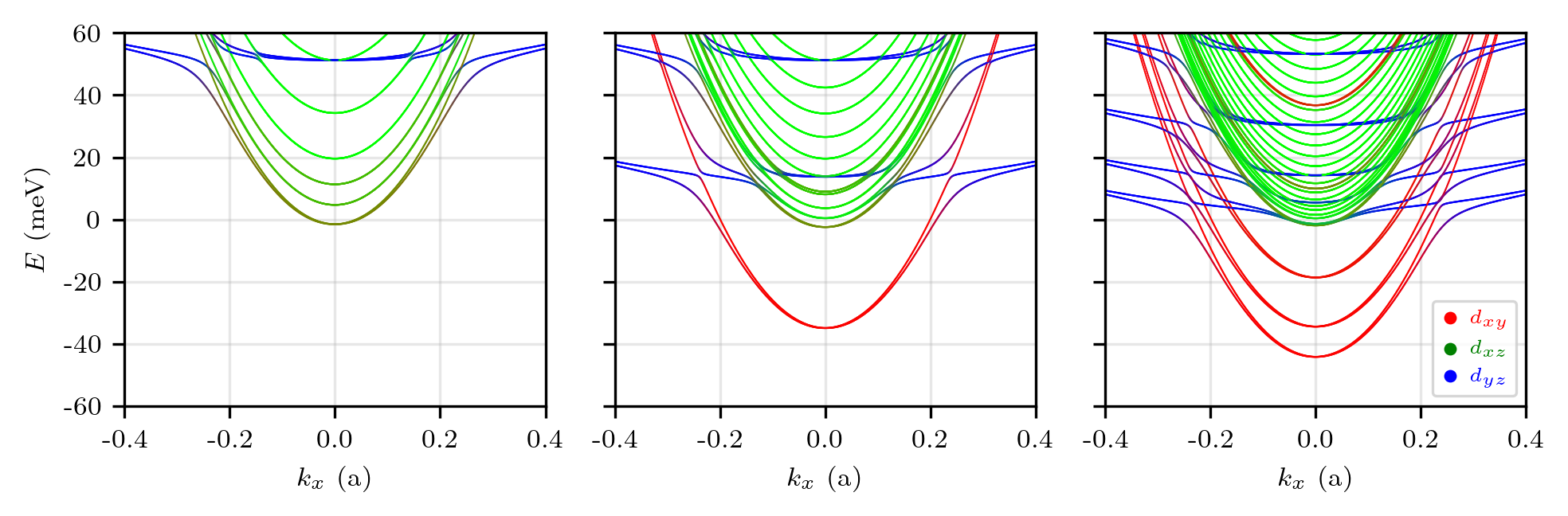}
\put(-438, 152){(a)}
\put(-290, 152){(b)}
\put(-142, 152){(c)}
\caption{Energy dispersions $E(k_x)$ for LAO/STO nanowires in the normal state determined for the nanowire widths (a) $n_y=12$, (b) $n_y=25$, and (c) $n_y=50$ lattice sites. the contributions of the $d_{xy}$, $d_{xz}$, and $d_{yz}$ orbitals are represented using the RGB color scheme} 
\label{fig:disp}
\end{figure*}
\begin{figure}[!t]
\includegraphics[]{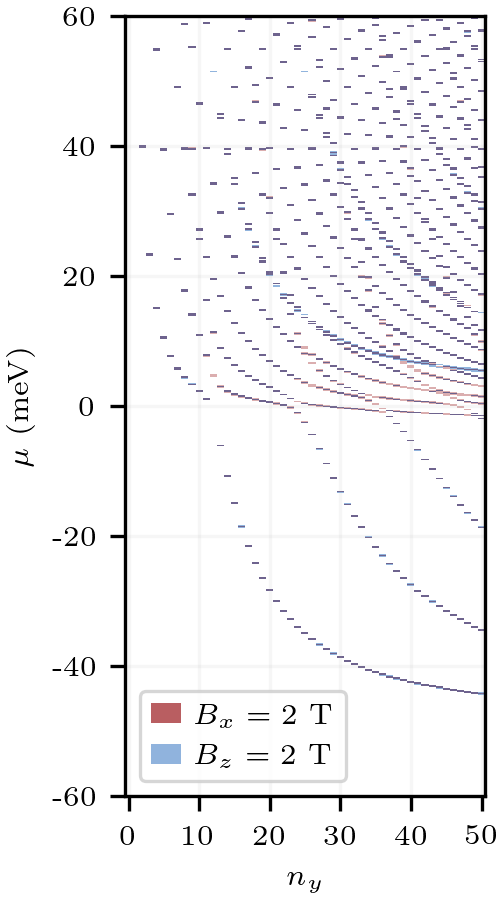}\hfill
\put(-92, 218){(a)}
\includegraphics[]{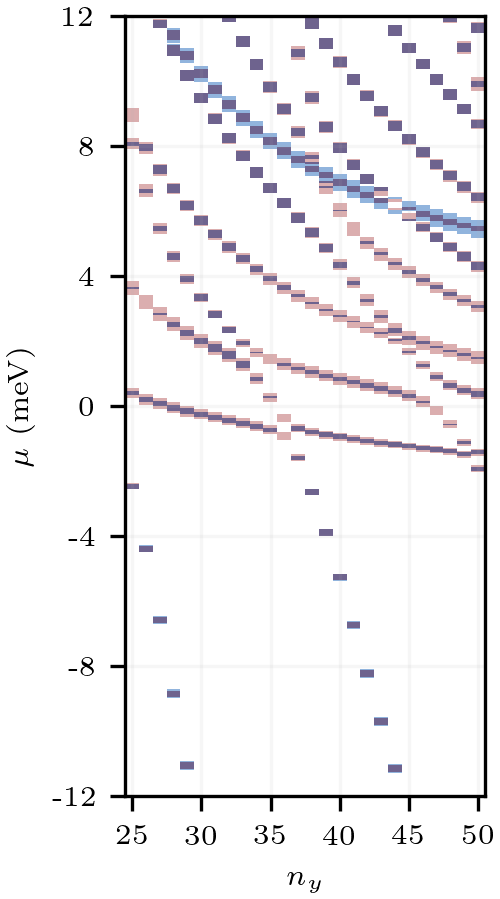}
\put(-92, 218){(b)}
\caption{(a) Phase diagram indicating the topological invariant $\mathbb{Z}_2 = -1$ as a function of the nanowire width $n_y$ and the chemical potential $\mu$. Panel (b) displays a magnified view of (a) to highlight the differences between the diagrams for the two considered orientations of the magnetic field. Results are shown for $B_x = 2$ T (red) and $B_z = 2$ T (blue).} 
\label{fig:diagram}
\end{figure}
 
Note that the propagation direction of the edge states associated with the bands $\gamma_1$ and $\gamma_2$ is opposite for a given orientation of the magnetic field, which results from the opposite chirality of these bands. Although the spectrum for $\mu_{1}$ is predominantly composed of electronic states with $d_{xy}$ orbital character [Fig.~\ref{fig:Ekx1D}(a-d)] , for $\mu_{2}$ [Fig.~\ref{fig:Ekx1D}(e-h)], the edge mode consists of an almost equal mixture of the $d_{yz}$ and $d_{xz}$ orbitals and coexists with a fully gapped spectrum of states composed of $d_{xy}$ and $d_{yz}$ states. For the two considered chemical potentials, $\mu_{1}$ and $\mu_{2}$, no discernible edge modes are observed when the magnetic field is oriented along the $x$ axis [Fig.~\ref{fig:Ekx1D} (b,f)]; consistently, the Wannier center does not exhibit any flow between unit cells. Figure~\ref{fig:Ekx1D} (j-l) demonstrates that the absence of edge states is also characteristic for the chemical potential $\mu_{3}$, which corresponds to the minimum of the third electronic band in Fig.~\ref{fig:disp_NM}(a). In this case, the system is gapless when $B$ is applied along the $x$ axis, whereas it remains fully gapped when the magnetic field is aligned with the confinement direction $y$. The absence of edge states for the band $\gamma_3$ in this case arises from the weak strength of the SO coupling [Fig.~\ref{fig:disp_NM}(b)], which prevents their formation.

The linear character of the edge-state dispersions driven by $B_y$ and $B_z$ indicates the presence of a high-velocity Majorana channel, which differs between the two bands, $\gamma_{1}$ and $\gamma_{2}$. In panels (c,g) of Fig.~\ref{fig:Ekx1D}, we indicate two points (dots) corresponding to the edge modes for which the modulus squared of the wave function is displayed in Fig.~\ref{fig:waves} (a,b). It clearly demonstrates a pronounced spatial localization of these states at the boundaries, characterized by a band-dependent decay length that is substantially shorter for the band $\gamma_2$ than for the band $\gamma_1$. 
The decay length is governed by the Fermi velocity and the effective energy gap through the relation
\[
\xi = \frac{\hbar V_F}{\Delta_{\mathrm{eff}}}\,,
\]
which, for the bands under consideration, takes the following values: for the band $\gamma_{1}$, the Fermi velocity is $V_F = 17.4~\text{meV}a/\hbar$ and the effective gap is $\Delta_{\mathrm{eff}} \approx 0.02$~meV, yielding $\xi = 870a$, while for the band $\gamma_2$ the corresponding quantities are $V_F = 6.4~\text{meV}a/\hbar$ and $\Delta_{\mathrm{eff}} \approx 0.02$~meV, giving $\xi = 320a$. These values are consistent with the edge state decay profiles presented in Fig.~\ref{fig:waves} (a,b).

The band $\gamma_3$ and a magnetic field oriented along the $z$ axis should be considered separately. Although the dispersion relation shown in Fig.~\ref{fig:Ekx1D} (l) does not exhibit any obvious edge states for this band, the analysis of the Wannier centers (panel below) reveals a flow between unit cells with a characteristic re-entrant behavior. Note that, in this configuration, the topological gap of the fully 2D system was found to be extremely small, on the order of $10^{-4}$~meV. This corresponds to a very large coherence length, $\xi = 41500a$ ($V_F = 16.6~\text{meV}a/\hbar$), which makes the edge states effectively undetectable for an assumed confinement with $n_y = 10000$. To unambiguously demonstrate the presence of edge states in this magnetic field orientation, we increase the system size in the $y$ direction to $n_y = 500000$ lattice sites and compute the wave functions of the two states closest to zero energy. Figure~\ref{fig:waves}(c) clearly demonstrates that these states exhibit spatial localization at the boundaries along the confinement direction, thereby corroborating the topological character of the system.
\begin{figure*}[!t]
\includegraphics[]{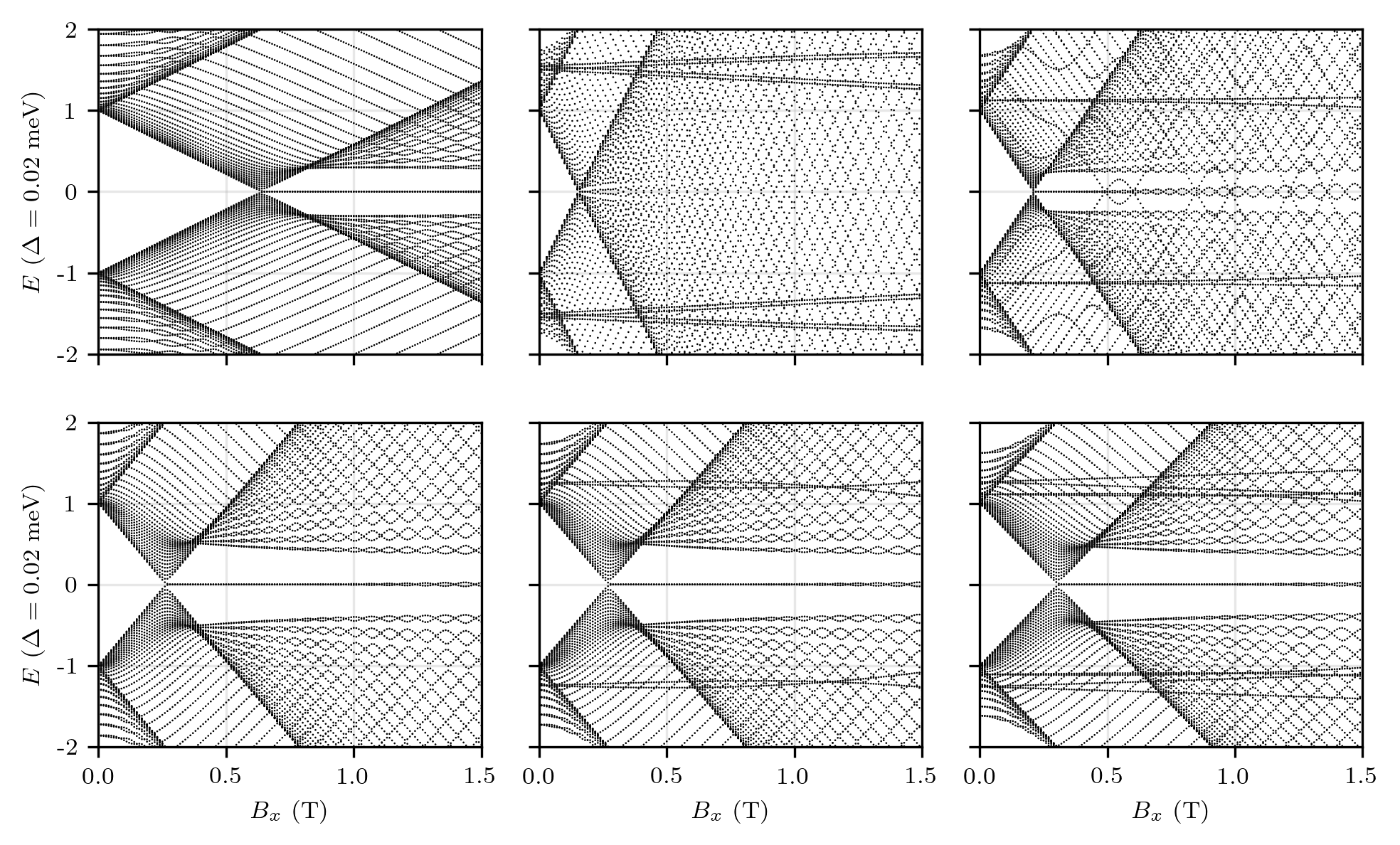}
\put(-440, 282){(a)}
\put(-291, 282){(b)}
\put(-142, 282){(c)}
\put(-440, 148){(d)}
\put(-291, 148){(e)}
\put(-142, 148){(f)}
\caption{Energy spectrum of the LAO/STO nanowire as a function of a magnetic field applied along the nanowire axis, $B_x$. Results are shown for the chemical potential fixed at the minimum of the three lowest subbands, for nanowire widths of (a–c) \(n_y = 12\) and (d–f) \(n_y = 50\) lattice sites, assuming $n_x=10000$. The subbands chosen for the analysis are shown in Fig.~\ref{fig:disp}(a, c).} 
\label{fig:MZM}
\end{figure*}
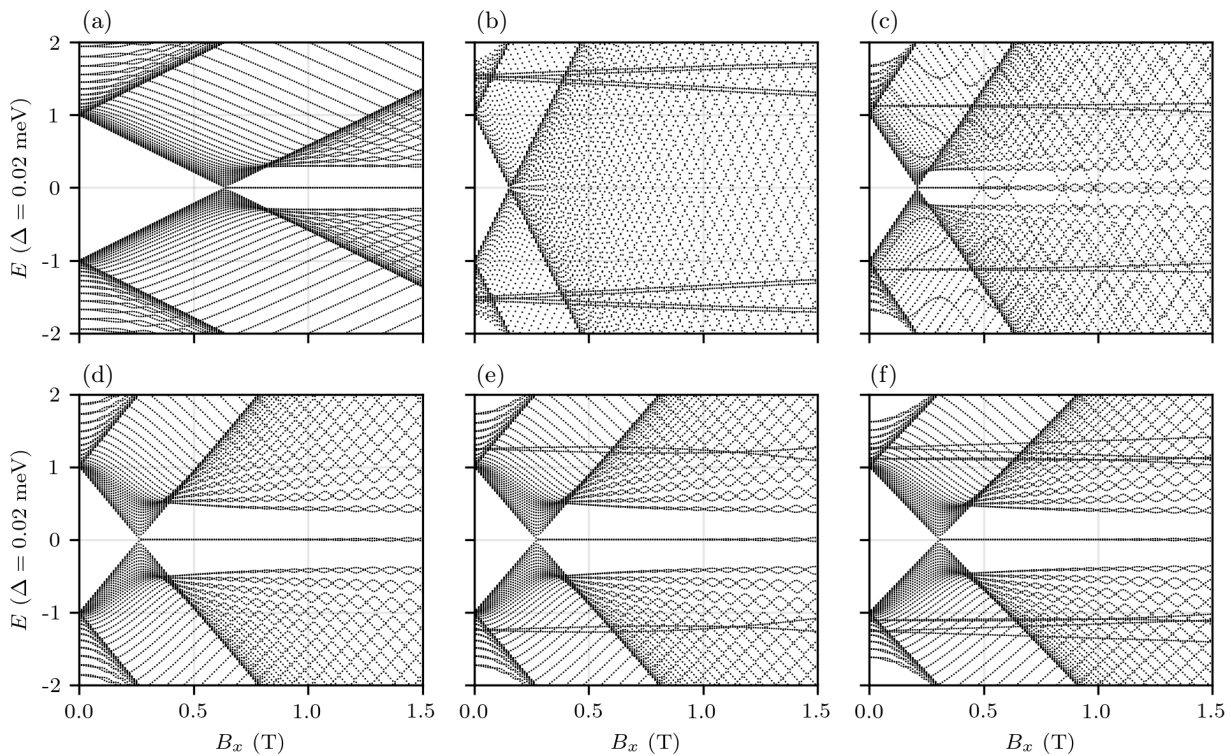










\subsection{Towards Majorana zero modes}
Now, we analyze the conditions for the existence of MZMs in LAO/STO nanowires. For this purpose, we consider a narrow quasi-1D nanoribbon with width \( n_y \leq 50 \). Fig.~\ref{fig:disp} presents the energy dispersions in the normal state determined for nanowires with different widths $n_y=12, 25$, and $50$ lattice sites.  The color scale indicates the contribution from the individual \(d\) orbitals. As shown in Fig.~\ref{fig:disp}, while in the wider nanowire with \(n_y = 50\) the low-energy bands are almost entirely determined by the \(d_{xy}\) orbitals, a reduction of the nanowire width enhances the role of the higher-energy \(d_{xz}/d_{yz}\) states. In particular, for \(n_y = 12\) [Fig.~\ref{fig:disp}(a)] only the lowest subband has a significant \(d_{xy}\) character, whereas the higher-energy subbands are characterized by a significant contribution from the \(d_{xz}\) orbitals.  

Upon reduction to one dimension, the \(\mathbb{Z}_2\) topological index is the unique invariant that characterizes the superconducting phase, in accordance with the standard topological classification
\begin{equation}
    \mathbb{Z}_2 = \mathrm{sgn}\!\left[\mathrm{Pf}\,(A(0))\,\mathrm{Pf}\,(A(\pi))\right],
    \label{eq:1D_Z2_invariant}
\end{equation}
where $\mathrm{Pf}(\dots)$ denotes the Pfaffian and $A(k)$ is a quasi-1D Hamiltonian $H(k_x)$ in Majorana basis 
\begin{equation}
A(k_x) = U H(k_x) U^\dagger,
\end{equation}
where
\begin{equation}
      U = \mathbb{1}_{n_y\times n_y} \otimes \frac{1}{\sqrt{2}}\mqty(\mathbb{1}_{6\times 6} & \mathbb{1}_{6\times 6} \\ -i\mathbb{1}_{6\times 6} & i\mathbb{1}_{6\times 6}).
\end{equation}
Negative sign of Pfaffians product ($\mathbb{Z}_2=-1$) indicates a topological phase.\\

In Fig.~\ref{fig:diagram}(a), we present the phase diagram for LAO/STO nanowires, as a function of the chemical potential and nanowire width, calculated for a magnetic field of $2$T oriented along the $x$- and $z$-axes. The orientation along the confinement direction $y$ is not included here, as it is characterized exclusively by the trivial phase. Although the extents of the topological phase in Fig.~\ref{fig:diagram}(a) appear similar for the two field orientations, they differ quantitatively. A direct comparison in the range of $n_y$ where these discrepancies are most pronounced is shown in Fig.~\ref{fig:diagram}(b), which clearly displays the parameter regime in which the topological phase is induced solely by the $B_x$ component (aligned with the nanowire axis) and is absent when the magnetic field is applied along the $z$-axis. The possibility of inducing MZMs in quasi-one-dimensional LAO/STO nanowires by means of the longitudinal magnetic field $B_x$ is consistent with previous analyses~\cite{Perroni2019,Fukaya2018}.
\begin{figure*}[!t]
\includegraphics[]{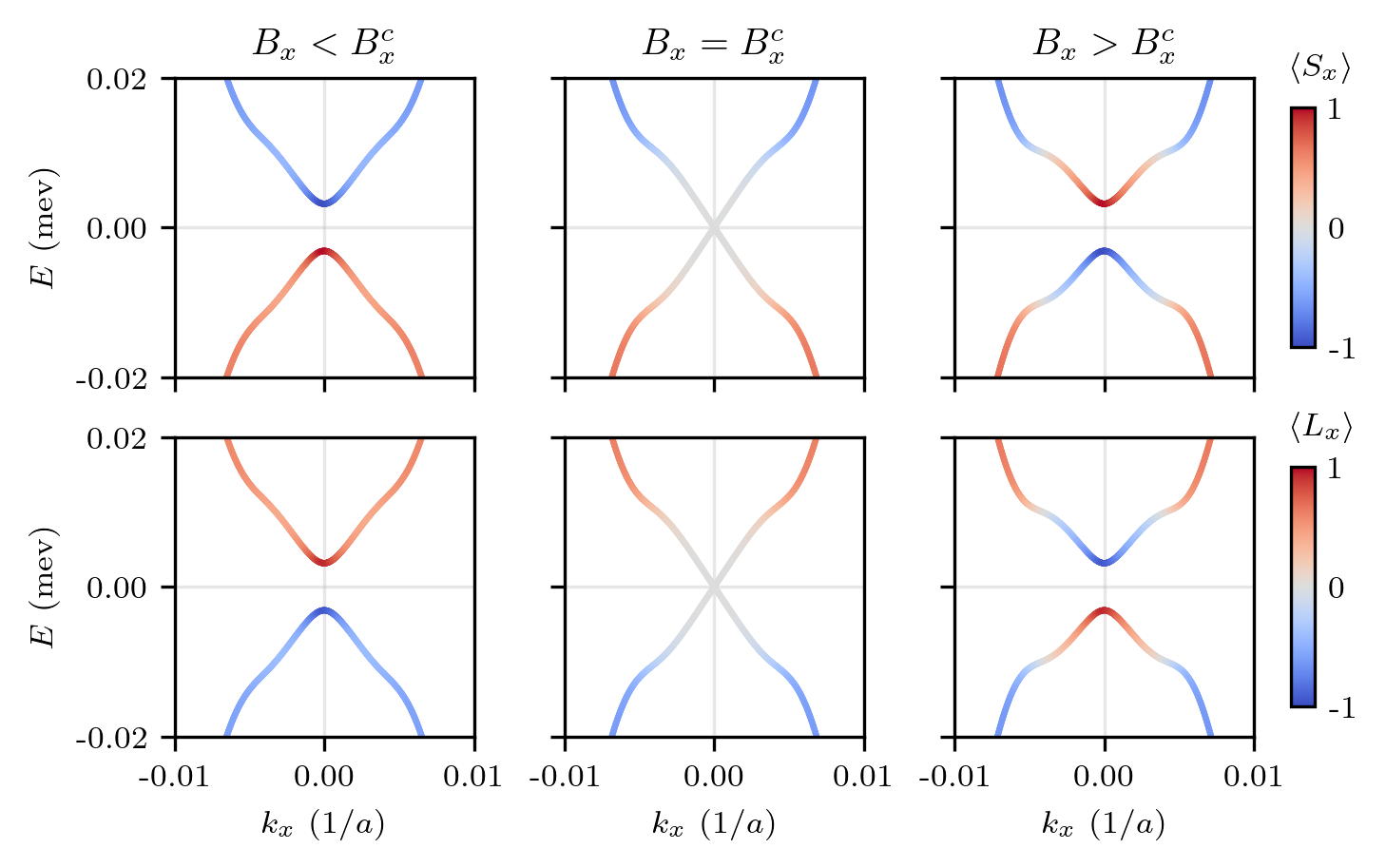}
\put(-305, 160){(a)}
\put(-205, 160){(b)}
\put(-107, 160){(c)}
\put(-305, 70){(d)}
\put(-205, 70){(e)}
\put(-107, 70){(f)}
\caption{Dispersion relations $E(k_x)$ of the superconducting LAO/STO nanowire with width \(n_y = 12\) and a magnetic field applied along the nanowire axis, \(B_x\), with a magnitude corresponding to the trivial phase (a,d), the topological transition (b,e), and the topological phase (c,f). The color scale represents the average value of the \(x\)-component of the spin (a-c) and the orbital momentum (d-f). Chemical potential was set at the lowest subbands minimum level visible in Fig.~\ref{fig:disp}(a).}
\label{fig:SL}
\end{figure*}
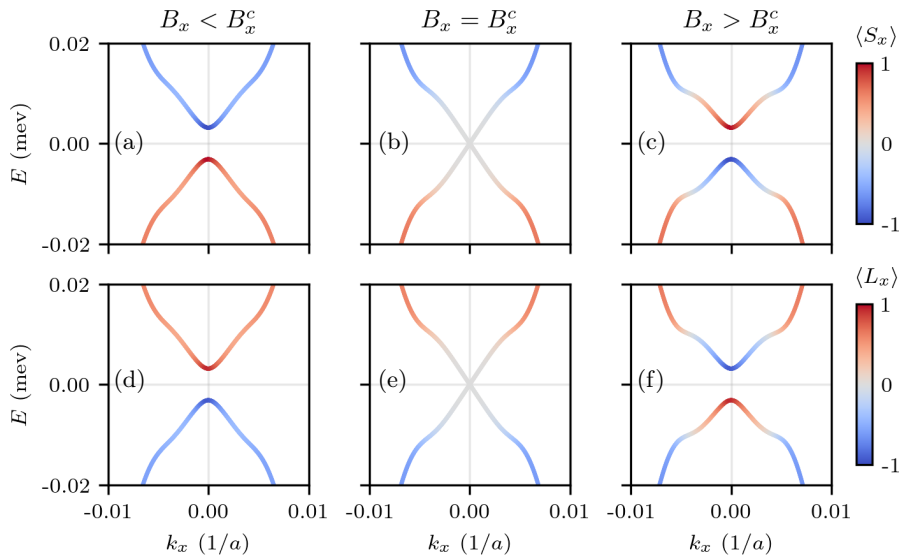

To elucidate the existence of MZMs in LAO/STO nanowires, we impose open boundary conditions also along the $x$-axis, choosing $n_x = 10000$ lattice sites to reduce hybridization between spatially separated Majorana states. The corresponding energy spectrum, for two representative nanowire widths: (i) a narrow wire $n_y = 12$, for which the $d_{xz}$ orbitals contribute to the low-energy electronic structure, and (ii) a wide wire with $n_y = 50$, where the lowest-energy states are dominated by the $d_{xy}$ orbital, is presented in Fig.~\ref{fig:MZM}. The subsequent panels display the three lowest subbands; in each case, the chemical potential $\mu$ is chosen to coincide with its minimum.

For $n_y = 50$ [Fig.~\ref{fig:MZM} (d,e,f)], the energy spectrum closely resembles that typically obtained for Rashba nanowires, consistent with expectations, as the $d_{xy}$ orbital, which mainly participates in the considered subband, is characterized by an effective Rashba type SO coupling — see Appendix~\ref{sec:A1}. The energy spectrum is different for a narrow nanowire, in which  $d_{xz}$ orbitals contribute to the subbands. In Fig.~\ref{fig:MZM} (a–c), one can see that, although the energy spectrum associated with the lowest subband remains similar to the standard Rashba nanowire with well-pronounced MZMs - reflecting the significant contribution of the $d_{xy}$ orbital to this subband - the spectra of the higher bands, as well as the occurrence of MZMs therein, are strongly modified. 
For some of these higher bands, zero-energy MZMs are not observed at all, or their hybridization is so strong (due to their large decay length relative to the considered nanowire length) that they cannot be clearly resolved. 
This scenario occurs for the second band, where in Fig.~\ref{fig:MZM}(b) we observe only a weakly developed topological gap, which increases as the nanowire length is increased (data not shown). A qualitatively different behavior for this higher-energy state results from the participation of $d_{xz}$ orbitals (higher helical bands), including one characterized by an extremely small topological gap and a long coherence length.

Finally, for a narrow nanowire with a well-defined topological transition occurring in the first subband, we study the inversion of the quasiparticle spin and orbital angular momentum (as the electrons are described by \(d\)-orbitals) at the topological phase transition. In Fig.~\ref{fig:SL}, we present the dispersion relations \(E(k_x)\) determined for a nanowire with width \(n_y = 12\) and three magnetic-field values \(B_x\) corresponding to the trivial phase, the topological transition, and the topological state. We consider only the \(S_x\) and \(L_x\) components, as the remaining ones provide a negligible contribution and are nearly zero. 
As demonstrated in Ref.~\cite{Szumniak} for Rashba nanowires, and confirmed by our results [Fig.~\ref{fig:SL}(a-c)] for LAO/STO nanowires, the spin of quasiparticle states around the topological gap closing reverses sign upon crossing the transition due to band inversion. Importantly, we find a similar behavior for the orbital momentum, which vanishes at the topological transition at \(E = 0\) and then reverses with respect to the trivial phase near \(k = 0\). This orbital-momentum reversal can be considered a signature of topological superconductivity in systems with an electronic structure defined by states with nonzero $L$, as in the case of LAO/STO, where the electons are described by \(d\) orbitals.

\section{Summary}
\label{sec:Summary}

In this paper we have systematically investigated the topological superconductivity in nanostructres based on 2DEG created at the (001) LAO/STO interface. 
The LAO/STO 2DEG is a multiband system characterized by strong SO coupling which lifts the spin degeneracy of the electronic \(d\)-orbitals, thereby splitting them into distinct Fermi-surface sheets with opposite spin helicities. In momentum space, these Fermi sheets host counter-propagating chiral spin textures, resulting in a mutual cancellation of the associated Berry phases and, consequently, a topologically trivial state. To realize topological superconductivity in the presence of conventional $s$-wave pairing, this cancellation must be lifted by applying an external magnetic field, which should be sufficiently strong to effectively depopulate one of the helical bands.  In this paper we consider LAO/STO nanostructures in the presence of the external magnetic field applied in different orientations. By employing a realistic tight-binding model that captures the intrinsic multiband character of the system, along with both atomic and Rashba spin-orbit couplings, we have studied the topological phase diagrams for fully two-dimensional, quasi-two-dimensional and one-dimensional geometries.

For the fully two-dimensional system, we have demonstrated that a purely in-plane magnetic field is insufficient to induce a topological transition, as the in-plane spin alignment, determined by the SO coupling, inhibits the formation of the helical gap. In this case, a finite out-of-plane component of the magnetic field is required to drive the topological phase transition. 
The corresponding critical field exhibits a pronounced dependence on the electronic band, governed not only by the spin Zeeman splitting but also by the orbital Zeeman effect and atomic SO coupling. We have demonstrated that the critical magnetic-field magnitude for the lowest-lying band, predominantly of \(d_{xy}\) character, is approximately isotropic with respect to the magnetic field orientation. By comparison, the higher-energy bands, arising from hybridized \(d_{xz}\) and \(d_{yz}\) orbitals, display a pronounced anisotropy of the critical field.

Restricting the system to a quasi-two-dimensional geometry - by appropriate OBC in the $y$ direction - relaxes the directional constraints on the applied magnetic field. In this regime, purely in-plane magnetic fields become sufficient to drive the system into a topological superconducting phase. In this case, we have shown that the nature of the resulting edge modes is strongly dictated by the field orientation: an out-of-plane field generates standard counter-propagating chiral modes, while a transverse in-plane field can induce unconventional co-propagating antichiral edge modes. 

Finally, we have analyzed the emergence of topological superconductivity and MZMs in LAO/STO nanowires (1D systems) with different widths, demonstrating that the induction of these exotic states in subbands composed of \(d_{xz/yz}\) orbitals is characterized by an extremely long decay length, making the creation of such states impossible in some subbands for realistic nanowire lengths. Moreover, we have found that the orbital momentum reverses its sign at the topological transition, making it a clear signature of topological superconductivity.

Note that the analysis presented in this paper was performed under the assumption of an s‑wave pairing symmetry, in accordance with experimental measurements indicating a nodeless superconducting gap in LAO/STO 2DEG\cite{Monteiro2017}. It should be emphasized, however, that the issue of the gap symmetry in LAO/STO remains unresolved and is still under active debate~\cite{Zegrodnik_2020,Wojcik_dome_schrodinger_poisson_sc_2024}. In our recent theoretical work, we demonstrated that most of the experimentally observed features, such as the characteristic dome-shaped dependence of the critical temperature on the electron density, can be explained as a consequence of an extended s‑wave gap symmetry. Nevertheless, our additional analysis carried out for this type of symmetry does not reveal any significant differences relative to the results presented in this paper. 
The absence of substantial discrepancies arises from the fact that the regime with an enhanced superconducting gap corresponds to low electron concentrations, for which the Fermi surface is localized in the vicinity of the $\Gamma$ point of the Brillouin zone. In this region, the momentum dependence of the extended s‑wave order parameter is weak, rendering it effectively similar to a conventional s‑wave symmetry.

\begin{acknowledgments}
This research was partially supported by a subsidy from the Polish
Ministry of Science and Higher Education and the program „Excellence initiative – research university” for the AGH University. Computing infrastructure PLGrid (HPC Centers: ACK Cyfronet AGH) was used within computational grant no. PLG/2024/017744.
 
\end{acknowledgments}

\section*{Data Availability}
The data that support the findings of this article \cite{LAO_STO_data_2025} and the implementation of numerical scheme \cite{LAO_STO_repo} are openly available.


\begin{thebibliography}{69}%
\makeatletter
\providecommand \@ifxundefined [1]{%
 \@ifx{#1\undefined}
}%
\providecommand \@ifnum [1]{%
 \ifnum #1\expandafter \@firstoftwo
 \else \expandafter \@secondoftwo
 \fi
}%
\providecommand \@ifx [1]{%
 \ifx #1\expandafter \@firstoftwo
 \else \expandafter \@secondoftwo
 \fi
}%
\providecommand \natexlab [1]{#1}%
\providecommand \enquote  [1]{``#1''}%
\providecommand \bibnamefont  [1]{#1}%
\providecommand \bibfnamefont [1]{#1}%
\providecommand \citenamefont [1]{#1}%
\providecommand \href@noop [0]{\@secondoftwo}%
\providecommand \href [0]{\begingroup \@sanitize@url \@href}%
\providecommand \@href[1]{\@@startlink{#1}\@@href}%
\providecommand \@@href[1]{\endgroup#1\@@endlink}%
\providecommand \@sanitize@url [0]{\catcode `\\12\catcode `\$12\catcode
  `\&12\catcode `\#12\catcode `\^12\catcode `\_12\catcode `\%12\relax}%
\providecommand \@@startlink[1]{}%
\providecommand \@@endlink[0]{}%
\providecommand \url  [0]{\begingroup\@sanitize@url \@url }%
\providecommand \@url [1]{\endgroup\@href {#1}{\urlprefix }}%
\providecommand \urlprefix  [0]{URL }%
\providecommand \Eprint [0]{\href }%
\providecommand \doibase [0]{https://doi.org/}%
\providecommand \selectlanguage [0]{\@gobble}%
\providecommand \bibinfo  [0]{\@secondoftwo}%
\providecommand \bibfield  [0]{\@secondoftwo}%
\providecommand \translation [1]{[#1]}%
\providecommand \BibitemOpen [0]{}%
\providecommand \bibitemStop [0]{}%
\providecommand \bibitemNoStop [0]{.\EOS\space}%
\providecommand \EOS [0]{\spacefactor3000\relax}%
\providecommand \BibitemShut  [1]{\csname bibitem#1\endcsname}%
\let\auto@bib@innerbib\@empty
\bibitem [{\citenamefont {Sato}\ and\ \citenamefont {Ando}(2017)}]{Sato_2017}%
  \BibitemOpen
  \bibfield  {author} {\bibinfo {author} {\bibfnamefont {M.}~\bibnamefont
  {Sato}}\ and\ \bibinfo {author} {\bibfnamefont {Y.}~\bibnamefont {Ando}},\
  }\bibfield  {title} {\bibinfo {title} {Topological superconductors: a
  review},\ }\href {https://doi.org/10.1088/1361-6633/aa6ac7} {\bibfield
  {journal} {\bibinfo  {journal} {Reports on Progress in Physics}\ }\textbf
  {\bibinfo {volume} {80}},\ \bibinfo {pages} {076501} (\bibinfo {year}
  {2017})}\BibitemShut {NoStop}%
\bibitem [{\citenamefont {Bernevig}\ and\ \citenamefont
  {Hughes}(2013)}]{BernevigHughes2013}%
  \BibitemOpen
  \bibfield  {author} {\bibinfo {author} {\bibfnamefont {B.~A.}\ \bibnamefont
  {Bernevig}}\ and\ \bibinfo {author} {\bibfnamefont {T.~L.}\ \bibnamefont
  {Hughes}},\ }\href@noop {} {\emph {\bibinfo {title} {Topological Insulators
  and Topological Superconductors}}}\ (\bibinfo  {publisher} {Princeton
  University Press},\ \bibinfo {address} {Princeton, NJ},\ \bibinfo {year}
  {2013})\BibitemShut {NoStop}%
\bibitem [{\citenamefont {Fu}\ and\ \citenamefont {Kane}(2008)}]{FuKane2008}%
  \BibitemOpen
  \bibfield  {author} {\bibinfo {author} {\bibfnamefont {L.}~\bibnamefont
  {Fu}}\ and\ \bibinfo {author} {\bibfnamefont {C.~L.}\ \bibnamefont {Kane}},\
  }\bibfield  {title} {\bibinfo {title} {Superconducting proximity effect and
  majorana fermions at the surface of a topological insulator},\ }\href@noop {}
  {\bibfield  {journal} {\bibinfo  {journal} {Physical Review Letters}\
  }\textbf {\bibinfo {volume} {100}},\ \bibinfo {pages} {096407} (\bibinfo
  {year} {2008})}\BibitemShut {NoStop}%
\bibitem [{\citenamefont {Pientka}\ \emph {et~al.}(2013)\citenamefont
  {Pientka}, \citenamefont {Glazman},\ and\ \citenamefont {von
  Oppen}}]{Pientka2013}%
  \BibitemOpen
  \bibfield  {author} {\bibinfo {author} {\bibfnamefont {F.}~\bibnamefont
  {Pientka}}, \bibinfo {author} {\bibfnamefont {L.~I.}\ \bibnamefont
  {Glazman}},\ and\ \bibinfo {author} {\bibfnamefont {F.}~\bibnamefont {von
  Oppen}},\ }\bibfield  {title} {\bibinfo {title} {Topological superconducting
  phase in helical shiba chains},\ }\href@noop {} {\bibfield  {journal}
  {\bibinfo  {journal} {Physical Review B}\ }\textbf {\bibinfo {volume} {88}},\
  \bibinfo {pages} {155420} (\bibinfo {year} {2013})}\BibitemShut {NoStop}%
\bibitem [{\citenamefont {Alicea}(2010)}]{Alicea2010}%
  \BibitemOpen
  \bibfield  {author} {\bibinfo {author} {\bibfnamefont {J.}~\bibnamefont
  {Alicea}},\ }\bibfield  {title} {\bibinfo {title} {Majorana fermions in a
  tunable semiconductor device},\ }\href@noop {} {\bibfield  {journal}
  {\bibinfo  {journal} {Physical Review B}\ }\textbf {\bibinfo {volume} {81}},\
  \bibinfo {pages} {125318} (\bibinfo {year} {2010})}\BibitemShut {NoStop}%
\bibitem [{\citenamefont {Sau}\ \emph {et~al.}(2010)\citenamefont {Sau},
  \citenamefont {Lutchyn}, \citenamefont {Tewari},\ and\ \citenamefont
  {Das~Sarma}}]{Sau2010}%
  \BibitemOpen
  \bibfield  {author} {\bibinfo {author} {\bibfnamefont {J.~D.}\ \bibnamefont
  {Sau}}, \bibinfo {author} {\bibfnamefont {R.~M.}\ \bibnamefont {Lutchyn}},
  \bibinfo {author} {\bibfnamefont {S.}~\bibnamefont {Tewari}},\ and\ \bibinfo
  {author} {\bibfnamefont {S.}~\bibnamefont {Das~Sarma}},\ }\bibfield  {title}
  {\bibinfo {title} {Generic new platform for topological quantum computation
  using semiconductor heterostructures},\ }\href@noop {} {\bibfield  {journal}
  {\bibinfo  {journal} {Physical Review Letters}\ }\textbf {\bibinfo {volume}
  {104}},\ \bibinfo {pages} {040502} (\bibinfo {year} {2010})}\BibitemShut
  {NoStop}%
\bibitem [{\citenamefont {Kitaev}(2003)}]{Kitaev2003}%
  \BibitemOpen
  \bibfield  {author} {\bibinfo {author} {\bibfnamefont {A.~Y.}\ \bibnamefont
  {Kitaev}},\ }\bibfield  {title} {\bibinfo {title} {Fault-tolerant quantum
  computation by anyons},\ }\href@noop {} {\bibfield  {journal} {\bibinfo
  {journal} {Annals of Physics}\ }\textbf {\bibinfo {volume} {303}},\ \bibinfo
  {pages} {2} (\bibinfo {year} {2003})}\BibitemShut {NoStop}%
\bibitem [{\citenamefont {Lutchyn}\ \emph {et~al.}(2010)\citenamefont
  {Lutchyn}, \citenamefont {Sau},\ and\ \citenamefont
  {Das~Sarma}}]{Lutchyn2010}%
  \BibitemOpen
  \bibfield  {author} {\bibinfo {author} {\bibfnamefont {R.~M.}\ \bibnamefont
  {Lutchyn}}, \bibinfo {author} {\bibfnamefont {J.~D.}\ \bibnamefont {Sau}},\
  and\ \bibinfo {author} {\bibfnamefont {S.}~\bibnamefont {Das~Sarma}},\
  }\bibfield  {title} {\bibinfo {title} {Majorana fermions and a topological
  phase transition in semiconductor-superconductor heterostructures},\
  }\href@noop {} {\bibfield  {journal} {\bibinfo  {journal} {Physical Review
  Letters}\ }\textbf {\bibinfo {volume} {105}},\ \bibinfo {pages} {077001}
  (\bibinfo {year} {2010})}\BibitemShut {NoStop}%
\bibitem [{\citenamefont {Oreg}\ \emph {et~al.}(2010)\citenamefont {Oreg},
  \citenamefont {Refael},\ and\ \citenamefont {von Oppen}}]{Oreg2010}%
  \BibitemOpen
  \bibfield  {author} {\bibinfo {author} {\bibfnamefont {Y.}~\bibnamefont
  {Oreg}}, \bibinfo {author} {\bibfnamefont {G.}~\bibnamefont {Refael}},\ and\
  \bibinfo {author} {\bibfnamefont {F.}~\bibnamefont {von Oppen}},\ }\bibfield
  {title} {\bibinfo {title} {Helical liquids and majorana bound states in
  quantum wires},\ }\href@noop {} {\bibfield  {journal} {\bibinfo  {journal}
  {Physical Review Letters}\ }\textbf {\bibinfo {volume} {105}},\ \bibinfo
  {pages} {177002} (\bibinfo {year} {2010})}\BibitemShut {NoStop}%
\bibitem [{\citenamefont {Mourik}\ \emph {et~al.}(2012)\citenamefont {Mourik},
  \citenamefont {Zuo}, \citenamefont {Frolov}, \citenamefont {Plissard},
  \citenamefont {Bakkers},\ and\ \citenamefont {Kouwenhoven}}]{Mourik2012}%
  \BibitemOpen
  \bibfield  {author} {\bibinfo {author} {\bibfnamefont {V.}~\bibnamefont
  {Mourik}}, \bibinfo {author} {\bibfnamefont {K.}~\bibnamefont {Zuo}},
  \bibinfo {author} {\bibfnamefont {S.~M.}\ \bibnamefont {Frolov}}, \bibinfo
  {author} {\bibfnamefont {S.~R.}\ \bibnamefont {Plissard}}, \bibinfo {author}
  {\bibfnamefont {E.~P. A.~M.}\ \bibnamefont {Bakkers}},\ and\ \bibinfo
  {author} {\bibfnamefont {L.~P.}\ \bibnamefont {Kouwenhoven}},\ }\bibfield
  {title} {\bibinfo {title} {Signatures of majorana fermions in hybrid
  superconductor-semiconductor nanowire devices},\ }\href@noop {} {\bibfield
  {journal} {\bibinfo  {journal} {Science}\ }\textbf {\bibinfo {volume}
  {336}},\ \bibinfo {pages} {1003} (\bibinfo {year} {2012})}\BibitemShut
  {NoStop}%
\bibitem [{\citenamefont {Fu}\ and\ \citenamefont {Kane}(2009)}]{FuKane2009}%
  \BibitemOpen
  \bibfield  {author} {\bibinfo {author} {\bibfnamefont {L.}~\bibnamefont
  {Fu}}\ and\ \bibinfo {author} {\bibfnamefont {C.~L.}\ \bibnamefont {Kane}},\
  }\bibfield  {title} {\bibinfo {title} {Josephson current and noise at a
  superconductor–quantum-spin-hall-insulator–superconductor junction},\
  }\href {https://doi.org/10.1103/PhysRevB.79.161408} {\bibfield  {journal}
  {\bibinfo  {journal} {Physical Review B}\ }\textbf {\bibinfo {volume} {79}},\
  \bibinfo {pages} {161408} (\bibinfo {year} {2009})}\BibitemShut {NoStop}%
\bibitem [{\citenamefont {Hell}\ \emph {et~al.}(2017)\citenamefont {Hell},
  \citenamefont {Flensberg},\ and\ \citenamefont {Leijnse}}]{Hell2017}%
  \BibitemOpen
  \bibfield  {author} {\bibinfo {author} {\bibfnamefont {M.}~\bibnamefont
  {Hell}}, \bibinfo {author} {\bibfnamefont {K.}~\bibnamefont {Flensberg}},\
  and\ \bibinfo {author} {\bibfnamefont {M.}~\bibnamefont {Leijnse}},\
  }\bibfield  {title} {\bibinfo {title} {Two-dimensional platform for networks
  of majorana bound states},\ }\href
  {https://doi.org/10.1103/PhysRevLett.118.107701} {\bibfield  {journal}
  {\bibinfo  {journal} {Physical Review Letters}\ }\textbf {\bibinfo {volume}
  {118}},\ \bibinfo {pages} {107701} (\bibinfo {year} {2017})}\BibitemShut
  {NoStop}%
\bibitem [{\citenamefont {Pientka}\ \emph {et~al.}(2017)\citenamefont
  {Pientka}, \citenamefont {Keselman}, \citenamefont {Berg}, \citenamefont
  {Oreg}, \citenamefont {Stern},\ and\ \citenamefont {Halperin}}]{Pientka2017}%
  \BibitemOpen
  \bibfield  {author} {\bibinfo {author} {\bibfnamefont {F.}~\bibnamefont
  {Pientka}}, \bibinfo {author} {\bibfnamefont {A.}~\bibnamefont {Keselman}},
  \bibinfo {author} {\bibfnamefont {E.}~\bibnamefont {Berg}}, \bibinfo {author}
  {\bibfnamefont {Y.}~\bibnamefont {Oreg}}, \bibinfo {author} {\bibfnamefont
  {A.}~\bibnamefont {Stern}},\ and\ \bibinfo {author} {\bibfnamefont {B.~I.}\
  \bibnamefont {Halperin}},\ }\bibfield  {title} {\bibinfo {title} {Topological
  superconductivity in a planar josephson junction},\ }\href
  {https://doi.org/10.1103/PhysRevX.7.021032} {\bibfield  {journal} {\bibinfo
  {journal} {Physical Review X}\ }\textbf {\bibinfo {volume} {7}},\ \bibinfo
  {pages} {021032} (\bibinfo {year} {2017})}\BibitemShut {NoStop}%
\bibitem [{\citenamefont {Ohtomo}\ and\ \citenamefont
  {Hwang}(2004)}]{Ohtomo_Hwang_2004}%
  \BibitemOpen
  \bibfield  {author} {\bibinfo {author} {\bibfnamefont {A.}~\bibnamefont
  {Ohtomo}}\ and\ \bibinfo {author} {\bibfnamefont {H.~Y.}\ \bibnamefont
  {Hwang}},\ }\bibfield  {title} {\bibinfo {title} {A high-mobility electron
  gas at the $\mathrm{LaAlO_3/SrTiO_3}$ heterointerface},\ }\href
  {https://doi.org/10.1038/nature02308} {\bibfield  {journal} {\bibinfo
  {journal} {Nature}\ }\textbf {\bibinfo {volume} {427}},\ \bibinfo {pages}
  {423} (\bibinfo {year} {2004})}\BibitemShut {NoStop}%
\bibitem [{\citenamefont {Diez}\ \emph {et~al.}(2015)\citenamefont {Diez},
  \citenamefont {Monteiro}, \citenamefont {Mattoni}, \citenamefont {Cobanera},
  \citenamefont {Hyart}, \citenamefont {Mulazimoglu}, \citenamefont {Bovenzi},
  \citenamefont {Beenakker},\ and\ \citenamefont {Caviglia}}]{Diez2015}%
  \BibitemOpen
  \bibfield  {author} {\bibinfo {author} {\bibfnamefont {M.}~\bibnamefont
  {Diez}}, \bibinfo {author} {\bibfnamefont {A.~M. R. V.~L.}\ \bibnamefont
  {Monteiro}}, \bibinfo {author} {\bibfnamefont {G.}~\bibnamefont {Mattoni}},
  \bibinfo {author} {\bibfnamefont {E.}~\bibnamefont {Cobanera}}, \bibinfo
  {author} {\bibfnamefont {T.}~\bibnamefont {Hyart}}, \bibinfo {author}
  {\bibfnamefont {E.}~\bibnamefont {Mulazimoglu}}, \bibinfo {author}
  {\bibfnamefont {N.}~\bibnamefont {Bovenzi}}, \bibinfo {author} {\bibfnamefont
  {C.~W.~J.}\ \bibnamefont {Beenakker}},\ and\ \bibinfo {author} {\bibfnamefont
  {A.~D.}\ \bibnamefont {Caviglia}},\ }\bibfield  {title} {\bibinfo {title}
  {Giant negative magnetoresistance driven by spin-orbit coupling at the
  $\mathrm{LaAlO_3/SrTiO_3}$ interface},\ }\href
  {https://doi.org/10.1103/PhysRevLett.115.016803} {\bibfield  {journal}
  {\bibinfo  {journal} {Phys. Rev. Lett.}\ }\textbf {\bibinfo {volume} {115}},\
  \bibinfo {pages} {016803} (\bibinfo {year} {2015})}\BibitemShut {NoStop}%
\bibitem [{\citenamefont {Rout}\ \emph {et~al.}(2017)\citenamefont {Rout},
  \citenamefont {Maniv}, \citenamefont {Goldstein},\ and\ \citenamefont
  {Dagan}}]{Rout2017}%
  \BibitemOpen
  \bibfield  {author} {\bibinfo {author} {\bibfnamefont {P.~K.}\ \bibnamefont
  {Rout}}, \bibinfo {author} {\bibfnamefont {E.}~\bibnamefont {Maniv}},
  \bibinfo {author} {\bibfnamefont {M.}~\bibnamefont {Goldstein}},\ and\
  \bibinfo {author} {\bibfnamefont {Y.}~\bibnamefont {Dagan}},\ }\bibfield
  {title} {\bibinfo {title} {Link between the superconducting dome and
  spin-orbit interaction in the (111) $\mathrm{LaAlO_3/SrTiO_3}$ interface},\
  }\href {https://doi.org/10.1103/PhysRevLett.119.237002} {\bibfield  {journal}
  {\bibinfo  {journal} {Phys. Rev. Lett.}\ }\textbf {\bibinfo {volume} {119}},\
  \bibinfo {pages} {237002} (\bibinfo {year} {2017})}\BibitemShut {NoStop}%
\bibitem [{\citenamefont {Caviglia}\ \emph {et~al.}(2010)\citenamefont
  {Caviglia}, \citenamefont {Gariglio}, \citenamefont {Cancellieri},
  \citenamefont {Sac{\'e}p{\'e}}, \citenamefont {F{\^e}te}, \citenamefont
  {Reyren}, \citenamefont {Gabay}, \citenamefont {Morpurgo},\ and\
  \citenamefont {Triscone}}]{Caviglia2010}%
  \BibitemOpen
  \bibfield  {author} {\bibinfo {author} {\bibfnamefont {A.~D.}\ \bibnamefont
  {Caviglia}}, \bibinfo {author} {\bibfnamefont {S.}~\bibnamefont {Gariglio}},
  \bibinfo {author} {\bibfnamefont {C.}~\bibnamefont {Cancellieri}}, \bibinfo
  {author} {\bibfnamefont {B.}~\bibnamefont {Sac{\'e}p{\'e}}}, \bibinfo
  {author} {\bibfnamefont {A.}~\bibnamefont {F{\^e}te}}, \bibinfo {author}
  {\bibfnamefont {N.}~\bibnamefont {Reyren}}, \bibinfo {author} {\bibfnamefont
  {M.}~\bibnamefont {Gabay}}, \bibinfo {author} {\bibfnamefont {A.~F.}\
  \bibnamefont {Morpurgo}},\ and\ \bibinfo {author} {\bibfnamefont {J.-M.}\
  \bibnamefont {Triscone}},\ }\bibfield  {title} {\bibinfo {title} {Tunable
  rashba spin-orbit interaction at oxide interfaces},\ }\href
  {https://doi.org/10.1103/PhysRevLett.104.126803} {\bibfield  {journal}
  {\bibinfo  {journal} {Phys. Rev. Lett.}\ }\textbf {\bibinfo {volume} {104}},\
  \bibinfo {pages} {126803} (\bibinfo {year} {2010})}\BibitemShut {NoStop}%
\bibitem [{\citenamefont {Ben~Shalom}\ \emph {et~al.}(2010)\citenamefont
  {Ben~Shalom}, \citenamefont {Sachs}, \citenamefont {Rakhmilevitch},
  \citenamefont {Palevski},\ and\ \citenamefont {Dagan}}]{Shalom2010}%
  \BibitemOpen
  \bibfield  {author} {\bibinfo {author} {\bibfnamefont {M.}~\bibnamefont
  {Ben~Shalom}}, \bibinfo {author} {\bibfnamefont {M.}~\bibnamefont {Sachs}},
  \bibinfo {author} {\bibfnamefont {D.}~\bibnamefont {Rakhmilevitch}}, \bibinfo
  {author} {\bibfnamefont {A.}~\bibnamefont {Palevski}},\ and\ \bibinfo
  {author} {\bibfnamefont {Y.}~\bibnamefont {Dagan}},\ }\bibfield  {title}
  {\bibinfo {title} {Tuning spin-orbit coupling and superconductivity at the
  $\mathrm{LaAlO_3/SrTiO_3}$ interface},\ }\href
  {https://doi.org/10.1103/PhysRevLett.104.126802} {\bibfield  {journal}
  {\bibinfo  {journal} {Phys. Rev. Lett.}\ }\textbf {\bibinfo {volume} {104}},\
  \bibinfo {pages} {126802} (\bibinfo {year} {2010})}\BibitemShut {NoStop}%
\bibitem [{\citenamefont {Yin}\ \emph {et~al.}(2020)\citenamefont {Yin},
  \citenamefont {Seiler}, \citenamefont {Tang}, \citenamefont {Leermakers},
  \citenamefont {Lebedev}, \citenamefont {Zeitler},\ and\ \citenamefont
  {Aarts}}]{Yin2020}%
  \BibitemOpen
  \bibfield  {author} {\bibinfo {author} {\bibfnamefont {C.}~\bibnamefont
  {Yin}}, \bibinfo {author} {\bibfnamefont {P.}~\bibnamefont {Seiler}},
  \bibinfo {author} {\bibfnamefont {L.~M.~K.}\ \bibnamefont {Tang}}, \bibinfo
  {author} {\bibfnamefont {I.}~\bibnamefont {Leermakers}}, \bibinfo {author}
  {\bibfnamefont {N.}~\bibnamefont {Lebedev}}, \bibinfo {author} {\bibfnamefont
  {U.}~\bibnamefont {Zeitler}},\ and\ \bibinfo {author} {\bibfnamefont
  {J.}~\bibnamefont {Aarts}},\ }\bibfield  {title} {\bibinfo {title} {Tuning
  rashba spin-orbit coupling at $\mathrm{LaAlO_3/SrTiO_3}$ interfaces by band
  filling},\ }\href {https://doi.org/10.1103/PhysRevB.101.245114} {\bibfield
  {journal} {\bibinfo  {journal} {Phys. Rev. B}\ }\textbf {\bibinfo {volume}
  {101}},\ \bibinfo {pages} {245114} (\bibinfo {year} {2020})}\BibitemShut
  {NoStop}%
\bibitem [{\citenamefont {Singh}\ \emph {et~al.}(2017)\citenamefont {Singh},
  \citenamefont {Jouan}, \citenamefont {Hurand}, \citenamefont
  {Feuillet-Palma}, \citenamefont {Kumar}, \citenamefont {Dogra}, \citenamefont
  {Budhani}, \citenamefont {Lesueur},\ and\ \citenamefont
  {Bergeal}}]{Singh2017}%
  \BibitemOpen
  \bibfield  {author} {\bibinfo {author} {\bibfnamefont {G.}~\bibnamefont
  {Singh}}, \bibinfo {author} {\bibfnamefont {A.}~\bibnamefont {Jouan}},
  \bibinfo {author} {\bibfnamefont {S.}~\bibnamefont {Hurand}}, \bibinfo
  {author} {\bibfnamefont {C.}~\bibnamefont {Feuillet-Palma}}, \bibinfo
  {author} {\bibfnamefont {P.}~\bibnamefont {Kumar}}, \bibinfo {author}
  {\bibfnamefont {A.}~\bibnamefont {Dogra}}, \bibinfo {author} {\bibfnamefont
  {R.}~\bibnamefont {Budhani}}, \bibinfo {author} {\bibfnamefont
  {J.}~\bibnamefont {Lesueur}},\ and\ \bibinfo {author} {\bibfnamefont
  {N.}~\bibnamefont {Bergeal}},\ }\bibfield  {title} {\bibinfo {title} {Effect
  of disorder on superconductivity and rashba spin-orbit coupling in
  $\mathrm{LaAlO_3/SrTiO_3}$ interfaces},\ }\href
  {https://doi.org/10.1103/PhysRevB.96.024509} {\bibfield  {journal} {\bibinfo
  {journal} {Phys. Rev. B}\ }\textbf {\bibinfo {volume} {96}},\ \bibinfo
  {pages} {024509} (\bibinfo {year} {2017})}\BibitemShut {NoStop}%
\bibitem [{\citenamefont {Hurand}\ \emph {et~al.}(2015)\citenamefont {Hurand},
  \citenamefont {Jouan}, \citenamefont {Feuillet-Palma}, \citenamefont {Singh},
  \citenamefont {Kumar}, \citenamefont {Dogra}, \citenamefont {Budhani},
  \citenamefont {Caprara}, \citenamefont {Grilli}, \citenamefont {Lesueur},\
  and\ \citenamefont {Bergeal}}]{Hurand2015}%
  \BibitemOpen
  \bibfield  {author} {\bibinfo {author} {\bibfnamefont {S.}~\bibnamefont
  {Hurand}}, \bibinfo {author} {\bibfnamefont {A.}~\bibnamefont {Jouan}},
  \bibinfo {author} {\bibfnamefont {C.}~\bibnamefont {Feuillet-Palma}},
  \bibinfo {author} {\bibfnamefont {G.}~\bibnamefont {Singh}}, \bibinfo
  {author} {\bibfnamefont {S.}~\bibnamefont {Kumar}}, \bibinfo {author}
  {\bibfnamefont {A.}~\bibnamefont {Dogra}}, \bibinfo {author} {\bibfnamefont
  {R.~C.}\ \bibnamefont {Budhani}}, \bibinfo {author} {\bibfnamefont
  {S.}~\bibnamefont {Caprara}}, \bibinfo {author} {\bibfnamefont
  {M.}~\bibnamefont {Grilli}}, \bibinfo {author} {\bibfnamefont
  {J.}~\bibnamefont {Lesueur}},\ and\ \bibinfo {author} {\bibfnamefont
  {N.}~\bibnamefont {Bergeal}},\ }\bibfield  {title} {\bibinfo {title}
  {Field-effect control of superconductivity and rashba spin-orbit coupling},\
  }\href {https://doi.org/10.1038/srep12751} {\bibfield  {journal} {\bibinfo
  {journal} {Sci. Rep.}\ }\textbf {\bibinfo {volume} {5}},\ \bibinfo {pages}
  {12751} (\bibinfo {year} {2015})}\BibitemShut {NoStop}%
\bibitem [{\citenamefont {Reyren}\ \emph {et~al.}(2007)\citenamefont {Reyren},
  \citenamefont {Thiel}, \citenamefont {Caviglia}, \citenamefont
  {Fitting~Kourkoutis}, \citenamefont {Hammerl}, \citenamefont {Richter},
  \citenamefont {Schneider}, \citenamefont {Kopp}, \citenamefont {Rüetschi},
  \citenamefont {Jaccard}, \citenamefont {Gabay}, \citenamefont {Muller},
  \citenamefont {Triscone},\ and\ \citenamefont {Mannhart}}]{Reyren2007}%
  \BibitemOpen
  \bibfield  {author} {\bibinfo {author} {\bibfnamefont {N.}~\bibnamefont
  {Reyren}}, \bibinfo {author} {\bibfnamefont {S.}~\bibnamefont {Thiel}},
  \bibinfo {author} {\bibfnamefont {A.~D.}\ \bibnamefont {Caviglia}}, \bibinfo
  {author} {\bibfnamefont {L.}~\bibnamefont {Fitting~Kourkoutis}}, \bibinfo
  {author} {\bibfnamefont {G.}~\bibnamefont {Hammerl}}, \bibinfo {author}
  {\bibfnamefont {C.}~\bibnamefont {Richter}}, \bibinfo {author} {\bibfnamefont
  {C.~W.}\ \bibnamefont {Schneider}}, \bibinfo {author} {\bibfnamefont
  {T.}~\bibnamefont {Kopp}}, \bibinfo {author} {\bibfnamefont {A.-S.}\
  \bibnamefont {Rüetschi}}, \bibinfo {author} {\bibfnamefont {D.}~\bibnamefont
  {Jaccard}}, \bibinfo {author} {\bibfnamefont {M.}~\bibnamefont {Gabay}},
  \bibinfo {author} {\bibfnamefont {D.~A.}\ \bibnamefont {Muller}}, \bibinfo
  {author} {\bibfnamefont {J.-M.}\ \bibnamefont {Triscone}},\ and\ \bibinfo
  {author} {\bibfnamefont {J.}~\bibnamefont {Mannhart}},\ }\bibfield  {title}
  {\bibinfo {title} {Superconducting interfaces between insulating oxides},\
  }\href {https://doi.org/10.1126/science.1146006} {\bibfield  {journal}
  {\bibinfo  {journal} {Science}\ }\textbf {\bibinfo {volume} {317}},\ \bibinfo
  {pages} {1196} (\bibinfo {year} {2007})}\BibitemShut {NoStop}%
\bibitem [{\citenamefont {Joshua}\ \emph {et~al.}(2012)\citenamefont {Joshua},
  \citenamefont {Pecker}, \citenamefont {Ruhman}, \citenamefont {Altman},\ and\
  \citenamefont {Ilani}}]{joshua2012universal}%
  \BibitemOpen
  \bibfield  {author} {\bibinfo {author} {\bibfnamefont {A.}~\bibnamefont
  {Joshua}}, \bibinfo {author} {\bibfnamefont {S.}~\bibnamefont {Pecker}},
  \bibinfo {author} {\bibfnamefont {J.}~\bibnamefont {Ruhman}}, \bibinfo
  {author} {\bibfnamefont {E.}~\bibnamefont {Altman}},\ and\ \bibinfo {author}
  {\bibfnamefont {S.}~\bibnamefont {Ilani}},\ }\bibfield  {title} {\bibinfo
  {title} {A universal critical density underlying the physics of electrons at
  the $\mathrm{LaAlO_3/SrTiO_3}$ interface},\ }\href
  {https://doi.org/10.1038/ncomms2116} {\bibfield  {journal} {\bibinfo
  {journal} {Nat. Commun.}\ }\textbf {\bibinfo {volume} {3}},\ \bibinfo {pages}
  {1129} (\bibinfo {year} {2012})}\BibitemShut {NoStop}%
\bibitem [{\citenamefont {Maniv}\ \emph {et~al.}(2015)\citenamefont {Maniv},
  \citenamefont {Ben~Shalom}, \citenamefont {Ron}, \citenamefont {Mograbi},
  \citenamefont {Palevski}, \citenamefont {Goldstein},\ and\ \citenamefont
  {Dagan}}]{maniv2015strong}%
  \BibitemOpen
  \bibfield  {author} {\bibinfo {author} {\bibfnamefont {E.}~\bibnamefont
  {Maniv}}, \bibinfo {author} {\bibfnamefont {M.}~\bibnamefont {Ben~Shalom}},
  \bibinfo {author} {\bibfnamefont {A.}~\bibnamefont {Ron}}, \bibinfo {author}
  {\bibfnamefont {M.}~\bibnamefont {Mograbi}}, \bibinfo {author} {\bibfnamefont
  {A.}~\bibnamefont {Palevski}}, \bibinfo {author} {\bibfnamefont
  {M.}~\bibnamefont {Goldstein}},\ and\ \bibinfo {author} {\bibfnamefont
  {Y.}~\bibnamefont {Dagan}},\ }\bibfield  {title} {\bibinfo {title} {Strong
  correlations elucidate the electronic structure and phase diagram of
  $\mathrm{LaAlO_3/SrTiO_3}$},\ }\href {https://doi.org/10.1038/ncomms9239}
  {\bibfield  {journal} {\bibinfo  {journal} {Nat. Commun.}\ }\textbf {\bibinfo
  {volume} {6}},\ \bibinfo {pages} {8239} (\bibinfo {year} {2015})}\BibitemShut
  {NoStop}%
\bibitem [{\citenamefont {Biscaras}\ \emph {et~al.}(2012)\citenamefont
  {Biscaras}, \citenamefont {Bergeal}, \citenamefont {Hurand}, \citenamefont
  {Grossetête}, \citenamefont {Rastogi}, \citenamefont {Budhani},
  \citenamefont {LeBoeuf}, \citenamefont {Proust},\ and\ \citenamefont
  {Lesueur}}]{Biscaras}%
  \BibitemOpen
  \bibfield  {author} {\bibinfo {author} {\bibfnamefont {J.}~\bibnamefont
  {Biscaras}}, \bibinfo {author} {\bibfnamefont {N.}~\bibnamefont {Bergeal}},
  \bibinfo {author} {\bibfnamefont {S.}~\bibnamefont {Hurand}}, \bibinfo
  {author} {\bibfnamefont {C.}~\bibnamefont {Grossetête}}, \bibinfo {author}
  {\bibfnamefont {A.}~\bibnamefont {Rastogi}}, \bibinfo {author} {\bibfnamefont
  {R.~C.}\ \bibnamefont {Budhani}}, \bibinfo {author} {\bibfnamefont
  {D.}~\bibnamefont {LeBoeuf}}, \bibinfo {author} {\bibfnamefont
  {C.}~\bibnamefont {Proust}},\ and\ \bibinfo {author} {\bibfnamefont
  {J.}~\bibnamefont {Lesueur}},\ }\bibfield  {title} {\bibinfo {title}
  {Two-dimensional superconducting phase in $\mathrm{LaTiO_3/SrTiO_3}$
  heterostructures induced by high‑mobility carrier doping},\ }\href
  {https://doi.org/10.1103/PhysRevLett.108.247004} {\bibfield  {journal}
  {\bibinfo  {journal} {Phys. Rev. Lett.}\ }\textbf {\bibinfo {volume} {108}},\
  \bibinfo {pages} {247004} (\bibinfo {year} {2012})}\BibitemShut {NoStop}%
\bibitem [{\citenamefont {Monteiro}\ \emph {et~al.}(2019)\citenamefont
  {Monteiro}, \citenamefont {Vivek}, \citenamefont {Groenendijk}, \citenamefont
  {Bruneel}, \citenamefont {Leermakers}, \citenamefont {Zeitler}, \citenamefont
  {Gabay},\ and\ \citenamefont {Caviglia}}]{Monteiro2019}%
  \BibitemOpen
  \bibfield  {author} {\bibinfo {author} {\bibfnamefont {A.~M. R. V.~L.}\
  \bibnamefont {Monteiro}}, \bibinfo {author} {\bibfnamefont {M.}~\bibnamefont
  {Vivek}}, \bibinfo {author} {\bibfnamefont {D.~J.}\ \bibnamefont
  {Groenendijk}}, \bibinfo {author} {\bibfnamefont {P.}~\bibnamefont
  {Bruneel}}, \bibinfo {author} {\bibfnamefont {I.}~\bibnamefont {Leermakers}},
  \bibinfo {author} {\bibfnamefont {U.}~\bibnamefont {Zeitler}}, \bibinfo
  {author} {\bibfnamefont {M.}~\bibnamefont {Gabay}},\ and\ \bibinfo {author}
  {\bibfnamefont {A.~D.}\ \bibnamefont {Caviglia}},\ }\bibfield  {title}
  {\bibinfo {title} {Band inversion driven by electronic correlations at the
  (111) $\mathrm{LaAlO_3/SrTiO_3}$ interface},\ }\href
  {https://doi.org/10.1103/PhysRevB.99.201102} {\bibfield  {journal} {\bibinfo
  {journal} {Phys. Rev. B}\ }\textbf {\bibinfo {volume} {99}},\ \bibinfo
  {pages} {201102} (\bibinfo {year} {2019})}\BibitemShut {NoStop}%
\bibitem [{\citenamefont {Monteiro}\ \emph {et~al.}(2017)\citenamefont
  {Monteiro}, \citenamefont {Groenendijk}, \citenamefont {Groen}, \citenamefont
  {de~Bruijckere}, \citenamefont {Gaudenzi}, \citenamefont {van~der Zant},\
  and\ \citenamefont {Caviglia}}]{Monteiro2017}%
  \BibitemOpen
  \bibfield  {author} {\bibinfo {author} {\bibfnamefont {A.~M. R. V.~L.}\
  \bibnamefont {Monteiro}}, \bibinfo {author} {\bibfnamefont {D.~J.}\
  \bibnamefont {Groenendijk}}, \bibinfo {author} {\bibfnamefont
  {I.}~\bibnamefont {Groen}}, \bibinfo {author} {\bibfnamefont
  {J.}~\bibnamefont {de~Bruijckere}}, \bibinfo {author} {\bibfnamefont
  {R.}~\bibnamefont {Gaudenzi}}, \bibinfo {author} {\bibfnamefont {H.~S.~J.}\
  \bibnamefont {van~der Zant}},\ and\ \bibinfo {author} {\bibfnamefont {A.~D.}\
  \bibnamefont {Caviglia}},\ }\bibfield  {title} {\bibinfo {title}
  {Two-dimensional superconductivity at the (111) $\mathrm{LaAlO_3/SrTiO_3}$
  interface},\ }\href {https://doi.org/10.1103/PhysRevB.96.020504} {\bibfield
  {journal} {\bibinfo  {journal} {Phys. Rev. B}\ }\textbf {\bibinfo {volume}
  {96}},\ \bibinfo {pages} {020504} (\bibinfo {year} {2017})}\BibitemShut
  {NoStop}%
\bibitem [{\citenamefont {Wójcik}\ \emph {et~al.}(2024)\citenamefont
  {Wójcik}, \citenamefont {Szafran}, \citenamefont {Czarnecki}, \citenamefont
  {Citro},\ and\ \citenamefont
  {Zegrodnik}}]{Wojcik_dome_schrodinger_poisson_sc_2024}%
  \BibitemOpen
  \bibfield  {author} {\bibinfo {author} {\bibfnamefont {P.}~\bibnamefont
  {Wójcik}}, \bibinfo {author} {\bibfnamefont {B.}~\bibnamefont {Szafran}},
  \bibinfo {author} {\bibfnamefont {J.}~\bibnamefont {Czarnecki}}, \bibinfo
  {author} {\bibfnamefont {R.}~\bibnamefont {Citro}},\ and\ \bibinfo {author}
  {\bibfnamefont {M.}~\bibnamefont {Zegrodnik}},\ }\bibfield  {title} {\bibinfo
  {title} {Effect of electrostatic confinement on the dome-shaped
  superconducting phase diagram at the $\mathrm{LaAlO_3/SrTiO_3}$ interface},\
  }\href {https://doi.org/10.1038/s41598-024-77460-0} {\bibfield  {journal}
  {\bibinfo  {journal} {Scientific Reports}\ }\textbf {\bibinfo {volume}
  {14}},\ \bibinfo {pages} {26177} (\bibinfo {year} {2024})}\BibitemShut
  {NoStop}%
\bibitem [{\citenamefont {Michaeli}\ \emph {et~al.}(2012)\citenamefont
  {Michaeli}, \citenamefont {Potter},\ and\ \citenamefont {Lee}}]{Karen2012}%
  \BibitemOpen
  \bibfield  {author} {\bibinfo {author} {\bibfnamefont {K.}~\bibnamefont
  {Michaeli}}, \bibinfo {author} {\bibfnamefont {A.~C.}\ \bibnamefont
  {Potter}},\ and\ \bibinfo {author} {\bibfnamefont {P.~A.}\ \bibnamefont
  {Lee}},\ }\bibfield  {title} {\bibinfo {title} {Superconducting and
  ferromagnetic phases in $\mathrm{LaAlO_3/SrTiO_3}$: Finite momentum
  pairing},\ }\href {https://doi.org/10.1103/PhysRevLett.108.117003} {\bibfield
   {journal} {\bibinfo  {journal} {Phys. Rev. Lett.}\ }\textbf {\bibinfo
  {volume} {108}},\ \bibinfo {pages} {117003} (\bibinfo {year}
  {2012})}\BibitemShut {NoStop}%
\bibitem [{\citenamefont {Li}\ \emph {et~al.}(2011)\citenamefont {Li},
  \citenamefont {Richter}, \citenamefont {Mannhart},\ and\ \citenamefont
  {Ashoori}}]{Li2011}%
  \BibitemOpen
  \bibfield  {author} {\bibinfo {author} {\bibfnamefont {L.}~\bibnamefont
  {Li}}, \bibinfo {author} {\bibfnamefont {C.}~\bibnamefont {Richter}},
  \bibinfo {author} {\bibfnamefont {J.}~\bibnamefont {Mannhart}},\ and\
  \bibinfo {author} {\bibfnamefont {R.~C.}\ \bibnamefont {Ashoori}},\
  }\bibfield  {title} {\bibinfo {title} {Coexistence of magnetic order and
  two‑dimensional superconductivity at the $\mathrm{LaAlO_3/SrTiO_3}$
  interface},\ }\href {https://doi.org/10.1038/nphys2079} {\bibfield  {journal}
  {\bibinfo  {journal} {Nature Phys.}\ }\textbf {\bibinfo {volume} {7}},\
  \bibinfo {pages} {762} (\bibinfo {year} {2011})}\BibitemShut {NoStop}%
\bibitem [{\citenamefont {Brinkman}\ \emph {et~al.}(2007)\citenamefont
  {Brinkman}, \citenamefont {Huijben}, \citenamefont {van Zalk}, \citenamefont
  {Huijben}, \citenamefont {Zeitler}, \citenamefont {Maan}, \citenamefont
  {van~der Wiel}, \citenamefont {Rijnders}, \citenamefont {Blank},\ and\
  \citenamefont {Hilgenkamp}}]{Brinkman2007}%
  \BibitemOpen
  \bibfield  {author} {\bibinfo {author} {\bibfnamefont {A.}~\bibnamefont
  {Brinkman}}, \bibinfo {author} {\bibfnamefont {M.}~\bibnamefont {Huijben}},
  \bibinfo {author} {\bibfnamefont {M.}~\bibnamefont {van Zalk}}, \bibinfo
  {author} {\bibfnamefont {J.}~\bibnamefont {Huijben}}, \bibinfo {author}
  {\bibfnamefont {U.}~\bibnamefont {Zeitler}}, \bibinfo {author} {\bibfnamefont
  {J.~C.}\ \bibnamefont {Maan}}, \bibinfo {author} {\bibfnamefont {W.~G.}\
  \bibnamefont {van~der Wiel}}, \bibinfo {author} {\bibfnamefont
  {G.}~\bibnamefont {Rijnders}}, \bibinfo {author} {\bibfnamefont {D.~H.~A.}\
  \bibnamefont {Blank}},\ and\ \bibinfo {author} {\bibfnamefont
  {H.}~\bibnamefont {Hilgenkamp}},\ }\bibfield  {title} {\bibinfo {title}
  {Magnetic effects at the interface between non-magnetic oxides},\ }\href
  {https://doi.org/10.1038/nmat1931} {\bibfield  {journal} {\bibinfo  {journal}
  {Nature Mater.}\ }\textbf {\bibinfo {volume} {6}},\ \bibinfo {pages} {493}
  (\bibinfo {year} {2007})}\BibitemShut {NoStop}%
\bibitem [{\citenamefont {Bert}\ \emph {et~al.}(2011)\citenamefont {Bert},
  \citenamefont {Kalisky}, \citenamefont {Bell}, \citenamefont {Kim},
  \citenamefont {Hikita}, \citenamefont {Hwang},\ and\ \citenamefont
  {Moler}}]{Bert2011}%
  \BibitemOpen
  \bibfield  {author} {\bibinfo {author} {\bibfnamefont {J.~A.}\ \bibnamefont
  {Bert}}, \bibinfo {author} {\bibfnamefont {B.}~\bibnamefont {Kalisky}},
  \bibinfo {author} {\bibfnamefont {C.}~\bibnamefont {Bell}}, \bibinfo {author}
  {\bibfnamefont {M.}~\bibnamefont {Kim}}, \bibinfo {author} {\bibfnamefont
  {Y.}~\bibnamefont {Hikita}}, \bibinfo {author} {\bibfnamefont {H.~Y.}\
  \bibnamefont {Hwang}},\ and\ \bibinfo {author} {\bibfnamefont {K.~A.}\
  \bibnamefont {Moler}},\ }\bibfield  {title} {\bibinfo {title} {Direct imaging
  of the coexistence of ferromagnetism and superconductivity at the
  $\mathrm{LaAlO_3/SrTiO_3}$ interface},\ }\href
  {https://doi.org/10.1038/nphys2079} {\bibfield  {journal} {\bibinfo
  {journal} {Nature Phys.}\ }\textbf {\bibinfo {volume} {7}},\ \bibinfo {pages}
  {767} (\bibinfo {year} {2011})}\BibitemShut {NoStop}%
\bibitem [{\citenamefont {Gastiasoro}\ \emph {et~al.}(2020)\citenamefont
  {Gastiasoro} \emph {et~al.}}]{Gastiasoro2020}%
  \BibitemOpen
  \bibfield  {author} {\bibinfo {author} {\bibfnamefont {M.~N.}\ \bibnamefont
  {Gastiasoro}} \emph {et~al.},\ }\bibfield  {title} {\bibinfo {title}
  {Anisotropic superconductivity mediated by ferroelectric fluctuations},\
  }\href@noop {} {\bibfield  {journal} {\bibinfo  {journal} {Phys. Rev. B}\
  }\textbf {\bibinfo {volume} {101}},\ \bibinfo {pages} {174501} (\bibinfo
  {year} {2020})}\BibitemShut {NoStop}%
\bibitem [{\citenamefont {Kanasugi}\ \emph {et~al.}(2018)\citenamefont
  {Kanasugi}, \citenamefont {Yuan},\ and\ \citenamefont
  {Yanase}}]{Kanasugi2018}%
  \BibitemOpen
  \bibfield  {author} {\bibinfo {author} {\bibfnamefont {S.}~\bibnamefont
  {Kanasugi}}, \bibinfo {author} {\bibfnamefont {N.}~\bibnamefont {Yuan}},\
  and\ \bibinfo {author} {\bibfnamefont {Y.}~\bibnamefont {Yanase}},\
  }\bibfield  {title} {\bibinfo {title} {Spin-orbit-coupled ferroelectric
  superconductivity},\ }\href {https://doi.org/10.1103/PhysRevB.98.024521}
  {\bibfield  {journal} {\bibinfo  {journal} {Phys. Rev. B}\ }\textbf {\bibinfo
  {volume} {98}},\ \bibinfo {pages} {024521} (\bibinfo {year}
  {2018})}\BibitemShut {NoStop}%
\bibitem [{\citenamefont {Kanasugi}\ \emph {et~al.}(2019)\citenamefont
  {Kanasugi}, \citenamefont {Kawakami},\ and\ \citenamefont
  {Yanase}}]{Kanasugi2019}%
  \BibitemOpen
  \bibfield  {author} {\bibinfo {author} {\bibfnamefont {S.}~\bibnamefont
  {Kanasugi}}, \bibinfo {author} {\bibfnamefont {T.}~\bibnamefont {Kawakami}},\
  and\ \bibinfo {author} {\bibfnamefont {Y.}~\bibnamefont {Yanase}},\
  }\bibfield  {title} {\bibinfo {title} {Multiorbital ferroelectric
  superconductivity in doped srtio$_3$},\ }\href
  {https://doi.org/10.1103/PhysRevB.100.094504} {\bibfield  {journal} {\bibinfo
   {journal} {Phys. Rev. B}\ }\textbf {\bibinfo {volume} {100}},\ \bibinfo
  {pages} {094504} (\bibinfo {year} {2019})}\BibitemShut {NoStop}%
\bibitem [{\citenamefont {Honig}\ \emph {et~al.}(2013)\citenamefont {Honig},
  \citenamefont {Sulpizio}, \citenamefont {Drori}, \citenamefont {Joshua},
  \citenamefont {Zeldov},\ and\ \citenamefont {Ilani}}]{Honig2013}%
  \BibitemOpen
  \bibfield  {author} {\bibinfo {author} {\bibfnamefont {M.}~\bibnamefont
  {Honig}}, \bibinfo {author} {\bibfnamefont {J.~A.}\ \bibnamefont {Sulpizio}},
  \bibinfo {author} {\bibfnamefont {J.}~\bibnamefont {Drori}}, \bibinfo
  {author} {\bibfnamefont {A.}~\bibnamefont {Joshua}}, \bibinfo {author}
  {\bibfnamefont {E.}~\bibnamefont {Zeldov}},\ and\ \bibinfo {author}
  {\bibfnamefont {S.}~\bibnamefont {Ilani}},\ }\bibfield  {title} {\bibinfo
  {title} {Local electrostatic imaging of striped domain order in
  $\mathrm{LaAlO_3/SrTiO_3}$},\ }\href {https://doi.org/10.1038/nmat3810}
  {\bibfield  {journal} {\bibinfo  {journal} {Nature Mater.}\ }\textbf
  {\bibinfo {volume} {12}},\ \bibinfo {pages} {1112} (\bibinfo {year}
  {2013})}\BibitemShut {NoStop}%
\bibitem [{\citenamefont {Singh}\ \emph {et~al.}(2024)\citenamefont {Singh},
  \citenamefont {Guzman}, \citenamefont {Saïz}, \citenamefont {Zhou},
  \citenamefont {Gazquez}, \citenamefont {Masoudinia}, \citenamefont {Winkler},
  \citenamefont {Claeson}, \citenamefont {Fraxedas}, \citenamefont {Bergeal},
  \citenamefont {Herranz},\ and\ \citenamefont {Kalaboukhov}}]{Singh2024}%
  \BibitemOpen
  \bibfield  {author} {\bibinfo {author} {\bibfnamefont {G.}~\bibnamefont
  {Singh}}, \bibinfo {author} {\bibfnamefont {R.}~\bibnamefont {Guzman}},
  \bibinfo {author} {\bibfnamefont {G.}~\bibnamefont {Saïz}}, \bibinfo
  {author} {\bibfnamefont {W.}~\bibnamefont {Zhou}}, \bibinfo {author}
  {\bibfnamefont {J.}~\bibnamefont {Gazquez}}, \bibinfo {author} {\bibfnamefont
  {F.}~\bibnamefont {Masoudinia}}, \bibinfo {author} {\bibfnamefont
  {D.}~\bibnamefont {Winkler}}, \bibinfo {author} {\bibfnamefont
  {T.}~\bibnamefont {Claeson}}, \bibinfo {author} {\bibfnamefont
  {J.}~\bibnamefont {Fraxedas}}, \bibinfo {author} {\bibfnamefont
  {N.}~\bibnamefont {Bergeal}}, \bibinfo {author} {\bibfnamefont
  {G.}~\bibnamefont {Herranz}},\ and\ \bibinfo {author} {\bibfnamefont
  {A.}~\bibnamefont {Kalaboukhov}},\ }\bibfield  {title} {\bibinfo {title}
  {Stoichiometric control of electron mobility and 2d superconductivity at
  laalo$_3$–srtio$_3$ interfaces},\ }\href
  {https://doi.org/10.1038/s42005-024-01644-3} {\bibfield  {journal} {\bibinfo
  {journal} {Communications Physics}\ }\textbf {\bibinfo {volume} {7}},\
  \bibinfo {pages} {149} (\bibinfo {year} {2024})}\BibitemShut {NoStop}%
\bibitem [{\citenamefont {Ren}\ \emph {et~al.}(2022)\citenamefont {Ren},
  \citenamefont {Li}, \citenamefont {Sun}, \citenamefont {Ju}, \citenamefont
  {Liu}, \citenamefont {Hong}, \citenamefont {Sun}, \citenamefont {Tao},
  \citenamefont {Zhou}, \citenamefont {Xu},\ and\ \citenamefont
  {Xie}}]{Ren2022}%
  \BibitemOpen
  \bibfield  {author} {\bibinfo {author} {\bibfnamefont {T.}~\bibnamefont
  {Ren}}, \bibinfo {author} {\bibfnamefont {M.}~\bibnamefont {Li}}, \bibinfo
  {author} {\bibfnamefont {X.}~\bibnamefont {Sun}}, \bibinfo {author}
  {\bibfnamefont {L.}~\bibnamefont {Ju}}, \bibinfo {author} {\bibfnamefont
  {Y.}~\bibnamefont {Liu}}, \bibinfo {author} {\bibfnamefont {S.}~\bibnamefont
  {Hong}}, \bibinfo {author} {\bibfnamefont {Y.}~\bibnamefont {Sun}}, \bibinfo
  {author} {\bibfnamefont {Q.}~\bibnamefont {Tao}}, \bibinfo {author}
  {\bibfnamefont {Y.}~\bibnamefont {Zhou}}, \bibinfo {author} {\bibfnamefont
  {Z.-A.}\ \bibnamefont {Xu}},\ and\ \bibinfo {author} {\bibfnamefont
  {Y.}~\bibnamefont {Xie}},\ }\bibfield  {title} {\bibinfo {title}
  {Two-dimensional superconductivity at the surfaces of $\mathrm{KTaO_3}$ gated
  with ionic liquid},\ }\href {https://doi.org/10.1126/sciadv.abn4273}
  {\bibfield  {journal} {\bibinfo  {journal} {Sci. Adv.}\ }\textbf {\bibinfo
  {volume} {8}},\ \bibinfo {pages} {4273} (\bibinfo {year} {2022})}\BibitemShut
  {NoStop}%
\bibitem [{\citenamefont {Liu}\ \emph {et~al.}(2021)\citenamefont {Liu},
  \citenamefont {Yan}, \citenamefont {Jin}, \citenamefont {Ma}, \citenamefont
  {Hsiao}, \citenamefont {Lin}, \citenamefont {Bretz-Sullivan}, \citenamefont
  {Zhou}, \citenamefont {Pearson}, \citenamefont {Fisher}, \citenamefont
  {Jiang}, \citenamefont {Han}, \citenamefont {Zuo}, \citenamefont {Wen},
  \citenamefont {Fong}, \citenamefont {Sun}, \citenamefont {Zhou},\ and\
  \citenamefont {Bhattacharya}}]{Liu2021}%
  \BibitemOpen
  \bibfield  {author} {\bibinfo {author} {\bibfnamefont {C.}~\bibnamefont
  {Liu}}, \bibinfo {author} {\bibfnamefont {X.}~\bibnamefont {Yan}}, \bibinfo
  {author} {\bibfnamefont {D.}~\bibnamefont {Jin}}, \bibinfo {author}
  {\bibfnamefont {Y.}~\bibnamefont {Ma}}, \bibinfo {author} {\bibfnamefont
  {H.-W.}\ \bibnamefont {Hsiao}}, \bibinfo {author} {\bibfnamefont
  {Y.}~\bibnamefont {Lin}}, \bibinfo {author} {\bibfnamefont {T.~M.}\
  \bibnamefont {Bretz-Sullivan}}, \bibinfo {author} {\bibfnamefont
  {X.}~\bibnamefont {Zhou}}, \bibinfo {author} {\bibfnamefont {J.}~\bibnamefont
  {Pearson}}, \bibinfo {author} {\bibfnamefont {B.}~\bibnamefont {Fisher}},
  \bibinfo {author} {\bibfnamefont {J.~S.}\ \bibnamefont {Jiang}}, \bibinfo
  {author} {\bibfnamefont {W.}~\bibnamefont {Han}}, \bibinfo {author}
  {\bibfnamefont {J.-M.}\ \bibnamefont {Zuo}}, \bibinfo {author} {\bibfnamefont
  {J.}~\bibnamefont {Wen}}, \bibinfo {author} {\bibfnamefont {D.~D.}\
  \bibnamefont {Fong}}, \bibinfo {author} {\bibfnamefont {J.}~\bibnamefont
  {Sun}}, \bibinfo {author} {\bibfnamefont {H.}~\bibnamefont {Zhou}},\ and\
  \bibinfo {author} {\bibfnamefont {A.}~\bibnamefont {Bhattacharya}},\
  }\bibfield  {title} {\bibinfo {title} {Two-dimensional superconductivity and
  anisotropic transport at $\mathrm{KTaO_3}$ (111) interfaces},\ }\href
  {https://doi.org/10.1126/science.aba5511} {\bibfield  {journal} {\bibinfo
  {journal} {Science}\ }\textbf {\bibinfo {volume} {371}},\ \bibinfo {pages}
  {716} (\bibinfo {year} {2021})}\BibitemShut {NoStop}%
\bibitem [{\citenamefont {Chen}\ \emph
  {et~al.}(2021{\natexlab{a}})\citenamefont {Chen}, \citenamefont {Liu},
  \citenamefont {Zhang}, \citenamefont {Liu}, \citenamefont {Tian},
  \citenamefont {Sun}, \citenamefont {Zhang}, \citenamefont {Zhou},
  \citenamefont {Sun},\ and\ \citenamefont {Xie}}]{Chen2021}%
  \BibitemOpen
  \bibfield  {author} {\bibinfo {author} {\bibfnamefont {Z.}~\bibnamefont
  {Chen}}, \bibinfo {author} {\bibfnamefont {Y.}~\bibnamefont {Liu}}, \bibinfo
  {author} {\bibfnamefont {H.}~\bibnamefont {Zhang}}, \bibinfo {author}
  {\bibfnamefont {Z.}~\bibnamefont {Liu}}, \bibinfo {author} {\bibfnamefont
  {H.}~\bibnamefont {Tian}}, \bibinfo {author} {\bibfnamefont {Y.}~\bibnamefont
  {Sun}}, \bibinfo {author} {\bibfnamefont {M.}~\bibnamefont {Zhang}}, \bibinfo
  {author} {\bibfnamefont {Y.}~\bibnamefont {Zhou}}, \bibinfo {author}
  {\bibfnamefont {J.}~\bibnamefont {Sun}},\ and\ \bibinfo {author}
  {\bibfnamefont {Y.}~\bibnamefont {Xie}},\ }\bibfield  {title} {\bibinfo
  {title} {Electric field control of superconductivity at the
  $\mathrm{LaAlO_3/KTaO_3}$ (111) interface},\ }\href
  {https://doi.org/10.1126/science.abb3848} {\bibfield  {journal} {\bibinfo
  {journal} {Science}\ }\textbf {\bibinfo {volume} {372}},\ \bibinfo {pages}
  {721} (\bibinfo {year} {2021}{\natexlab{a}})}\BibitemShut {NoStop}%
\bibitem [{\citenamefont {Liu}\ \emph {et~al.}(2023)\citenamefont {Liu},
  \citenamefont {Zhou}, \citenamefont {Hong}, \citenamefont {Fisher},
  \citenamefont {Zheng}, \citenamefont {Pearson}, \citenamefont {Jiang},
  \citenamefont {Jin}, \citenamefont {Norman},\ and\ \citenamefont
  {Bhattacharya}}]{Liu2023}%
  \BibitemOpen
  \bibfield  {author} {\bibinfo {author} {\bibfnamefont {C.}~\bibnamefont
  {Liu}}, \bibinfo {author} {\bibfnamefont {X.}~\bibnamefont {Zhou}}, \bibinfo
  {author} {\bibfnamefont {D.}~\bibnamefont {Hong}}, \bibinfo {author}
  {\bibfnamefont {B.}~\bibnamefont {Fisher}}, \bibinfo {author} {\bibfnamefont
  {H.}~\bibnamefont {Zheng}}, \bibinfo {author} {\bibfnamefont
  {J.}~\bibnamefont {Pearson}}, \bibinfo {author} {\bibfnamefont {J.~S.}\
  \bibnamefont {Jiang}}, \bibinfo {author} {\bibfnamefont {D.}~\bibnamefont
  {Jin}}, \bibinfo {author} {\bibfnamefont {M.~R.}\ \bibnamefont {Norman}},\
  and\ \bibinfo {author} {\bibfnamefont {A.}~\bibnamefont {Bhattacharya}},\
  }\bibfield  {title} {\bibinfo {title} {Tunable superconductivity and its
  origin at $\mathrm{KTaO_3}$ interfaces},\ }\href
  {https://doi.org/10.1038/s41467-023-36309-2} {\bibfield  {journal} {\bibinfo
  {journal} {Nat. Commun.}\ }\textbf {\bibinfo {volume} {14}},\ \bibinfo
  {pages} {951} (\bibinfo {year} {2023})}\BibitemShut {NoStop}%
\bibitem [{\citenamefont {Zhang}\ \emph {et~al.}(2025)\citenamefont {Zhang},
  \citenamefont {Qin}, \citenamefont {Sun}, \citenamefont {Hong}, \citenamefont
  {Zhou},\ and\ \citenamefont {Xie}}]{Zhang2025}%
  \BibitemOpen
  \bibfield  {author} {\bibinfo {author} {\bibfnamefont {M.}~\bibnamefont
  {Zhang}}, \bibinfo {author} {\bibfnamefont {M.}~\bibnamefont {Qin}}, \bibinfo
  {author} {\bibfnamefont {Y.}~\bibnamefont {Sun}}, \bibinfo {author}
  {\bibfnamefont {S.}~\bibnamefont {Hong}}, \bibinfo {author} {\bibfnamefont
  {Y.}~\bibnamefont {Zhou}},\ and\ \bibinfo {author} {\bibfnamefont
  {Y.}~\bibnamefont {Xie}},\ }\bibfield  {title} {\bibinfo {title} {Enhanced
  superconductivity and coexisting ferroelectricity at oxide interfaces},\
  }\href {https://doi.org/10.1038/s41467-025-66903-5} {\bibfield  {journal}
  {\bibinfo  {journal} {Nat. Commun.}\ }\textbf {\bibinfo {volume} {17}},\
  \bibinfo {pages} {219} (\bibinfo {year} {2025})}\BibitemShut {NoStop}%
\bibitem [{\citenamefont {Chen}\ \emph
  {et~al.}(2021{\natexlab{b}})\citenamefont {Chen}, \citenamefont {Liu},
  \citenamefont {Sun}, \citenamefont {Chen}, \citenamefont {Liu}, \citenamefont
  {Zhang}, \citenamefont {Li}, \citenamefont {Zhang}, \citenamefont {Hong},
  \citenamefont {Ren}, \citenamefont {Zhang}, \citenamefont {Tian},
  \citenamefont {Zhou}, \citenamefont {Sun},\ and\ \citenamefont
  {Xie}}]{Chen2021_110}%
  \BibitemOpen
  \bibfield  {author} {\bibinfo {author} {\bibfnamefont {Z.}~\bibnamefont
  {Chen}}, \bibinfo {author} {\bibfnamefont {Z.}~\bibnamefont {Liu}}, \bibinfo
  {author} {\bibfnamefont {Y.}~\bibnamefont {Sun}}, \bibinfo {author}
  {\bibfnamefont {X.}~\bibnamefont {Chen}}, \bibinfo {author} {\bibfnamefont
  {Y.}~\bibnamefont {Liu}}, \bibinfo {author} {\bibfnamefont {H.}~\bibnamefont
  {Zhang}}, \bibinfo {author} {\bibfnamefont {H.}~\bibnamefont {Li}}, \bibinfo
  {author} {\bibfnamefont {M.}~\bibnamefont {Zhang}}, \bibinfo {author}
  {\bibfnamefont {S.}~\bibnamefont {Hong}}, \bibinfo {author} {\bibfnamefont
  {T.}~\bibnamefont {Ren}}, \bibinfo {author} {\bibfnamefont {C.}~\bibnamefont
  {Zhang}}, \bibinfo {author} {\bibfnamefont {H.}~\bibnamefont {Tian}},
  \bibinfo {author} {\bibfnamefont {Y.}~\bibnamefont {Zhou}}, \bibinfo {author}
  {\bibfnamefont {J.}~\bibnamefont {Sun}},\ and\ \bibinfo {author}
  {\bibfnamefont {Y.}~\bibnamefont {Xie}},\ }\bibfield  {title} {\bibinfo
  {title} {Two-dimensional superconductivity at the $\mathrm{LaAlO_3/KTaO_3}$
  (110) heterointerface},\ }\href
  {https://doi.org/10.1103/PhysRevLett.126.026802} {\bibfield  {journal}
  {\bibinfo  {journal} {Phys. Rev. Lett.}\ }\textbf {\bibinfo {volume} {126}},\
  \bibinfo {pages} {026802} (\bibinfo {year} {2021}{\natexlab{b}})}\BibitemShut
  {NoStop}%
\bibitem [{\citenamefont {Hua}\ \emph {et~al.}(2022)\citenamefont {Hua},
  \citenamefont {Meng}, \citenamefont {Huang}, \citenamefont {Li},
  \citenamefont {Wang}, \citenamefont {Ge}, \citenamefont {Xiang},\ and\
  \citenamefont {Chen}}]{Hua2022}%
  \BibitemOpen
  \bibfield  {author} {\bibinfo {author} {\bibfnamefont {X.}~\bibnamefont
  {Hua}}, \bibinfo {author} {\bibfnamefont {F.}~\bibnamefont {Meng}}, \bibinfo
  {author} {\bibfnamefont {Z.}~\bibnamefont {Huang}}, \bibinfo {author}
  {\bibfnamefont {Z.}~\bibnamefont {Li}}, \bibinfo {author} {\bibfnamefont
  {S.}~\bibnamefont {Wang}}, \bibinfo {author} {\bibfnamefont {B.}~\bibnamefont
  {Ge}}, \bibinfo {author} {\bibfnamefont {Z.}~\bibnamefont {Xiang}},\ and\
  \bibinfo {author} {\bibfnamefont {X.}~\bibnamefont {Chen}},\ }\bibfield
  {title} {\bibinfo {title} {Tunable two-dimensional superconductivity and
  spin-orbit coupling at the $\mathrm{EuO/KTaO_3}$ (110) interface},\ }\href
  {https://doi.org/10.1038/s41535-022-00506-x} {\bibfield  {journal} {\bibinfo
  {journal} {npj Quantum Mater.}\ }\textbf {\bibinfo {volume} {7}},\ \bibinfo
  {pages} {97} (\bibinfo {year} {2022})}\BibitemShut {NoStop}%
\bibitem [{\citenamefont {Bjørlig}\ \emph {et~al.}(2020)\citenamefont
  {Bjørlig}, \citenamefont {Carrad}, \citenamefont {Prawiroatmodjo},
  \citenamefont {von Soosten}, \citenamefont {Gan}, \citenamefont {Chen},
  \citenamefont {Pryds}, \citenamefont {Paaske},\ and\ \citenamefont
  {Jespersen}}]{Jespersen2020}%
  \BibitemOpen
  \bibfield  {author} {\bibinfo {author} {\bibfnamefont {A.~V.}\ \bibnamefont
  {Bjørlig}}, \bibinfo {author} {\bibfnamefont {D.~J.}\ \bibnamefont
  {Carrad}}, \bibinfo {author} {\bibfnamefont {G.~E. D.~K.}\ \bibnamefont
  {Prawiroatmodjo}}, \bibinfo {author} {\bibfnamefont {M.}~\bibnamefont {von
  Soosten}}, \bibinfo {author} {\bibfnamefont {Y.}~\bibnamefont {Gan}},
  \bibinfo {author} {\bibfnamefont {Y.}~\bibnamefont {Chen}}, \bibinfo {author}
  {\bibfnamefont {N.}~\bibnamefont {Pryds}}, \bibinfo {author} {\bibfnamefont
  {J.}~\bibnamefont {Paaske}},\ and\ \bibinfo {author} {\bibfnamefont {T.~S.}\
  \bibnamefont {Jespersen}},\ }\bibfield  {title} {\bibinfo {title} {g-factors
  in $\mathrm{LaAlO_3/SrTiO_3}$ quantum dots},\ }\href
  {https://doi.org/10.1103/PhysRevMaterials.4.122001} {\bibfield  {journal}
  {\bibinfo  {journal} {Phys. Rev. Mater.}\ }\textbf {\bibinfo {volume} {4}},\
  \bibinfo {pages} {122001} (\bibinfo {year} {2020})}\BibitemShut {NoStop}%
\bibitem [{\citenamefont {Jouan}\ \emph {et~al.}(2020)\citenamefont {Jouan},
  \citenamefont {Singh}, \citenamefont {Lesne}, \citenamefont {Vaz},
  \citenamefont {Bibes}, \citenamefont {Barthélémy}, \citenamefont {Ulysse},
  \citenamefont {Stornaiuolo}, \citenamefont {Salluzzo}, \citenamefont
  {Hurand}, \citenamefont {Lesueur}, \citenamefont {Feuillet-Palma},\ and\
  \citenamefont {Bergeal}}]{Jouan2020}%
  \BibitemOpen
  \bibfield  {author} {\bibinfo {author} {\bibfnamefont {A.}~\bibnamefont
  {Jouan}}, \bibinfo {author} {\bibfnamefont {G.}~\bibnamefont {Singh}},
  \bibinfo {author} {\bibfnamefont {E.}~\bibnamefont {Lesne}}, \bibinfo
  {author} {\bibfnamefont {D.~C.}\ \bibnamefont {Vaz}}, \bibinfo {author}
  {\bibfnamefont {M.}~\bibnamefont {Bibes}}, \bibinfo {author} {\bibfnamefont
  {A.}~\bibnamefont {Barthélémy}}, \bibinfo {author} {\bibfnamefont
  {C.}~\bibnamefont {Ulysse}}, \bibinfo {author} {\bibfnamefont
  {D.}~\bibnamefont {Stornaiuolo}}, \bibinfo {author} {\bibfnamefont
  {M.}~\bibnamefont {Salluzzo}}, \bibinfo {author} {\bibfnamefont
  {S.}~\bibnamefont {Hurand}}, \bibinfo {author} {\bibfnamefont
  {J.}~\bibnamefont {Lesueur}}, \bibinfo {author} {\bibfnamefont
  {C.}~\bibnamefont {Feuillet-Palma}},\ and\ \bibinfo {author} {\bibfnamefont
  {N.}~\bibnamefont {Bergeal}},\ }\bibfield  {title} {\bibinfo {title}
  {Quantized conductance in a one-dimensional ballistic oxide nanodevice},\
  }\href {https://doi.org/10.1038/s41928-020-0383-2} {\bibfield  {journal}
  {\bibinfo  {journal} {Nat. Electron.}\ }\textbf {\bibinfo {volume} {3}},\
  \bibinfo {pages} {201} (\bibinfo {year} {2020})}\BibitemShut {NoStop}%
\bibitem [{\citenamefont {Fukaya}\ \emph {et~al.}(2018)\citenamefont {Fukaya},
  \citenamefont {Lado}, \citenamefont {Black‑Schaffer},\ and\ \citenamefont
  {Alvarez}}]{Fukaya2018}%
  \BibitemOpen
  \bibfield  {author} {\bibinfo {author} {\bibfnamefont {Y.}~\bibnamefont
  {Fukaya}}, \bibinfo {author} {\bibfnamefont {J.~L.}\ \bibnamefont {Lado}},
  \bibinfo {author} {\bibfnamefont {A.~M.}\ \bibnamefont {Black‑Schaffer}},\
  and\ \bibinfo {author} {\bibfnamefont {G.}~\bibnamefont {Alvarez}},\
  }\bibfield  {title} {\bibinfo {title} {Interorbital topological
  superconductivity in spin‑orbit coupled superconductors},\ }\href
  {https://doi.org/10.1103/PhysRevB.97.174522} {\bibfield  {journal} {\bibinfo
  {journal} {Phys. Rev. B}\ }\textbf {\bibinfo {volume} {97}},\ \bibinfo
  {pages} {174522} (\bibinfo {year} {2018})}\BibitemShut {NoStop}%
\bibitem [{\citenamefont {Maiellaro}\ \emph {et~al.}(2023)\citenamefont
  {Maiellaro}, \citenamefont {Settino}, \citenamefont {Guarcello},
  \citenamefont {Romeo},\ and\ \citenamefont {Citro}}]{Maiellaro2023}%
  \BibitemOpen
  \bibfield  {author} {\bibinfo {author} {\bibfnamefont {A.}~\bibnamefont
  {Maiellaro}}, \bibinfo {author} {\bibfnamefont {J.}~\bibnamefont {Settino}},
  \bibinfo {author} {\bibfnamefont {C.}~\bibnamefont {Guarcello}}, \bibinfo
  {author} {\bibfnamefont {F.}~\bibnamefont {Romeo}},\ and\ \bibinfo {author}
  {\bibfnamefont {R.}~\bibnamefont {Citro}},\ }\bibfield  {title} {\bibinfo
  {title} {Hallmarks of orbital‑flavored majorana states in josephson
  junctions based on oxide nanochannels},\ }\href
  {https://doi.org/10.1103/PhysRevB.107.L201405} {\bibfield  {journal}
  {\bibinfo  {journal} {Phys. Rev. B}\ }\textbf {\bibinfo {volume} {107}},\
  \bibinfo {pages} {L201405} (\bibinfo {year} {2023})}\BibitemShut {NoStop}%
\bibitem [{\citenamefont {Stornaiuolo}\ \emph {et~al.}(2017)\citenamefont
  {Stornaiuolo}, \citenamefont {Lesne}, \citenamefont
  {Rodríguez‑Velamazán}, \citenamefont {Granger}, \citenamefont {Cantoni},
  \citenamefont {Reyren}, \citenamefont {Lesueur}, \citenamefont {Bibes},\ and\
  \citenamefont {Barthélémy}}]{Stornaiuolo2017}%
  \BibitemOpen
  \bibfield  {author} {\bibinfo {author} {\bibfnamefont {D.}~\bibnamefont
  {Stornaiuolo}}, \bibinfo {author} {\bibfnamefont {E.}~\bibnamefont {Lesne}},
  \bibinfo {author} {\bibfnamefont {J.~A.}\ \bibnamefont
  {Rodríguez‑Velamazán}}, \bibinfo {author} {\bibfnamefont
  {F.}~\bibnamefont {Granger}}, \bibinfo {author} {\bibfnamefont
  {C.}~\bibnamefont {Cantoni}}, \bibinfo {author} {\bibfnamefont
  {N.}~\bibnamefont {Reyren}}, \bibinfo {author} {\bibfnamefont
  {J.}~\bibnamefont {Lesueur}}, \bibinfo {author} {\bibfnamefont
  {M.}~\bibnamefont {Bibes}},\ and\ \bibinfo {author} {\bibfnamefont
  {A.}~\bibnamefont {Barthélémy}},\ }\bibfield  {title} {\bibinfo {title}
  {Signatures of unconventional superconductivity in the
  $\mathrm{LaAlO_3/SrTiO_3}$ two‑dimensional system},\ }\href
  {https://doi.org/10.1103/PhysRevB.95.140502} {\bibfield  {journal} {\bibinfo
  {journal} {Phys. Rev. B}\ }\textbf {\bibinfo {volume} {95}},\ \bibinfo
  {pages} {140502} (\bibinfo {year} {2017})}\BibitemShut {NoStop}%
\bibitem [{\citenamefont {Guarcello}\ \emph {et~al.}(2024)\citenamefont
  {Guarcello}, \citenamefont {Maiellaro}, \citenamefont {Settino},
  \citenamefont {Gaiardoni}, \citenamefont {Trama}, \citenamefont {Romeo},\
  and\ \citenamefont {Citro}}]{GUARCELLO2024115596}%
  \BibitemOpen
  \bibfield  {author} {\bibinfo {author} {\bibfnamefont {C.}~\bibnamefont
  {Guarcello}}, \bibinfo {author} {\bibfnamefont {A.}~\bibnamefont
  {Maiellaro}}, \bibinfo {author} {\bibfnamefont {J.}~\bibnamefont {Settino}},
  \bibinfo {author} {\bibfnamefont {I.}~\bibnamefont {Gaiardoni}}, \bibinfo
  {author} {\bibfnamefont {M.}~\bibnamefont {Trama}}, \bibinfo {author}
  {\bibfnamefont {F.}~\bibnamefont {Romeo}},\ and\ \bibinfo {author}
  {\bibfnamefont {R.}~\bibnamefont {Citro}},\ }\bibfield  {title} {\bibinfo
  {title} {Probing topological superconductivity of oxide nanojunctions using
  fractional shapiro steps},\ }\href
  {https://doi.org/https://doi.org/10.1016/j.chaos.2024.115596} {\bibfield
  {journal} {\bibinfo  {journal} {Chaos, Solitons and Fractals}\ }\textbf
  {\bibinfo {volume} {189}},\ \bibinfo {pages} {115596} (\bibinfo {year}
  {2024})}\BibitemShut {NoStop}%
\bibitem [{\citenamefont {Perroni}\ \emph {et~al.}(2019)\citenamefont
  {Perroni}, \citenamefont {Cataudella}, \citenamefont {Salluzzo},
  \citenamefont {Cuoco},\ and\ \citenamefont {Citro}}]{Perroni2019}%
  \BibitemOpen
  \bibfield  {author} {\bibinfo {author} {\bibfnamefont {C.~A.}\ \bibnamefont
  {Perroni}}, \bibinfo {author} {\bibfnamefont {V.}~\bibnamefont {Cataudella}},
  \bibinfo {author} {\bibfnamefont {M.}~\bibnamefont {Salluzzo}}, \bibinfo
  {author} {\bibfnamefont {M.}~\bibnamefont {Cuoco}},\ and\ \bibinfo {author}
  {\bibfnamefont {R.}~\bibnamefont {Citro}},\ }\bibfield  {title} {\bibinfo
  {title} {Evolution of topological superconductivity by orbital-selective
  confinement in oxide nanowires},\ }\href
  {https://doi.org/10.1103/PhysRevB.100.094526} {\bibfield  {journal} {\bibinfo
   {journal} {Phys. Rev. B}\ }\textbf {\bibinfo {volume} {100}},\ \bibinfo
  {pages} {094526} (\bibinfo {year} {2019})}\BibitemShut {NoStop}%
\bibitem [{\citenamefont {Zegrodnik}\ and\ \citenamefont
  {Wójcik}(2020)}]{Zegrodnik_2020}%
  \BibitemOpen
  \bibfield  {author} {\bibinfo {author} {\bibfnamefont {M.}~\bibnamefont
  {Zegrodnik}}\ and\ \bibinfo {author} {\bibfnamefont {P.}~\bibnamefont
  {Wójcik}},\ }\bibfield  {title} {\bibinfo {title} {Superconducting dome in
  $\mathrm{LaAlO_3/SrTiO_3}$ interfaces as a direct consequence of the extended
  s-wave symmetry of the gap},\ }\href
  {https://doi.org/10.1103/PhysRevB.102.085420} {\bibfield  {journal} {\bibinfo
   {journal} {Phys. Rev. B}\ }\textbf {\bibinfo {volume} {102}},\ \bibinfo
  {pages} {085420} (\bibinfo {year} {2020})}\BibitemShut {NoStop}%
\bibitem [{\citenamefont {Popovi\ifmmode~\acute{c}\else \'{c}\fi{}}\ \emph
  {et~al.}(2008)\citenamefont {Popovi\ifmmode~\acute{c}\else \'{c}\fi{}},
  \citenamefont {Satpathy},\ and\ \citenamefont {Martin}}]{Popovic2008}%
  \BibitemOpen
  \bibfield  {author} {\bibinfo {author} {\bibfnamefont {Z.~S.}\ \bibnamefont
  {Popovi\ifmmode~\acute{c}\else \'{c}\fi{}}}, \bibinfo {author} {\bibfnamefont
  {S.}~\bibnamefont {Satpathy}},\ and\ \bibinfo {author} {\bibfnamefont
  {R.~M.}\ \bibnamefont {Martin}},\ }\bibfield  {title} {\bibinfo {title}
  {Origin of the two-dimensional electron gas carrier density at the
  $\mathrm{LaAlO_3/SrTiO_3}$ interface},\ }\href
  {https://doi.org/10.1103/PhysRevLett.101.256801} {\bibfield  {journal}
  {\bibinfo  {journal} {Phys. Rev. Lett.}\ }\textbf {\bibinfo {volume} {101}},\
  \bibinfo {pages} {256801} (\bibinfo {year} {2008})}\BibitemShut {NoStop}%
\bibitem [{\citenamefont {Pavlenko}\ \emph {et~al.}(2012)\citenamefont
  {Pavlenko}, \citenamefont {Kopp}, \citenamefont {Tsvelik}, \citenamefont
  {Mannhart},\ and\ \citenamefont {Sawatzky}}]{Pavlenko2012}%
  \BibitemOpen
  \bibfield  {author} {\bibinfo {author} {\bibfnamefont {N.}~\bibnamefont
  {Pavlenko}}, \bibinfo {author} {\bibfnamefont {T.}~\bibnamefont {Kopp}},
  \bibinfo {author} {\bibfnamefont {A.}~\bibnamefont {Tsvelik}}, \bibinfo
  {author} {\bibfnamefont {J.}~\bibnamefont {Mannhart}},\ and\ \bibinfo
  {author} {\bibfnamefont {G.~A.}\ \bibnamefont {Sawatzky}},\ }\bibfield
  {title} {\bibinfo {title} {Magnetic and superconducting phases at the
  $\mathrm{LaAlO_3/SrTiO_3}$ interface},\ }\href
  {https://doi.org/10.1103/PhysRevB.85.020407} {\bibfield  {journal} {\bibinfo
  {journal} {Phys. Rev. B}\ }\textbf {\bibinfo {volume} {85}},\ \bibinfo
  {pages} {020407} (\bibinfo {year} {2012})}\BibitemShut {NoStop}%
\bibitem [{\citenamefont {Pentcheva}\ \emph {et~al.}(2006)\citenamefont
  {Pentcheva}, \citenamefont {Seitsonen}, \citenamefont {Uppstu}, \citenamefont
  {Miertuš}, \citenamefont {Zegenhagen},\ and\ \citenamefont
  {Scheffler}}]{Pentcheva2006}%
  \BibitemOpen
  \bibfield  {author} {\bibinfo {author} {\bibfnamefont {R.}~\bibnamefont
  {Pentcheva}}, \bibinfo {author} {\bibfnamefont {A.~P.}\ \bibnamefont
  {Seitsonen}}, \bibinfo {author} {\bibfnamefont {A.}~\bibnamefont {Uppstu}},
  \bibinfo {author} {\bibfnamefont {S.}~\bibnamefont {Miertuš}}, \bibinfo
  {author} {\bibfnamefont {J.}~\bibnamefont {Zegenhagen}},\ and\ \bibinfo
  {author} {\bibfnamefont {M.}~\bibnamefont {Scheffler}},\ }\bibfield  {title}
  {\bibinfo {title} {Charge localization or itineracy at
  $\mathrm{LaAlO_3/SrTiO_3}$ interfaces},\ }\href
  {https://doi.org/10.1103/PhysRevB.74.035112} {\bibfield  {journal} {\bibinfo
  {journal} {Phys. Rev. B}\ }\textbf {\bibinfo {volume} {74}},\ \bibinfo
  {pages} {035112} (\bibinfo {year} {2006})}\BibitemShut {NoStop}%
\bibitem [{\citenamefont {W\'ojcik}\ \emph {et~al.}(2021)\citenamefont
  {W\'ojcik}, \citenamefont {Nowak},\ and\ \citenamefont
  {Zegrodnik}}]{wojcik2021impact}%
  \BibitemOpen
  \bibfield  {author} {\bibinfo {author} {\bibfnamefont {P.}~\bibnamefont
  {W\'ojcik}}, \bibinfo {author} {\bibfnamefont {M.~P.}\ \bibnamefont
  {Nowak}},\ and\ \bibinfo {author} {\bibfnamefont {M.}~\bibnamefont
  {Zegrodnik}},\ }\bibfield  {title} {\bibinfo {title} {Impact of spin-orbit
  interaction on the phase diagram and anisotropy of the in-plane critical
  magnetic field at the superconducting $\mathrm{LaAlO_3/SrTiO_3}$ interface},\
  }\href {https://doi.org/10.1103/PhysRevB.104.174503} {\bibfield  {journal}
  {\bibinfo  {journal} {Phys. Rev. B}\ }\textbf {\bibinfo {volume} {104}},\
  \bibinfo {pages} {174503} (\bibinfo {year} {2021})}\BibitemShut {NoStop}%
\bibitem [{\citenamefont {Khalsa}\ \emph {et~al.}(2013)\citenamefont {Khalsa},
  \citenamefont {Lee},\ and\ \citenamefont {MacDonald}}]{Khalsa2013}%
  \BibitemOpen
  \bibfield  {author} {\bibinfo {author} {\bibfnamefont {G.}~\bibnamefont
  {Khalsa}}, \bibinfo {author} {\bibfnamefont {B.}~\bibnamefont {Lee}},\ and\
  \bibinfo {author} {\bibfnamefont {A.~H.}\ \bibnamefont {MacDonald}},\
  }\bibfield  {title} {\bibinfo {title} {Theory of t$_{2g}$ electron-gas rashba
  interactions},\ }\href {https://doi.org/10.1103/PhysRevB.88.041302}
  {\bibfield  {journal} {\bibinfo  {journal} {Phys. Rev. B}\ }\textbf {\bibinfo
  {volume} {88}},\ \bibinfo {pages} {041302(R)} (\bibinfo {year}
  {2013})}\BibitemShut {NoStop}%
\bibitem [{\citenamefont {Trevisan}\ \emph {et~al.}(2019)\citenamefont
  {Trevisan}, \citenamefont {Scharf}, \citenamefont {Fernandes},\ and\
  \citenamefont {Fabian}}]{Trevisan2019}%
  \BibitemOpen
  \bibfield  {author} {\bibinfo {author} {\bibfnamefont {T.~V.}\ \bibnamefont
  {Trevisan}}, \bibinfo {author} {\bibfnamefont {B.}~\bibnamefont {Scharf}},
  \bibinfo {author} {\bibfnamefont {R.~M.}\ \bibnamefont {Fernandes}},\ and\
  \bibinfo {author} {\bibfnamefont {J.}~\bibnamefont {Fabian}},\ }\bibfield
  {title} {\bibinfo {title} {Microscopic theory of electron–phonon coupling
  and pairing in $\mathrm{SrTiO_3}$ interfaces and surfaces},\ }\href
  {https://doi.org/10.1103/PhysRevResearch.1.013003} {\bibfield  {journal}
  {\bibinfo  {journal} {Physical Review Research}\ }\textbf {\bibinfo {volume}
  {1}},\ \bibinfo {pages} {013003} (\bibinfo {year} {2019})}\BibitemShut
  {NoStop}%
\bibitem [{\citenamefont {Trevisan}\ \emph {et~al.}(2020)\citenamefont
  {Trevisan}, \citenamefont {Scharf},\ and\ \citenamefont
  {Fernandes}}]{Trevisan2020}%
  \BibitemOpen
  \bibfield  {author} {\bibinfo {author} {\bibfnamefont {T.~V.}\ \bibnamefont
  {Trevisan}}, \bibinfo {author} {\bibfnamefont {B.}~\bibnamefont {Scharf}},\
  and\ \bibinfo {author} {\bibfnamefont {R.~M.}\ \bibnamefont {Fernandes}},\
  }\bibfield  {title} {\bibinfo {title} {Fluctuation-induced odd-frequency
  superconductivity at two-dimensional oxide interfaces},\ }\href
  {https://doi.org/10.1103/PhysRevResearch.2.033225} {\bibfield  {journal}
  {\bibinfo  {journal} {Phys. Rev. Research}\ }\textbf {\bibinfo {volume}
  {2}},\ \bibinfo {pages} {033225} (\bibinfo {year} {2020})}\BibitemShut
  {NoStop}%
\bibitem [{\citenamefont {Trevisan}\ \emph {et~al.}(2018)\citenamefont
  {Trevisan} \emph {et~al.}}]{Trevisan2018}%
  \BibitemOpen
  \bibfield  {author} {\bibinfo {author} {\bibfnamefont {T.~V.}\ \bibnamefont
  {Trevisan}} \emph {et~al.},\ }\bibfield  {title} {\bibinfo {title}
  {Unconventional multiband superconductivity in bulk srtio3 and interfaces},\
  }\href@noop {} {\bibfield  {journal} {\bibinfo  {journal} {Phys. Rev. Lett.}\
  }\textbf {\bibinfo {volume} {121}},\ \bibinfo {pages} {127002} (\bibinfo
  {year} {2018})}\BibitemShut {NoStop}%
\bibitem [{\citenamefont {Singh}\ \emph {et~al.}(2022)\citenamefont {Singh}
  \emph {et~al.}}]{singh2022}%
  \BibitemOpen
  \bibfield  {author} {\bibinfo {author} {\bibfnamefont {G.}~\bibnamefont
  {Singh}} \emph {et~al.},\ }\bibfield  {title} {\bibinfo {title} {Gate-tunable
  pairing channels in superconducting non-centrosymmetric oxides nanowires},\
  }\href@noop {} {\bibfield  {journal} {\bibinfo  {journal} {npj Quantum
  Mater.}\ }\textbf {\bibinfo {volume} {7}},\ \bibinfo {pages} {2} (\bibinfo
  {year} {2022})}\BibitemShut {NoStop}%
\bibitem [{\citenamefont {Altland}\ and\ \citenamefont
  {Zirnbauer}(1997)}]{Altland}%
  \BibitemOpen
  \bibfield  {author} {\bibinfo {author} {\bibfnamefont {A.}~\bibnamefont
  {Altland}}\ and\ \bibinfo {author} {\bibfnamefont {M.~R.}\ \bibnamefont
  {Zirnbauer}},\ }\bibfield  {title} {\bibinfo {title} {Nonstandard symmetry
  classes in mesoscopic normal-superconducting hybrid structures},\ }\href
  {https://doi.org/10.1103/PhysRevB.55.1142} {\bibfield  {journal} {\bibinfo
  {journal} {Phys. Rev. B}\ }\textbf {\bibinfo {volume} {55}},\ \bibinfo
  {pages} {1142} (\bibinfo {year} {1997})}\BibitemShut {NoStop}%
\bibitem [{\citenamefont {Fukui}\ \emph {et~al.}(2005)\citenamefont {Fukui},
  \citenamefont {Hatsugai},\ and\ \citenamefont {Suzuki}}]{Fukui2005}%
  \BibitemOpen
  \bibfield  {author} {\bibinfo {author} {\bibfnamefont {T.}~\bibnamefont
  {Fukui}}, \bibinfo {author} {\bibfnamefont {Y.}~\bibnamefont {Hatsugai}},\
  and\ \bibinfo {author} {\bibfnamefont {H.}~\bibnamefont {Suzuki}},\
  }\bibfield  {title} {\bibinfo {title} {Chern numbers in discretized brillouin
  zone: Efficient method of computing (spin) hall conductances},\ }\href
  {https://doi.org/10.1143/JPSJ.74.1674} {\bibfield  {journal} {\bibinfo
  {journal} {Journal of the Physical Society of Japan}\ }\textbf {\bibinfo
  {volume} {74}},\ \bibinfo {pages} {1674} (\bibinfo {year} {2005})},\ \Eprint
  {https://arxiv.org/abs/https://doi.org/10.1143/JPSJ.74.1674}
  {https://doi.org/10.1143/JPSJ.74.1674} \BibitemShut {NoStop}%
\bibitem [{\citenamefont {Ruiz}\ \emph {et~al.}(2025)\citenamefont {Ruiz},
  \citenamefont {Mateos}, \citenamefont {Tosi}, \citenamefont {Strunk},
  \citenamefont {Balseiro},\ and\ \citenamefont {Arrachea}}]{Ruiz2025}%
  \BibitemOpen
  \bibfield  {author} {\bibinfo {author} {\bibfnamefont {G.~F.~R.}\
  \bibnamefont {Ruiz}}, \bibinfo {author} {\bibfnamefont {J.~H.}\ \bibnamefont
  {Mateos}}, \bibinfo {author} {\bibfnamefont {L.}~\bibnamefont {Tosi}},
  \bibinfo {author} {\bibfnamefont {C.}~\bibnamefont {Strunk}}, \bibinfo
  {author} {\bibfnamefont {C.}~\bibnamefont {Balseiro}},\ and\ \bibinfo
  {author} {\bibfnamefont {L.}~\bibnamefont {Arrachea}},\ }\bibfield  {title}
  {\bibinfo {title} {Antichiral edge states and bogoliubov fermi surfaces in a
  two-dimensional proximity-induced superconductor},\ }\href
  {https://doi.org/10.1103/z1xc-trf1} {\bibfield  {journal} {\bibinfo
  {journal} {Phys. Rev. B}\ }\textbf {\bibinfo {volume} {112}},\ \bibinfo
  {pages} {L241409} (\bibinfo {year} {2025})}\BibitemShut {NoStop}%
\bibitem [{\citenamefont {Colom\'es}\ and\ \citenamefont
  {Franz}(2018)}]{Colomes2018}%
  \BibitemOpen
  \bibfield  {author} {\bibinfo {author} {\bibfnamefont {E.}~\bibnamefont
  {Colom\'es}}\ and\ \bibinfo {author} {\bibfnamefont {M.}~\bibnamefont
  {Franz}},\ }\bibfield  {title} {\bibinfo {title} {Antichiral edge states in a
  modified haldane nanoribbon},\ }\href
  {https://doi.org/10.1103/PhysRevLett.120.086603} {\bibfield  {journal}
  {\bibinfo  {journal} {Phys. Rev. Lett.}\ }\textbf {\bibinfo {volume} {120}},\
  \bibinfo {pages} {086603} (\bibinfo {year} {2018})}\BibitemShut {NoStop}%
\bibitem [{\citenamefont {Szumniak}\ \emph {et~al.}(2017)\citenamefont
  {Szumniak}, \citenamefont {Chevallier}, \citenamefont {Loss},\ and\
  \citenamefont {Klinovaja}}]{Szumniak}%
  \BibitemOpen
  \bibfield  {author} {\bibinfo {author} {\bibfnamefont {P.}~\bibnamefont
  {Szumniak}}, \bibinfo {author} {\bibfnamefont {D.}~\bibnamefont
  {Chevallier}}, \bibinfo {author} {\bibfnamefont {D.}~\bibnamefont {Loss}},\
  and\ \bibinfo {author} {\bibfnamefont {J.}~\bibnamefont {Klinovaja}},\
  }\bibfield  {title} {\bibinfo {title} {Spin and charge signatures of
  topological superconductivity in rashba nanowires},\ }\href
  {https://doi.org/10.1103/PhysRevB.96.041401} {\bibfield  {journal} {\bibinfo
  {journal} {Phys. Rev. B}\ }\textbf {\bibinfo {volume} {96}},\ \bibinfo
  {pages} {041401} (\bibinfo {year} {2017})}\BibitemShut {NoStop}%
\bibitem [{\citenamefont {Żeberek}\ and\ \citenamefont
  {Wójcik}({\natexlab{a}})}]{LAO_STO_data_2025}%
  \BibitemOpen
  \bibfield  {author} {\bibinfo {author} {\bibfnamefont {P.}~\bibnamefont
  {Żeberek}}\ and\ \bibinfo {author} {\bibfnamefont {P.}~\bibnamefont
  {Wójcik}},\ }\href {https://github.com/piotr-zeberek/LAOSTO_data_2025}
  {\bibinfo {title} {{LAOSTO-Data-2025}}} ({\natexlab{a}}),\ \bibinfo {note}
  {{GitHub repository}}\BibitemShut {NoStop}%
\bibitem [{\citenamefont {Żeberek}\ and\ \citenamefont
  {Wójcik}({\natexlab{b}})}]{LAO_STO_repo}%
  \BibitemOpen
  \bibfield  {author} {\bibinfo {author} {\bibfnamefont {P.}~\bibnamefont
  {Żeberek}}\ and\ \bibinfo {author} {\bibfnamefont {P.}~\bibnamefont
  {Wójcik}},\ }\href {https://github.com/piotr-zeberek/LAOSTO-Topology-MSc}
  {\bibinfo {title} {{LAOSTO-Topology-MSc}}} ({\natexlab{b}}),\ \bibinfo {note}
  {{GitHub repository}}\BibitemShut {NoStop}%
\bibitem [{\citenamefont {Rashba}\ and\ \citenamefont
  {Efros}(2003)}]{Rashba2003}%
  \BibitemOpen
  \bibfield  {author} {\bibinfo {author} {\bibfnamefont {E.~I.}\ \bibnamefont
  {Rashba}}\ and\ \bibinfo {author} {\bibfnamefont {A.~L.}\ \bibnamefont
  {Efros}},\ }\bibfield  {title} {\bibinfo {title} {Orbital mechanisms of
  electron-spin manipulation by an electric field},\ }\href
  {https://doi.org/10.1103/PhysRevLett.91.126405} {\bibfield  {journal}
  {\bibinfo  {journal} {Phys. Rev. Lett.}\ }\textbf {\bibinfo {volume} {91}},\
  \bibinfo {pages} {126405} (\bibinfo {year} {2003})}\BibitemShut {NoStop}%
\end{thebibliography}

%

\appendix

\section{Simplified Hamiltonian for $d_{xy}$ orbital}
\label{sec:A1}
Using the standard folding-down transformation, we can reduce the full Hamiltonian Eq.~(\ref{eq:Hamiltonian_general}) into the effective $2\times 2$ Hamiltonian for the $d_{xy}$ electrons. The reduced Hamiltonian is given by
\begin{equation}
    \hat{H}^{eff}_{xy} = \hat{H}_{xy} + \hat{H}_{c}(\hat{H}_{xz/yz} - E)^{-1}\hat{H}_{c}^\dagger.
    \label{Heff}
\end{equation}
where 
\begin{equation}
\hat{H}_{xy}=
\left(
\begin{array}{cc}
\epsilon^{xy}_{\mathbf{k}}  & 0\\
 0 & \epsilon^{xy}_{\mathbf{k}} 
\end{array} \right)+\frac{1}{2}g\mu_B \mathbf{B}\cdot \pmb{\sigma},
\end{equation}
and
\begin{equation}
\hat{H}_{xz/yz}=
\left(
\begin{array}{cccc}
\epsilon^{xz}_{\mathbf{k}} & 0 & 0 & 0 \\
 0 & \epsilon^{xz}_{\mathbf{k}} & 0 & 0 \\
 0 & 0 & \epsilon^{yz}_{\mathbf{k}} & 0 \\ 
 0 & 0 & 0 & \epsilon^{yz}_{\mathbf{k}} \\
\end{array} \right),
\end{equation}
where in the latter term we neglect the coupling of the bands $d_{xz} / d_{yz}$ to the magnetic field and their hybridization assuming that the kinetic and SO energy constitute the major contribution to the energy.
The coupling between the $d_{xy}$ and $d_{yz}/d_{xz}$ orbitals is given by
\begin{eqnarray}
\hat{H}_{c}&=&
\frac{\Delta_{SO}}{3}
\left(
\begin{array}{cccc}
0 & i & 0 & -1 \\
i & 0  & 1 & 0   
\end{array} \right)  \\
&+&i\Delta_{RSO}
\left(
\begin{array}{cccc}
\sin k_y & 0 & \sin k_x & 0 \\
0 & \sin k_y & 0 & \sin k_x   
\end{array} \right). \nonumber
\end{eqnarray}

Assuming that \(\Delta_E\) represents the largest energy scale in the system and taking the bottom of the \(d_{xy}\) band as the reference energy, we can expand $(\hat{H}_{xz/yz} - E)^{-1}$ from Eq.~(\ref{Heff}) in the Taylor series 
\begin{equation}
(\hat{H}_{xz/yz} - E)^{-1} = \frac{1}{\epsilon^{xz/yz}_{\mathbf{k}}+\Delta_E-E} \mathds{1}_{4\times4} \approx \frac{1}{\Delta_E}\mathds{1}_{4\times4}.
\end{equation}
Then,the effective Hamiltonian is reduced to the following form
\begin{eqnarray}
\hat{H}^{eff}_{xy}&=&\left ( \epsilon^{xz}_{\mathbf{k}} + \frac{2\Delta_{SO}\gamma}{3(1-\gamma)}\right ) \mathbf{1}_{2\times2}+\frac{1}{2}g\mu_B \mathbf{B}\cdot \pmb{\sigma} \nonumber \\ 
&+&\alpha (\sigma_y \sin k_x - \sigma _x \sin k_y)
\label{eq:Hamiltonian_k_xy}
\end{eqnarray}
where $\gamma=\Delta_{SO} / 3\Delta _E$ and $\alpha=\Delta_{SO}\Delta_{RSO}/3\Delta_E$. The last term in Eq.~(\ref{eq:Hamiltonian_k_xy}) is related to the SO coupling of the Rashba type \cite{Rashba2003} similar to that observed in semiconductors.

\section{Influence of the in-plane magnetic field}
\label{sec:A2}
\begin{figure}[!t]
\includegraphics[]{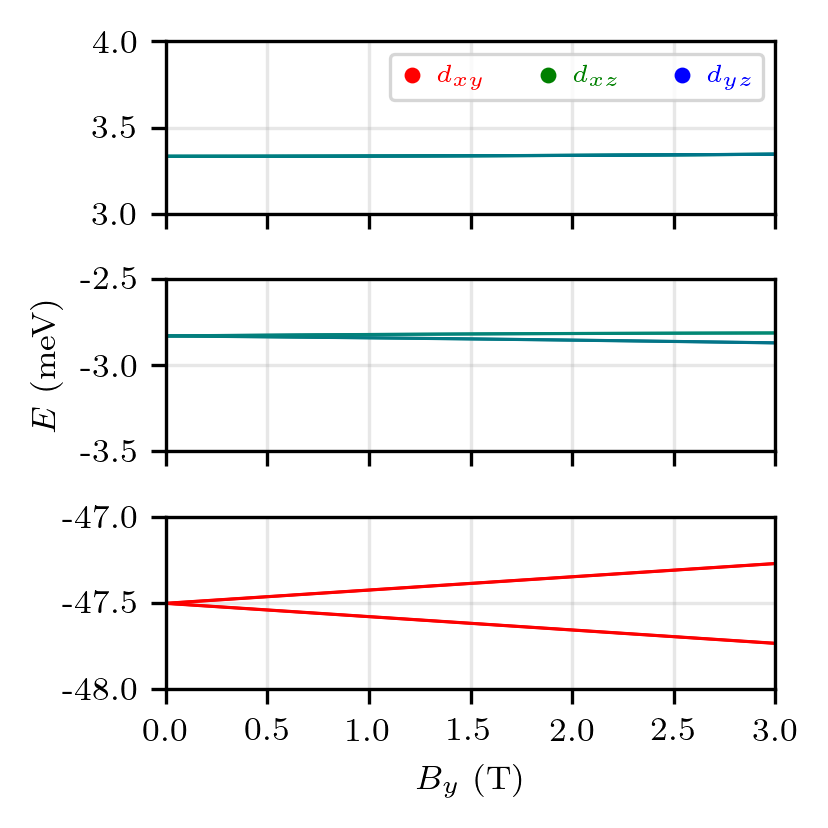}
\caption{Band minimum energy $E(k_x=0,k_y=0)$ of the LAO/STO 2DEG as a function of the in-plane magnetic field directed along the $y$-axis. The color scale represents the contributions of the individual $d$-orbitals.} 
\label{fig:B}
\end{figure}
Figure~\ref{fig:B} presents the band minimum energy $E(k_x=0,k_y=0)$ of LAO/STO 2DEG as a function of the in-plane magnetic field oriented along the $y$-axis. The color scale indicates the contributions of the individual $d$-orbitals. The response of the band minima at $\mathbf{k}=(0,0)$, where the gap closes at the topological phase transition, is clearly band dependent. A pronounced effect of the in-plane magnetic field is observed for the lowest-lying $\gamma_1$ bands, whereas the higher-lying band $\gamma_3$ exhibits almost no dependence on the magnetic field up to $B = 3$ T. 

\end{document}